\documentclass[twocolumn,preprintnumbers,superscriptaddress,longbibliography]{revtex4}

\usepackage{amssymb,amsmath,amsfonts,bm}

\usepackage{graphicx,amsmath,booktabs}%
\usepackage{color, array,amsmath,amsthm,amssymb,bm}
\usepackage{xcolor}
\usepackage[bookmarks=false]{hyperref}
\usepackage{subfiles}
\usepackage{braket}
\usepackage{hyperref}
\hypersetup{colorlinks=true,citecolor=blue,linkcolor=blue,urlcolor=blue,pdfstartview=FitH,bookmarksopen=true}

\makeatletter

\pdfpageheight\paperheight
\pdfpagewidth\paperwidth

\begin{document}

\title{Observation of discrete charge states of a coherent two-level system in a superconducting qubit}

\author{Bao-Jie Liu}
\altaffiliation{These authors contributed equally to this work}
\author{Ying-Ying Wang}
\altaffiliation{These authors contributed equally to this work}
\author{Tal Sheffer}
\altaffiliation[Present Address:~]{Departments of Physics, Yale University, New Haven, CT, USA}

\author{Chen Wang}
\email[Email: ]{wangc@umass.edu}

\affiliation{Department of Physics, University of Massachusetts-Amherst, Amherst, MA, USA}
\date{\today}%
\begin{abstract}

We report observations of discrete charge states of a coherent dielectric two-level system (TLS) that is strongly coupled to an offset-charge-sensitive superconducting transmon qubit.  We measure an offset charge of 0.072$e$ associated with the two TLS eigenstates, which have a transition frequency of 2.9 GHz and a relaxation time exceeding 3 ms. Combining measurements in the strong dispersive and resonant regime, we quantify both transverse and longitudinal couplings of the TLS-qubit interaction.  We further perform joint tracking of TLS transitions and quasiparticle tunneling dynamics but find no intrinsic correlations.  This study demonstrates microwave-frequency TLS as a source of low-frequency charge noise.

\end{abstract}
\maketitle
\def\thefootnote{*}

The ubiquitous presence of spurious microwave-frequency two-level systems (TLS) and low-frequency charge noise are two outstanding challenges for a number of solid-state quantum computing platforms~\cite{de2021materials,Paladino2014noise}.  In superconducting Josephson circuits, the TLS in aluminum oxide or other dielectric materials, broadly populating the GHz spectrum and sometimes individually resolvable, are known to cause resonant energy absorption and hence $T_1$ relaxation of the qubits~\cite{muller2019towards,martinis2005decoherence,Barends2013Coherent,carroll2022dynamics,wang2015surface}. On the other hand, the low-frequency charge noise, often with a 1/$f$ character~\cite{Weissman1988}, can be modeled by a bath of charged thermal fluctuators (TF) and can cause qubit dephasing 
at sub-Hz to kHz scale~\cite{Ku2005,Pourkabirian2014,simkins2009thermal,Gustafsson2013,de2018suppression}.  Although the phenomenological ``standard tunneling model"~\cite{phillips1987two,phillips1972tunneling} provides a tantalizing unifying picture of both effects, their microscopic origin remains elusive, and experimental evidence connecting the two effects separated by orders of magnitude in time scale remains lacking. 

From a practical point of view, a crucial advantage of superconducting circuits is the ability to design exponential insensitivity to low-frequency charge fluctuations, with the transmon being a prominent example~\cite{Koch2007Charge}.  
This leaves dielectric loss from resonant TLS defects the current bottleneck to superconducting qubit coherence and a focal point of studies~\cite{wang2015surface,mcrae2020materials,Crowley2023}.  Parameters of microwave TLS in qubits, namely their frequency, coupling strength, coherence times, etc., can be extracted from the splitting of avoided crossings observed in the qubit spectrum~\cite{neeley2008process,Shalibo2010Lifetime,Lisenfeld2010Measuring}, the qubit response to a Rabi drive~\cite{Thorbeck2023Two,Abdurakhimov222Identification}, or spectroscopic probes of the qubit $T_{1}$ times~\cite{martinis2005decoherence,Barends2013Coherent,spiecker2023two,Carruzzo2021}.  While it is often believed that most or all TLS are atomic-scale defects that couple to qubits via a charge dipole, many alternative models exist~\cite{muller2019towards,Barends2013Coherent,martinis2005decoherence,Constantin2009,Sarabi2016,Seongshik2006,Graaf2021Quantifying,burnett2014evidence}.  So far the strongest evidence on the electrical nature of TLS came from observations of their frequency-tunability via dc electric fields~\cite{Carruzzo2021,Sarabi2016,grabovskij2012strain,Hung2022}. 


Surprisingly, even for superconducting qubits understood as insensitive to both 1/$f$ charge and flux noises, slow frequency fluctuations remain prevalent~\cite{Paik2011,burnett2019decoherence,Steffen2019Correlating,levine2023demonstrating,gertler2021protecting}. This may be attributed to critical current fluctuations of the Josephson junctions~\cite{Ku2005,Constantin2007,Zaretskey2013,nugroho2013low,cole2010quantitative}. A more plausible explanation is the TLS-TF model, which effectively re-introduces low-frequency charge noise to charge-insensitive qubits through a two-step process:  The qubit is charge-coupled to a near-resonant TLS which experiences spectral diffusion via its coupling to the TF bath~\cite{burnett2019decoherence,Steffen2019Correlating,Seongshik2006,Graaf2021Quantifying,burnett2014evidence}.  A particularly striking type of noise widely encountered in otherwise highly coherent qubits is random telegraphing switching (RTS) between two precise frequencies separated typically by few to 100's kHz (see e.g.~\cite{Paik2011,Steffen2019Correlating,levine2023demonstrating,gertler2021protecting}).  
The RTS can be attributed to the dynamics of a dominant off-resonant TLS dispersively coupled to the qubit.  However, few studies identified such TLS to substantiate their correlation with the qubit noise~\cite{levine2023demonstrating}, and to the best of our knowledge no experiments so far have been able to quantify how a qubit and TLS interacts in the dispersive regime.  

In this Letter, we present unambiguous observations of the distinct charge states of a TLS in strong dispersive coupling to a superconducting qubit.  Employing a transmon qubit in the weakly-charge-sensitive regime previously used for tracking quasiparticle tunnelling and charge fluctuations~\cite{riste2013millisecond,connolly2023coexistence,Serniak2012Hot,wilen2021correlated}, our experiment identifies an offset charge shift of 0.072$e$ associated with the two TLS eigenstates.  The TLS shows an unusually long relaxation time exceeding 3 ms at 2.9 GHz, a record only matched in a recent experiment using phononic bandgap engineering~\cite{chen2023phonon}.  
From measurements of qubit-TLS coupling over up to $2$ GHz of detuning, we characterize the full Hamiltonian form of the qubit-TLS interaction. 
We further demonstrate simultaneous tracking of TLS transitions and charge-parity jumps, and find no intrinsic correlations between TLS and quasiparticle dynamics.  
Our observation of a single TLS capable of causing both GHz energy exchange and sub-kHz qubit frequency fluctuations suggests that 1/$f$ charge noise and microwave dielectric loss may indeed share a similar microscopic origin.

\begin{figure}[tbp]
	\centering
	\includegraphics[width=8.7cm]{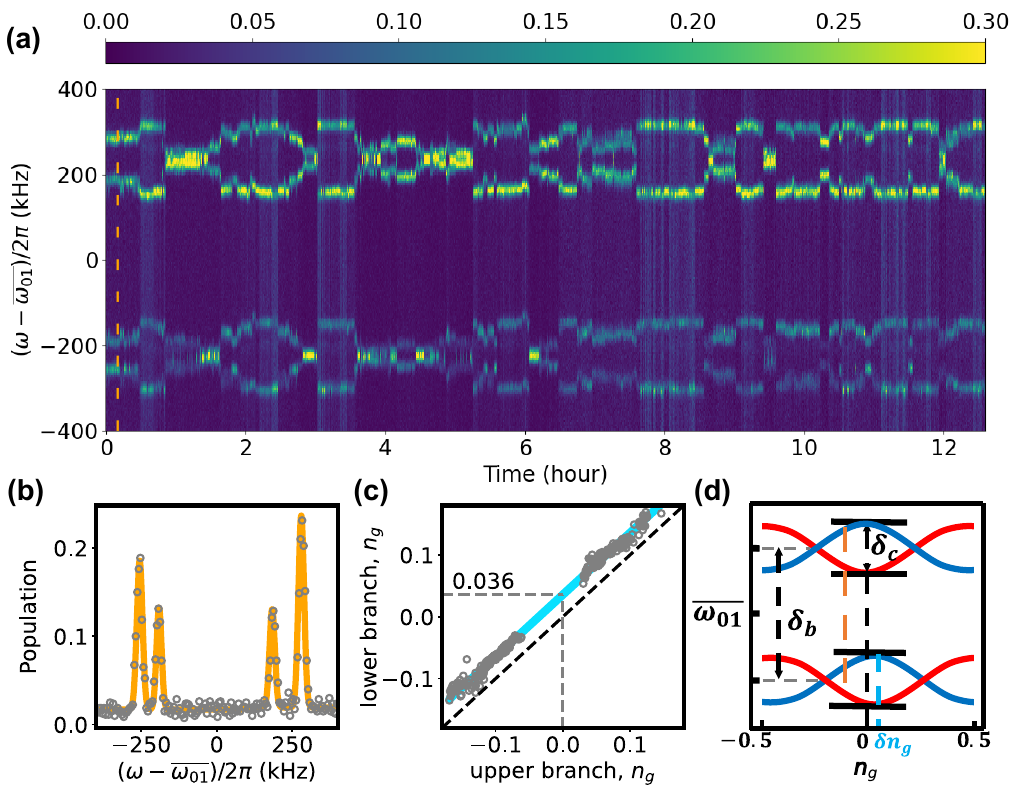}
	\caption{Observation of qubit transition frequency and offset charge fluctuations. \textbf{(a)} Spectroscopy data of the $|0\rangle$-$|1\rangle$ transition of the qubit measured repeatedly over time.  The pulse sequence is composed of a 48 $\mu$s 4$\sigma$-Gaussian $X_{\pi}$ excitation pulse, a 3 $\mu$s readout pulse and an immediate active qubit reset routine. \textbf{(b)} Qubit spectrum at a selected time in (a), fit to a sum of four Lorentzian peaks. 
    \textbf{(c)} Charge offsets $n_g$ of lower and upper branch states extracted from qubit spectroscopy, showing a difference 
    $\delta n_{g}\approx 0.036$ (in units of $2e$).  Data for $n_g$ close to 0 (mod $\frac{1}{4}$) are not included due to large uncertainties. The black dashed line is $\delta n_{g}=0$.
    \textbf{(d)} The transition-frequency diagram illustrates 
    four distinct qubit frequencies as a function of offset charge $n_{g}$.  The two frequency branches, each containing a charge dispersion of $\delta_c$ are vertically offset by $\delta_b$ and horizontally shifted by $\delta n_g$.  
	}\label{fig1}
\end{figure}

Our experiment is performed on a flux-tunable transmon qubit in a 3D copper cavity~\cite{Paik2011}. 
The average qubit frequency between $|0\rangle$ and $|1\rangle$ at zero external flux is $\overline{\omega_{01}}/2\pi=4.811$ GHz, and the qubit anharmonicity is $\alpha/2\pi=350$ MHz (see \cite{supp} for details).  
Due to a relatively small $E_{J}/E_{C}$ ratio ($35.9$) compared to typical transmons, the qubit frequency is weakly charge-dependent: $ {\omega_{01}} = \overline{\omega_{01}} + (-1)^\beta \frac{\delta_{c}}{2}\cos{(2\pi n_{g})}$, where $\beta=\pm1$ represents the charge parity, 
$\delta_{c}/2\pi=160$ kHz is the charge dispersion, and $n_{g}$ is the dimensionless offset charge (in units of $2e$), 
respectively~\cite{Koch2007Charge}. 
$n_g$ is a continuous variable that often fluctuates or drifts by order unity on the time scale of hours~\cite{riste2013millisecond} due to environmental charge shifts.  The discrete charge parity $\beta$ jumps between $1$ (even) and $-1$ (odd) whenever a quasiparticle tunnelling event occurs, typically on millisecond time scales.  Therefore, qubit spectroscopy measurements averaged over typical lab time scales of seconds to minutes are expected to show a pair of resonance frequencies, 
which fluctuate anti-symmetrically around a center-frequency $\overline{\omega_{01}}$ when repeated over longer time scales.  


The long coherence time of the qubit under study ($T_{2E}>150$ $\mu$s, compared to previous offset-charge-sensitive transmon studies~\cite{riste2013millisecond, connolly2023coexistence}) allows us to carry out high-resolution spectroscopy using long excitation pulses, 
with the results shown in Fig.~\ref{fig1}(a).  
Unexpectedly, we observe four distinct qubit frequencies in each spectroscopy sweep, which over time form two branches of the aforementioned even-odd-parity feature with the same $\delta_c=160$ kHz but a relative frequency offset of $\delta_b=460$ kHz.  In effect, the qubit frequency is shifted by a discrete amount depending on the state of a hidden switch (i.e.~a TLS) 
in addition to its dependence on offset charge and charge parity.  Crucially, we can extract the offset charge $n_g$ associated with each frequency branch from the separation of even and odd-parity peaks in the spectrum, and we observe that $n_g$ of the two branches are not equal, with an example line cut in Fig.~\ref{fig1}(b).  
Moreover, $n_g$ differs between the upper and lower branches consistently by $\delta n_g\approx0.036$ even when $n_g$ itself fluctuates significantly over time (Fig.~\ref{fig1}(c)). We illustrate this distinct $\delta n_g$ and the two branches of qubit frequency dependence on charge offset in Fig.~\ref{fig1}(d).

\begin{figure}[tbp]
	\centering
	\includegraphics[width=7.5cm]{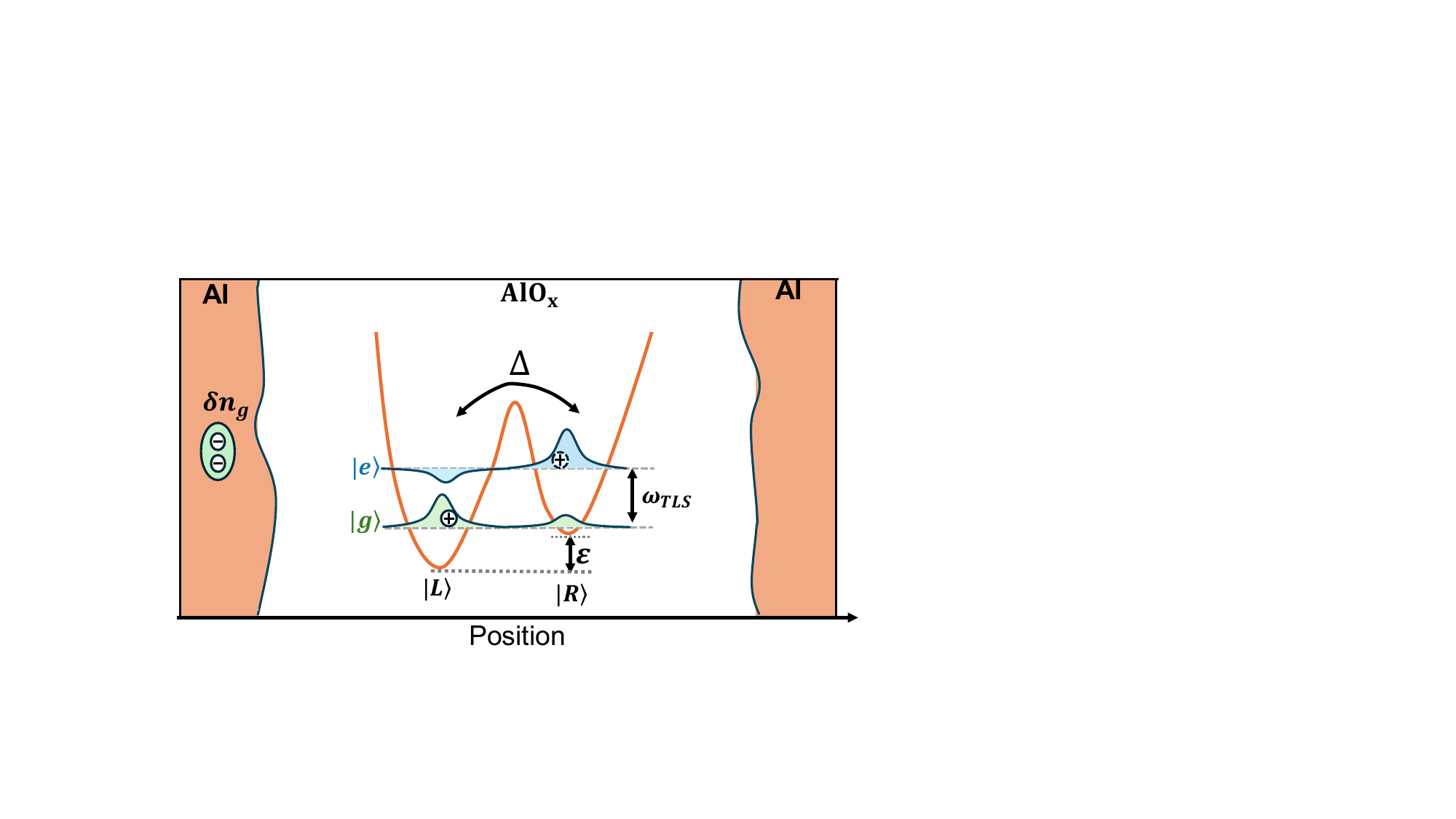}
	\caption{Schematic representation of the induced offset charge by a TLS in the Josephson junction tunnel barrier.  The TLS is modelled as a tunnelling charged particle in a double-well potential. States $|L,R \rangle$ and $|g,e \rangle$ of the TLS are in the position and the eigenstate basis, respectively. 
	}\label{fig2}
\end{figure}

The data in Fig.~\ref{fig1} provides clear evidence on the presence of two discrete charge states in the electrical environment of the qubit, but how can the charge states lead to a qubit frequency pull of $\delta_b$ greater than the charge dispersion $\delta_c$?  Below we show that the intra-branch offset-charge measurements $\delta n_g$ and the inter-branch average frequency shift $\delta_b$ provides distinct signatures of the longitudinal and transverse qubit-TLS coupling within the standard tunneling model of TLS, as illustrated in Fig.~\ref{fig2}.  
The TLS-qubit system Hamiltonian is given by, $\hat{H}=\hat{H}_{\mathrm{Q}}+\hat{H}_{\mathrm{TLS}}+\hat{H}_{\mathrm{C}} $, 
where $\frac{\hat{H}_{\mathrm{Q}}}{\hbar}=4 E_{C}(\hat{n}-n_g)^2+E_J \cos \hat{\varphi}$, 
$\frac{\hat{H}_{\mathrm{TLS}}}{\hbar}=\frac{1}{2}(\varepsilon\hat{\sigma}_z+\Delta \hat{\sigma}_x)$, and $\frac{\hat{H}_{\mathrm{C}}}{\hbar}=8 E_{C}\lambda\hat{n}\hat{\sigma}_{z}$, are the bare qubit, bare TLS and TLS-qubit interaction Hamiltonian, respectively. Here, $E_C$, $E_J$, $\hat{\varphi}$, $\hat{n}$, $\Delta$ and $\varepsilon$ are the charging energy, the Josephson energy, the phase and charge (Cooper-pair number) operators, the tunneling rate and asymmetry energy of TLS.  
The dimensionless parameter $\lambda$ characterizes the coupling between the charge dipole of the TLS and the electric field of the qubit.  Treating $H_\mathrm{C}$ as a perturbation, it is instructive to write the Hamiltonian in the eigenbasis of the TLS with transformation $\hat{\sigma}_x=\cos \theta \hat{\eta}_x+\sin \theta \hat{\eta}_z$ and $\hat{\sigma}_z=\cos \theta \hat{\eta}_z-\sin \theta \hat{\eta}_x$: 
\begin{equation}\label{eq3}
\begin{aligned}
\frac{\hat{H}}{\hbar}  = & 4E_{C}(\hat{n}-n_{g}+\lambda\cos\theta \hat{\eta}_z)^2+E_J \cos \hat{\varphi}+\frac{\omega_{\mathrm{TLS}} \hat{\eta}_z}{2} \\ 
 & -8\lambda \sin \theta E_{C}\hat{n}\hat{\eta}_x,
\end{aligned}
\end{equation}
where   $\omega_{\mathrm{TLS}}=\sqrt{\varepsilon^2+\Delta^2}$ and  $\theta=\arctan \left(\Delta / \varepsilon\right)$ are the TLS eigen-frequency and the mixing angle. 
The first term in Eq.~\eqref{eq3} indicates an offset charge shift of $\delta n_{g}=2\lambda\cos\theta$ associated with the TLS eigenstates. The last term represents a transverse  qubit-TLS coupling which allows resonant excitation swapping (if $\omega_{01}=\omega_{\mathrm{TLS}}$) as most previous studies focused on~\cite{Shalibo2010Lifetime, spiecker2023two,lisenfeld2019electric,Carruzzo2021}. Since the transmon is a multi-level system, dispersive frequency shift between the qubit and the TLS is also generally expected in the off-resonant limit, which can be estimated by $\delta_B\approx2g_{C}^2/(\overline{\omega_{01}}-\omega_{\mathrm{TLS}}-\alpha)$ under rotating wave approximation, where the Jaynes-Cumming-type coupling strength $g_{C}=8\lambda \sin \theta E_{C}\langle 0|\hat{n}|1\rangle$. The resultant qubit frequency consists of four distinct values corresponding to the ground/excited states of the TLS and the charge parity states of the qubit, as observed in our experiment.  
\begin{figure}[tbp]
	\centering
	\includegraphics[width=8.5cm]{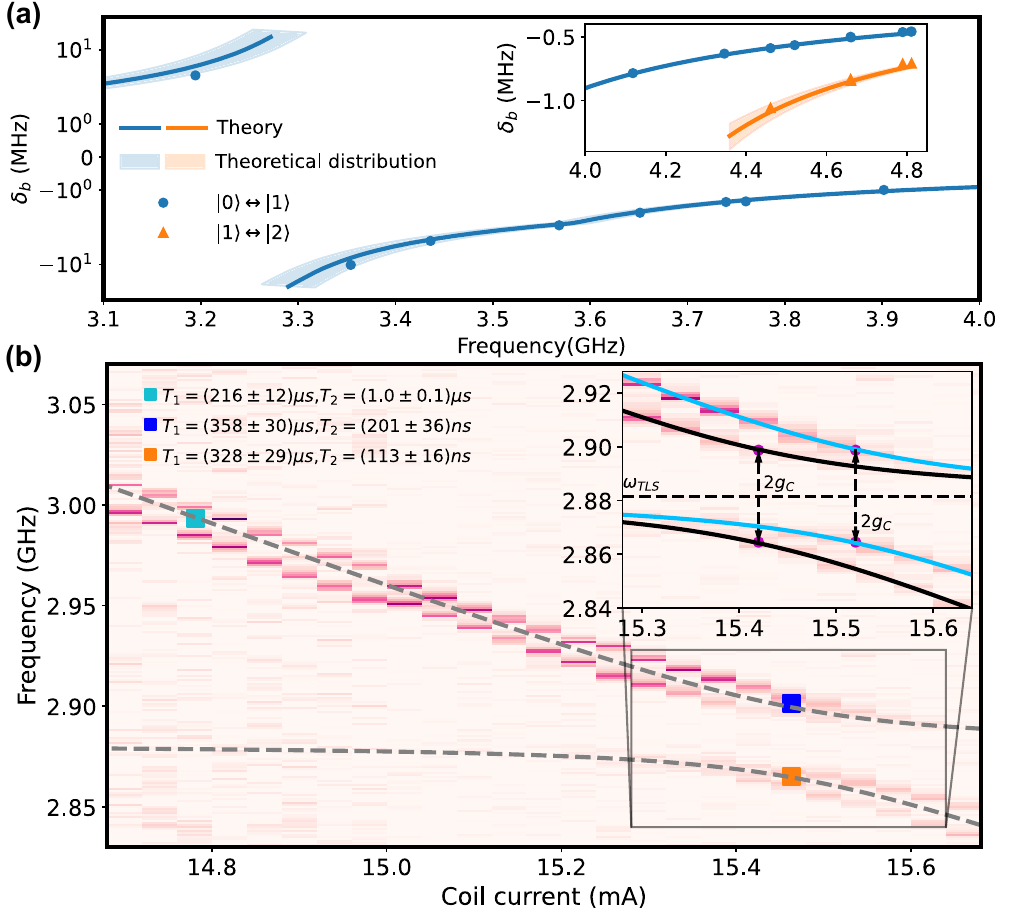}
	\caption{ \textbf{(a)} TLS-induced dispersive shift $\delta_{b}$ of $|0\rangle \leftrightarrow |1\rangle$ (circle dot) and $|1\rangle \leftrightarrow |2\rangle$ (triangle dot) transitions as a function of the transmon frequency $\overline{\omega_{01}}/2\pi$. Solid line is the theoretical model prediction (solid line), with the shadow around the line corresponding to the variation of $\delta_{b}$ considering the qubit charge offset distribution. \textbf{(b)} The qubit spectrum is measured as a function of the current applied to the magnetic flux coil. The TLS-qubit coupling $g_{C}/2\pi=17.5$ MHz and frequency of TLS $\omega_{\mathrm{TLS}}/2\pi = 2.881 $ GHz are extracted by fitting two set of the avoided level crossing, shown in the inset figure. Measured relaxation ($T_1$) and coherence ($T_{2}$) times are labeled for selected points of the spectrum.}\label{fig3}
\end{figure}
\begin{figure}[tbp]
	\centering
	\includegraphics[width=8.7cm]{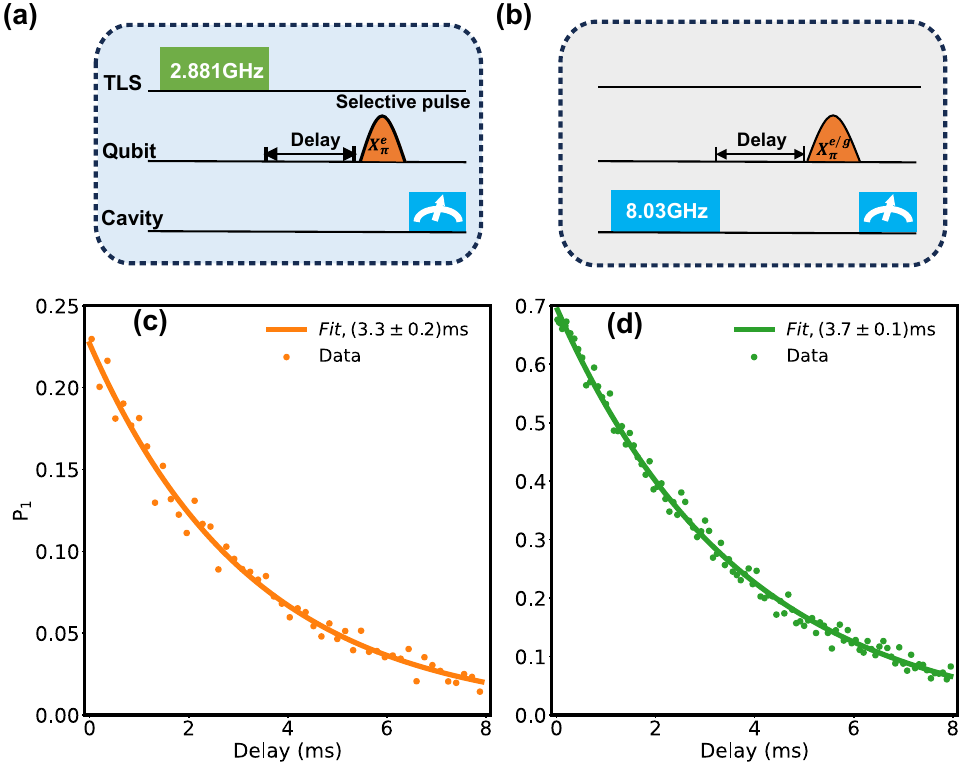}
	\caption{Pulse sequences of the TLS $T_{1}$ measurement, where two different excitation drives, \textbf{(a)} $2.881$ GHz and \textbf{(b)} $8.030$ GHz, are used.  Using the qubit as a sensor, a selective $X^{e}_{\pi}$ ($X^{g}_{\pi}$) pulse is applied to map the population of TLS state $|e\rangle$ ($|g\rangle$) to the population of qubit excited state.  Relaxation of the excited state probability of the TLS following the two excitation methods are shown in \textbf{(c)} and \textbf{(d)}, respectively.  The solid lines are exponential fits. 
	}\label{fig4}
\end{figure}

\begin{figure*}[tbp]
\centering
\includegraphics[width=17.5cm]{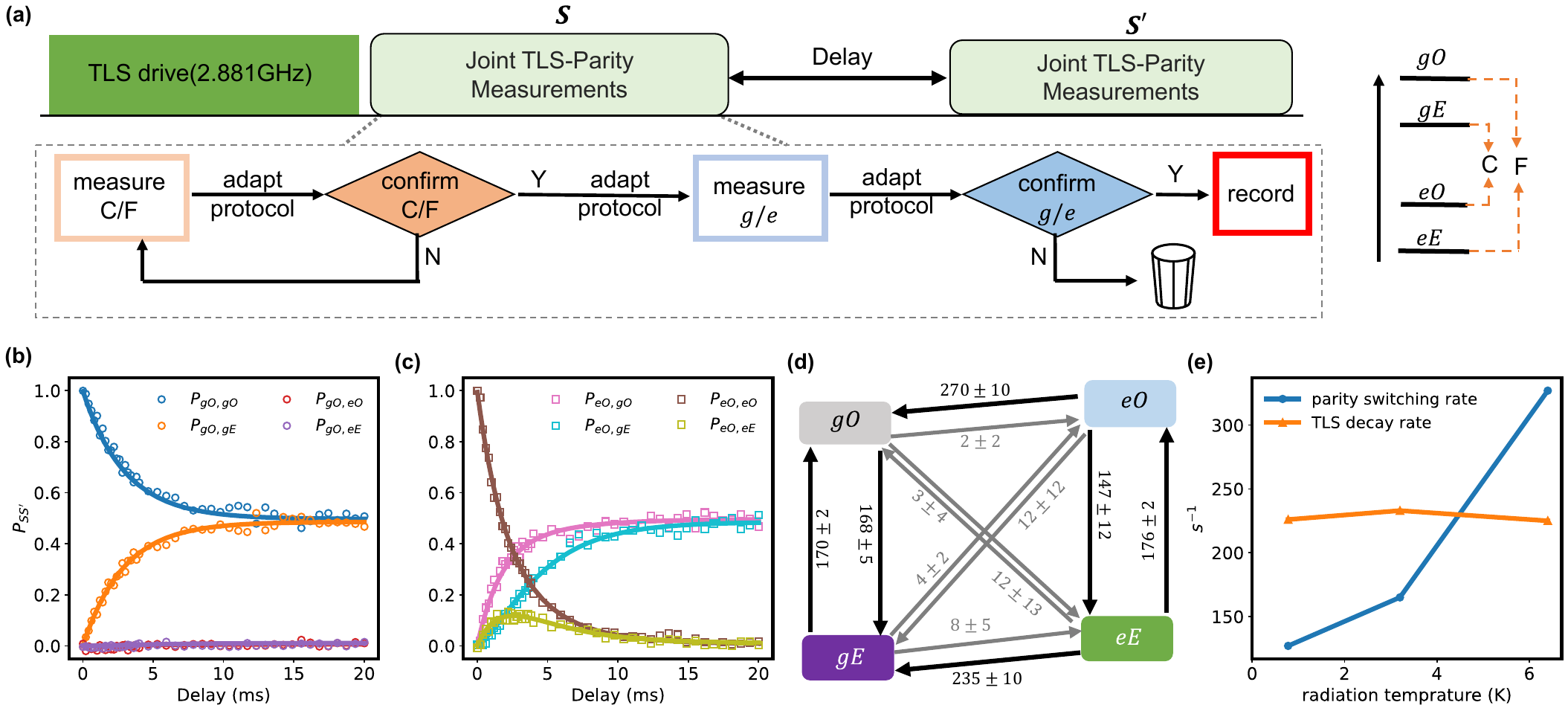}
\caption{\textbf{(a)} The joint TLS state and charge parity (TLS-parity) tracking protocol for a variable delay time. 
Each joint TLS-parity measurement is composed of two sets of FPGA-enabled adaptive Ramsey protocols, which first determines and confirms whether the system is in the `C' ($gE,eO$) or `F' ($gO,eE$) states (close or far from the center frequency), and then determines and confirms whether the system is in $g$ or $e$.  The protocol is only applied when $n_g$ is close to 0 (modulo $\frac{1}{2}$) and allows for about 90\% readout fidelity of the joint TLS-parity state.  A readout confusion matrix has been characterized and used to renormalize the measurement statistics.  See~\cite{supp} for details of the protocol.
\textbf{(b, c)} Probability $P_{SS'}$ of the joint TLS-parity being in state $S'\in\{gO, gE, eO, eE\}$ following a variable delay time after being initially in state $S=gO$ (b) and $S=eO$ (c).  (Results for $S=gE$ and $S=eE$ are similar~\cite{supp}.) 
\textbf{(d)} Transition rate diagram for the four joint TLS-parity states. Rates in unit of $s^{-1}$ are extracted from simultaneous fits of all $P_{SS'}$ with error bars obtained from bootstrapping. 
\textbf{(e)} Parity switching rate and TLS decay rate under different amount of external IR radiation as set by the temperature of a radiating resistor in the cryostat.}
\label{fig5}
\end{figure*}

To quantitatively compare our data to the TLS model above, 
we measured the qubit-TLS dispersive shift 
as a function of qubit frequency (Fig.~\ref{fig3}(a)). The increased dispersive shift and the divergent feature around $3.3$ GHz is a signature of the transmon ${1}$-${2}$ transition approaching the TLS resonance. As anticipated, we observe avoided level crossing in qubit spectroscopy at further lower frequencies, as shown in Fig.~\ref{fig3}(b). The inset shows the fit of two sets of avoided crossing structures corresponding to different charge-parity states, which gives the TLS frequency $\omega_\mathrm{TLS}/2\pi=2.881$ GHz and the coupling strength $g_C/2\pi=17.5$ MHz. 
By measuring the relaxation and coherence time of the hybridized mode (Fig.~\ref{fig3}(b)), we estimate the intrinsic $T_{2}$ of the TLS to be approximately $100$ ns and $T_{1}$ significantly longer than the qubit. 
From the measured $\delta n_g$ and $g_C$, we obtain the tunneling rate $\Delta/2\pi = 1.331$ GHz, asymmetry energy $\varepsilon /2\pi = 2.555$ GHz for this TLS.  
Our data quantitatively agrees with the theoretical prediction (solid line in Fig.~\ref{fig3}(a)) from the charge-coupled Hamiltonian (Eq.~(\ref{eq3})).  Importantly, the broad range of qubit-TLS detunings that we can access in Fig.~\ref{fig3}(a) allows us to distinguish different operator forms of qubit-TLS interaction.  The measured dispersive shift is inconsistent with flux-based TLS coupling models such as critical current fluctuations of the Josephson junctions~\cite{Constantin2007,Zaretskey2013,nugroho2013low} and magnetic flux fluctuations in the SQUID loop~\cite{muller2019towards}, and is also inconsistent with the TLS-TF model~\cite{Steffen2019Correlating,burnett2019decoherence} (see \cite{supp} for details).  Finally, the measured charge coupling strength strongly suggests that the TLS resides in the Josephson junction tunnel barrier, whose dipole moment along the electric field of the qubit mode is estimated as $p_z \approx 5.8$ Debye ($1.2$ $e\AA$) assuming a $1.5$-nm barrier thickness, consistent with the range of earlier measurements~\cite{Sarabi2016,Hung2022}.  We also note that our measurement of $\delta n_g$ is valid up to modulo 0.5, but taking any value other than 0.036 would result in unphysically large $p_z$ which is extremely unlikely. 

To measure the intrinsic $T_{1}$ of the TLS, we operate the transmon at the zero-flux point, where the coupling between the TLS and qubit is minimal. We have two methods to partially excite the TLS: (1) a strong drive (Fig.~\ref{fig4}(a)) at its resonance frequency, which remains $\omega_{\mathrm{TLS}}/2\pi=2.881$ GHz at zero external flux as verified through a TLS spectroscopy measurement; (2) a strong and long readout pulse (Fig.~\ref{fig4}(b)) (see~\cite{supp} for details). Using qubit as an ancilla, we are able to read out the TLS state by applying a selective $X_{\pi}^{g/e}$ pulse. The two methods give consistent TLS $T_{1}$ exceeding $3$ ms (Fig.~\ref{fig4}(c,d)), which surpassed typical TLS lifetime ($0.1 - 10$ $\mu$s)~\cite{neeley2008process,Shalibo2010Lifetime,Lisenfeld2010Measuring} by several orders of magnitude. This lifetime is only 
matched by the longest-living TLS in a recent study~\cite{chen2023phonon} employing phononic bandgap engineering. Since TLS relaxation is known to be dominated by phonon-mediated dissipation~\cite{phillips1972tunneling,chen2023phonon,odeh2023non}, our study shows that some charged TLS may naturally have suppressed coupling to the phonon baths, whose mechanisms will require future study. 

In addition to the phonon baths, 
quasiparticles (QP) in the electrode~\cite{Bilmes2017electronic} and QP tunneling through Josephson junctions~\cite{chen2023phonon} are also suggested as TLS relaxation channels.  Here we implement a technique to extract the correlation between QP tunneling and TLS decay/excitation through joint TLS-parity measurements. 
The TLS state and the qubit parity can be distinguished through the instantaneous qubit frequency, 
and determined from two bits of information obtained from Ramsey-like measurements (Fig.~\ref{fig5}(a) and see~\cite{supp} for details).
We partially excite the TLS with a resonant drive, and record the initial joint TLS-parity state $S$ right after the preparation and the state $S^\prime$ after a variable delay time as shown in Fig.~\ref{fig5}(a). 
The conditional probabilities $P_{SS'}$ of measuring $S'$ under initial state of $S$ after variable delay is plot in Fig.~\ref{fig5}(b,c).
To model the transitions between each states, we consider a master equation, $\dot{\boldsymbol{\rho}}=\boldsymbol{\Gamma} \boldsymbol{\rho}$, where $\boldsymbol{\Gamma}$ is a $4\times 4$ transition-rate matrix, 
and $\boldsymbol{\rho}$ is a 4-vector representing the population in $\{gO, gE, eO, eE\}$.
The fitting results are plotted as solid lines in Fig.~\ref{fig5}(b,c) and the extracted transition rates (Fig.~\ref{fig5}(d)) indicates the TLS parity switching time around 6 ms and TLS relaxing time around 4 ms. 
The low TLS-parity correlated jumping rates (diagonal arrows in Fig.~\ref{fig5}(d)) 
show no clear evidence of the QP tunneling induced TLS decay/excitation. 
We further applied IR radiation in the qubit environment through a resistive heater to intentionally increase the QP density by about 3 fold, 
and observed 
no change to the TLS decay rate, as shown in Fig.~\ref{fig5}(e) (see \cite{supp} for protocol and analysis).  This test shows that the TLS lifetime even at millisecond level is not limited by QPs in the electrodes. 

In conclusion, our measurements of both longitudinal and transverse coupling of a TLS to the qubit charge operator provide direct evidence to the existence of a coherent tunnelling charge within the Al/AlO$_x$/Al Josephson junction.  The distinctive offset-charge signature provides a self-calibrated measure of the microscopic charge displacement of the TLS, which traverses about $8\%$ ($=2\delta n_g/\cos\theta$) of the junction barrier thickness assuming a charge of $e$.  

The TLS is responsible for a sub-kHz RTS in the qubit frequency through coherent dispersive interaction.  
We note that the equilibrium excited-state population of the TLS in this study is below the measurement noise floor of $\sim1\%$ (Figs.~\ref{fig4} and \ref{fig5}(d), consistent with a temperature of $<30$ mK at $\sim$$2.9$ GHz).  Its salient signature in qubit spectroscopy (Fig.~\ref{fig1}) is related to spurious excitation of the TLS by qubit readout.  However, it is plausible that similar long-lived dispersively-coupled TLS also exists at lower frequencies (e.g.,~$0.1$-$1$ GHz), whose thermal excitations would directly lead to RTS of 10's of kHz without invoking TLS-TF interactions.  Due to the large TLS-qubit detuning, this type of RTS tends to be persistent and consistent for weeks or months, which is commonly observed but usually cured stochastically by thermal cycling in most studies (and hence left undocumented when the study is not noise-focused, with exceptions such as Refs.~\cite{Paik2011,levine2023demonstrating,gertler2021protecting}).  See~\cite{supp} for further discussions.  \\


{\it Acknowledgments}. This work was supported by the U.S. Department of Energy, Office of Science, National Quantum Information Science Research Centers, Co-design Center for Quantum Advantage (C$^2$QA) under contract number DE-SC0012704. Transmon qubits were fabricated and provided by the SQUILL Foundry at MIT Lincoln Laboratory, with funding from the Laboratory for Physical Sciences (LPS) Qubit Collaboratory.  We thank D. Rosenstock for experimental assistance.  We thank colleagues in the C$^2$QA Device thrust for helpful discussions.  We thank MIT Lincoln Laboratory for providing the Josephson travelling wave parametric amplifier for our measurements.

\clearpage
\onecolumngrid

\pagebreak
	\begin{center}
	\textbf{\large Supplementary Material for ``Observation of discrete charge states of a coherent two-level system in a superconducting qubit"}
\end{center}

\setcounter{equation}{0}
\setcounter{figure}{0}
\setcounter{table}{0}
\setcounter{page}{1}
\makeatletter
\renewcommand{\theequation}{S\arabic{equation}}
\renewcommand{\thefigure}{S\arabic{figure}}
\renewcommand{\bibnumfmt}[1]{[S#1]}
\renewcommand{\citenumfont}[1]{S#1}


\section{Experimental device and setup}
\begin{figure*}[!thtbp]
	\centering
	\includegraphics[width=17cm]{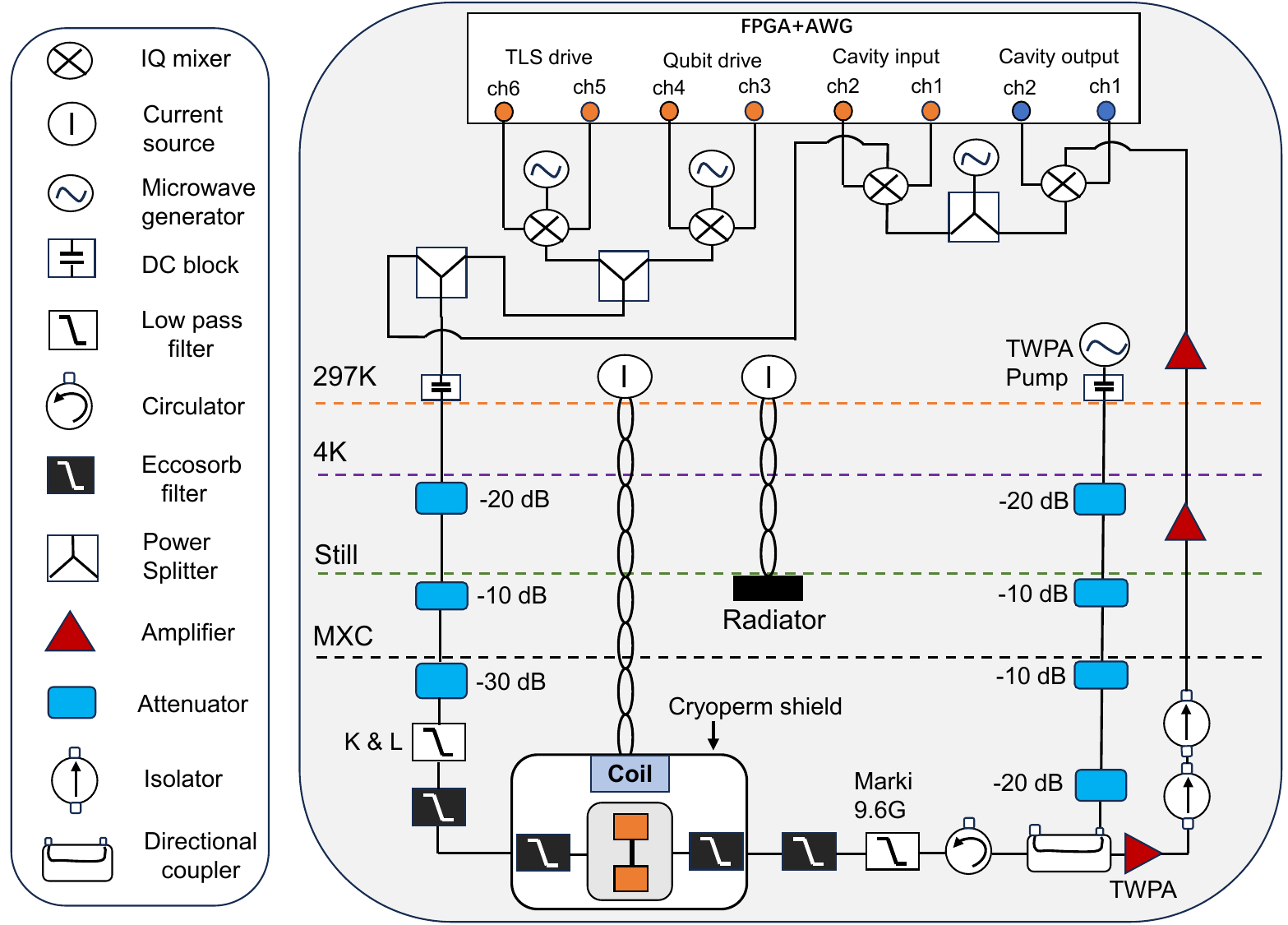}
	\caption{Diagram of the room-temperature microwave setup and cryogenic setup. }\label{Figs1}
\end{figure*}

The experimental setup incorporates a circuit quantum electrodynamics (cQED) architecture, featuring a 3D copper cavity for readout and a tunable transmon, similar to that in previous study~\cite{sPaik2011}. 
The transmon is fabricated by MIT Lincoln Laboratory on double-polished c-plane sapphire, with its aluminum junction is fabricated through Dolan bridge technology with a critical current density of 0.494 $\mu A/\mu m^{2}$. Room-temperature microwave and cryogenic setups are illustrated in Fig. \ref{Figs1}. Our experiments are conducted within a Bluefors LD250 dilution refrigerator, positioned at the mixing chamber (MXC) stage, maintaining a base temperature of approximately $10$ mK. IQ modulation generates coherent control signals for both the readout cavity and the transmon. We utilize a Traveling Wave Parametric Amplifier (TWPA) at the MXC stage and a High-Electron-Mobility Transistor (HEMT) amplifier at the $4$ K stage. To shield the experiment from external magnetic fields, it is housed in an Amuneal shield. RF lines are filtered using a K\&L $12$ GHz low-pass filter, a Marki $9.6$ GHz low-pass filter, and Eccosorb low-pass filters. A $10$ MHz reference signal is employed to phase-lock all radio-frequency (RF) instruments, including the ADC. For fast data acquisition, we use the OPX+ system in conjunction with microwave sources (Lab Brick and Windfreak) to generate microwave pulses. Magnetic flux control is achieved through a stabilized voltage source (Yokogawa $7651$). The control and coherent parameters of the device in this experiment are listed in Table~\ref{table1}.

\begin{table}[!thtbp]
\centering 
\caption{Summary of the parameters of measured transmon qubit, readout cavity, and TLS defect. The parameters of the qubit are measured at the zero-flux point.}
\begin{tabular}{|*{7}{>{\centering\arraybackslash}p{2.5cm}|}}
\hline
\multicolumn{7}{|c|}{Qubit parameters} \\
\hline
$\overline{\omega_{01}}/2\pi$ (GHz) & $\overline{\omega_{12}}/2\pi$ (GHz) & $E_J/h$ (GHz) & $E_C/2\pi$ (GHz) & $T_{1}$ ($\mu$s) & $T_{2}$ ($\mu s$) & $T_{2E}$ ($\mu$s) \\ 
\hline
4.811  & 4.461 & 10.88 & 0.303 & 147 & 80 & 157  \\
\hline
\multicolumn{7}{|c|}{TLS parameters} \\
\hline
$\omega_{\mathrm{TLS}}/2\pi$ (GHz) & $\Delta/2\pi$(GHz) & $\epsilon/2\pi$ (GHz) & $p_{z}$\footnote{assuming a $1.5$-nm barrier thickness} (D) & $g_{C}/2\pi$ (MHz) & $T_{1}$ (ms) & $T_{2}$ (ns) \\ 
\hline
2.881  & 1.331 & 2.555 & 5.8 & 17.5 & 3-4 &  $100$  \\
\hline
\multicolumn{7}{|c|}{Cavity parameters} \\
\hline
$\omega_{c}/2\pi$ (GHz)  & $ \chi/2\pi$ (MHz)  & $ \kappa/2\pi$ (MHz)&  \multicolumn{2}{|c|}{readout photon number $\overline{n}$ } & readout length ($\mu$s) & readout fidelity  \\
\hline
8.030 & 0.78 & 1.2 & \multicolumn{2}{|c|}{24} & 3 & 95-98\%  \\
\hline
\end{tabular}
\label{table1}
\end{table}

\section{Discrete dispersive frequency shift Fitting}

\begin{figure*}[tbp]
	\centering
	\includegraphics[width=18cm]{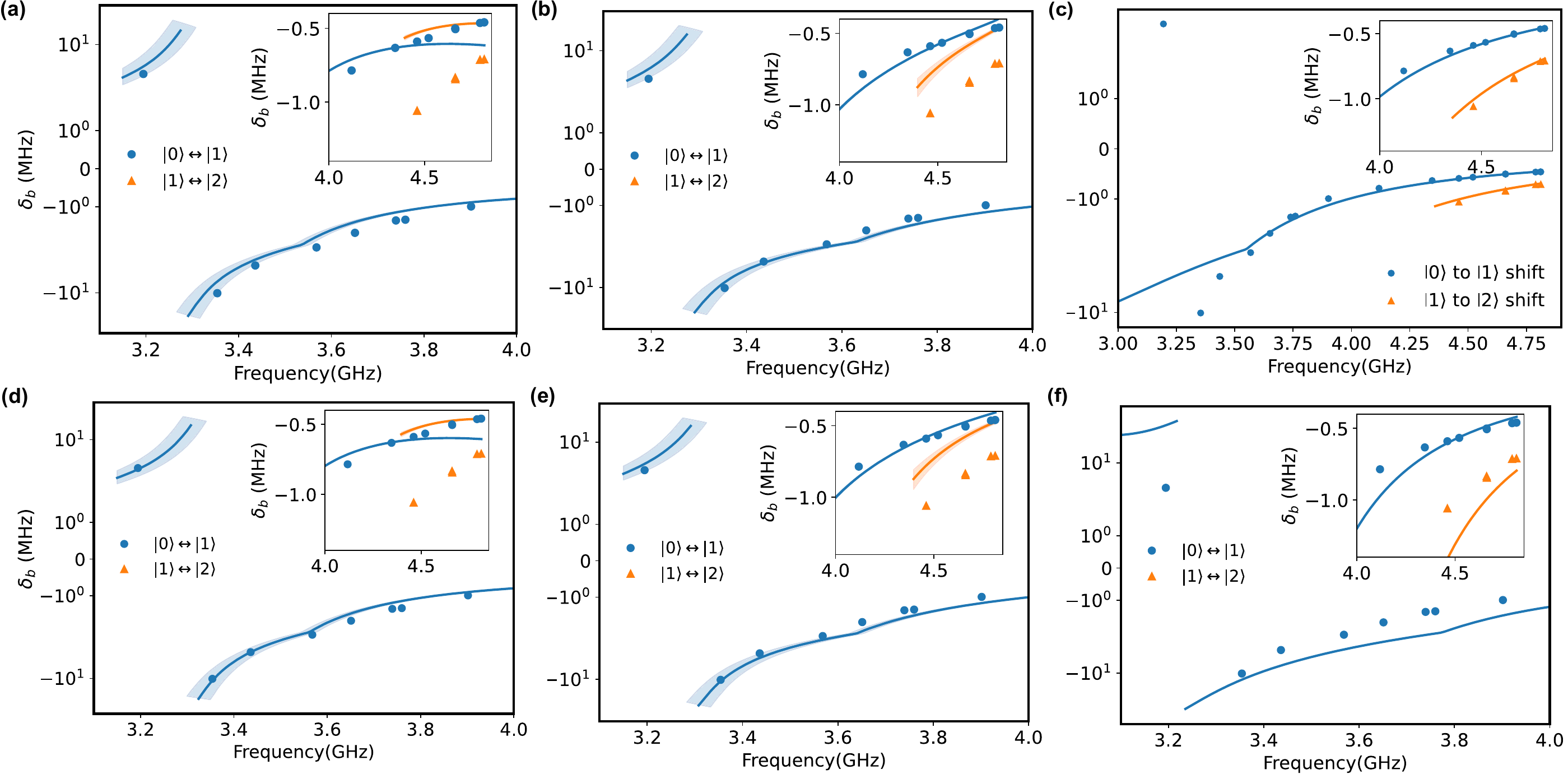}
	\caption{Models of \textbf{(a, d)} critical current fluctuation, \textbf{(b, e)} magnetic flux fluctuation, and \textbf{(c, f)} TLS-TF coupling are employed to fit dispersive frequency shifts of $\delta_{b}$ of $|0\rangle \leftrightarrow |1\rangle$ (circle dot) and $|1\rangle \leftrightarrow |2\rangle$ (trangle dot), with the shadow around the line corresponding to the variation of $\delta_{b}$ considering the qubit charge offset fluctuation caused by environment. The frequencies of the TLS fitting lines in \textbf{(a, b, c)} are fixed at $\omega_{\mathrm{TLS}}/2\pi=2881$ MHz. Detailed information regarding all fitting parameters is provided in Table \ref{table2}.} \label{Figs2}
\end{figure*}

In this section, we provide more information about theoretical predictions of charged qubit-TLS model in Sec.~\ref{chargedTLS}. On other hand, we show other mechanisms including critical current fluctuations in Josephson junctions (Sec.~\ref{current}), magnetic flux fluctuations in the SQUID (Sec.~\ref{flux}), and TLS-TF model (Sec.~\ref{TLSTF}), can't adequately fit the experimental data of the dispersive frequency shifts, as shown in Fig.~\ref{Figs2}. And the fitting parameters are illustrated in Table~\ref{table2}. In addition, only the transverse coupling (refer to $\theta=\pi/2$) between the qubit and TLS is considered, leading to a more accurate fitting of discrete dispersive frequency shifts for certain flux bias conditions than including both transverse and longitudinal couplings. Here, our protocol presents a more powerful tool that not only distinguishes between linear and non-linear coupling between qubit and TLS, but also provides insights into the identification of the charge operator $\hat{n}$ and flux operator $\hat{\varphi}$ type coupling of TLS, compared to the previous study~\cite{sAbdurakhimov222Identification}.

\subsection{Charge fluctuation model}\label{chargedTLS}

We have extensively discussed the charge fluctuations of the qubit-TLS coupling model in the main text. In this section, our focus is on providing a detailed introduction to utilizing this model for fitting discrete dispersive frequency shifts. The full Hamiltonian is given by,
\begin{equation}\label{SS1}
 \frac{\hat{H}}{\hbar}=\sum^{N}_{i=1}(\omega_{0i}+\delta^{i}_{C})|i\rangle\langle i|+\frac{\omega_{\mathrm{TLS}} \hat{\eta}_z}{2}+J_{C}\hat{n}\hat{\eta}_x, 
\end{equation}
where $\delta^{i}_{C}$ is charge dispersion of $i$ energy level, and $ J_{C}=8 \lambda \sin \theta E_C $ is the coupling strength 
connects with $g_{C}$ defined in main text through $g_{C} = J_{C} \bra{0}\hat{n}\ket{1}$.
Here, we calculate the charge operator $\hat{n}$ and high-level charge dispersion using SCqubit package~\cite{sgroszkowski2021scqubits} for theoretical prediction.

\subsection{Critical current fluctuation model}\label{current}

Critical current fluctuations of the Josephson junction will lead to a coupling between the TLS and a superconducting circuit through a modification of the Josephson energy in the Hamiltonian~\cite{smuller2019towards}. Based on these, the full Hamiltonian is given by,
 \begin{equation}\label{SS2}
 \frac{\hat{H}}{\hbar} =4 E_C\left(\hat{n}-n_g\right)^2-\left[E_{J 0}-\delta E_{J 0} (\cos \theta \hat{\eta}_{z} +\sin\theta \hat{\eta}_x)\right] \cos \left(\boldsymbol{\varphi}-\frac{\varphi_{e x t}}{2}\right)-E_{J 1} \cos \left(\boldsymbol{\varphi}+\frac{\varphi_{\text {ext }}}{2}\right) \ .
 \end{equation}
If we examine the simplest scenario involving a system with two junctions, each possessing identical Josephson energy ($E_{J0}=E_{J1}=E_{J}/2$), we can simplify the Hamiltonian.
 \begin{equation}\label{SS3}
 \frac{\hat{H}}{\hbar}=\sum^{N}_{i=1}(\omega_{0i}+\delta^{i}_{C})|i\rangle\langle i|+\frac{1}{2}\omega_\mathrm{TLS} \hat{\eta}_z-J_{C} (\cos \theta \hat{\eta}_{z} +\sin\theta \hat{\eta}_x)\left(\cos \varphi_{\text {ext }} \cos \boldsymbol{\varphi}+\sin \varphi_{\text {ext }} \sin \boldsymbol{\varphi}\right) \ ,
 \end{equation}
where $J_{C}=\delta E_{J}/2$, $\varphi_{\text {ext }}$ is the external flux, $\theta$ is mixing angle corresponding to the TLS asymmetry energy and tunnelling rate. 

\begin{table}[htb]
\centering 
\caption{Fitting parameters of dispersive frequency shifts in Fig.~\ref{Figs2}(a-f).}
\begin{tabular}{@{} p{3cm} *{6}{>{\centering\arraybackslash}p{2.3cm}} @{}}
\hline
\hline
Parameters & (a) & (b) & (c) & (d) & (e) & (f) \\
\hline
$\omega_{\mathrm{TLS}}/2\pi$ (GHz) & 2.881 & 2.881 & 2.881 & 2.914 & 2.900 & 3.232 \\
$J_{C}/2\pi$ (MHz) & 46.3 & 50.5 & 289.4 & 45.8 & 51 & 63.4 \\
$\theta$ & 1.573 & 1.577 & & 1.574 & 1.571 & \\
$\delta\omega_{\mathrm{TLS}}/2\pi$ (GHz) & & & -0.020 & & & -0.300 \\
\hline
\hline
\end{tabular}
\label{table2}
\end{table}
\subsection{Magnetic flux fluctuation model}\label{flux}

Fluctuations in the magnetic field threading SQUID loops within quantum circuits may arise from the presence of magnetic impurities on their surfaces. These loops, commonly employed in superconducting electronics, serve as an effective method for tuning the Josephson energy through the application of an external magnetic field.
After consideration of magnetic flux fluctuation, the Hamiltonian is given by,
\begin{equation}\label{SS4}
\begin{aligned}
 \frac{\hat{H}}{\hbar} & =4 E_C\left(\hat{n}-n_g\right)^2-E_{J 0} \cos [\hat{\varphi}-\frac{\varphi_{e x t}}{2}-\delta \varphi_{e x t} (\cos \theta \hat{\eta}_{z} +\sin\theta \hat{\eta}_x)]-E_{J 1} \cos(\varphi+\frac{\varphi_{e x t}}{2}) \\
& \approx 4 E_C(\hat{n}-n_g)^2-E_{J} \cos \varphi_{e x t}\cos\hat{\varphi}-\frac{E_{J}\delta \varphi_{e x t}}{2} (\cos \theta \hat{\eta}_{z} +\sin\theta \hat{\eta}_x) \sin (\hat{\varphi}-\frac{\varphi_{e x t}}{2}) \ .
\end{aligned}
\end{equation}
In the number state basis, we can express it as, 
\begin{equation}\label{SS5}
 \frac{\hat{H}}{\hbar}=\sum^{N}_{i=1}(\omega_{0i}+\delta^{i}_{C})|i\rangle\langle i|+\frac{1}{2}\omega_{T L S} \hat{\eta}_z-J_{C} \left(\cos \theta \hat{\eta}_{z} +\sin\theta \hat{\eta}_x\right) (\sin \frac{\varphi_{\text {ext }}}{2} \cos \hat{\varphi}-\cos \frac{\varphi_{\text {ext }}}{2} \sin \hat{\varphi}) \ ,
 \end{equation}
where the coupling strength is $J_{C}=E_{J}\delta \varphi_{e x t}/2$, and $\delta\varphi_{e x t}$ denotes the magnetic flux fluctuation.

\subsection{TLS-TF model}\label{TLSTF}

Considering a incoherent thermal fluctuators coupled to TLS, the Hamiltonian of qubit-TLS-TF model described by,
\begin{equation}\label{SS6}
 \frac{\hat{H}}{\hbar}=\sum^{N}_{i=1}(\omega_{0i}+\delta^{i}_{C})|i\rangle\langle i|+\frac{(\omega_{\mathrm{TLS}}+\delta\omega_{\mathrm{TLS}}\hat{v}_{z}\hat{\eta}_z)} {2}+J_{C}\hat{n}\hat{\eta}_x, 
\end{equation}
where $\delta\omega_{\mathrm{TLS}}\hat{v}_{z}\hat{\eta}_z$ is the coupling term between TLS and TF. For calculating the discrete dispersive frequency shift, we make the assumption that the TLS state is consistently in the ground state.  In accordance with this assumption, the dispersive frequency shift is given by, $\delta_{B}=(E_{1ge}-E_{0ge})-(E_{1gg}-E_{0gg})$, where $E_{1ge}$ represents the energy associated with the qubit $|1\rangle$ state, TLS ground state and TF excited state. 

\begin{figure*}[tbp]
	\centering
	\includegraphics[width=15cm]{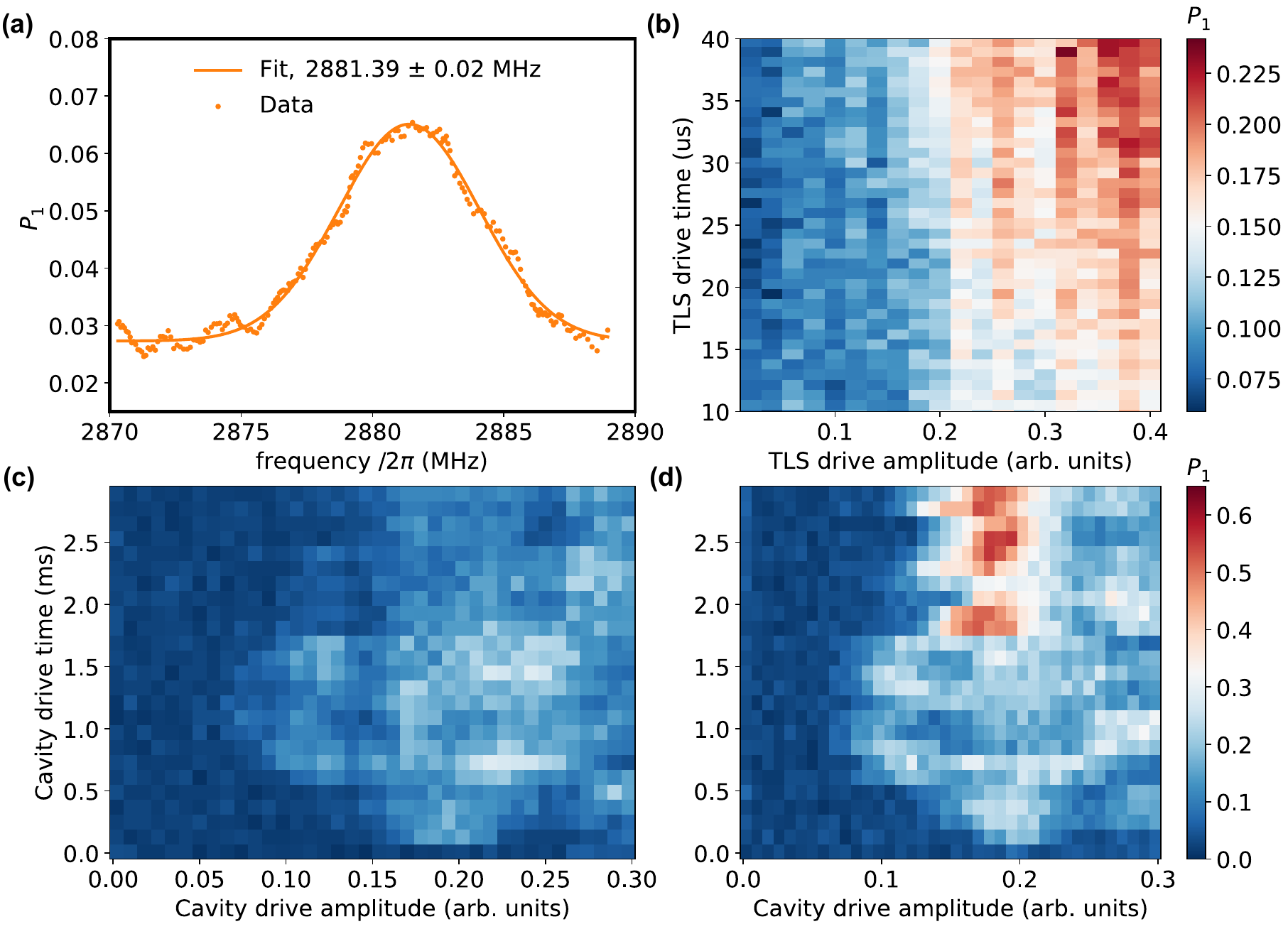}
	\caption{TLS control at zero-flux point. \textbf{(a)} TLS spectroscopy by sweeping the frequency of the drive pulse. The solid line is Lorenzian fit. Applying a selective $X_{\pi}^{g/e}$ pulse,  the TLS state can be read out by using a qubit as an ancilla.   \textbf{(b)} TLS excitation  as a function of TLS drive time and amplitude.  \textbf{(c)} Qubit excitation  and \textbf{(d)} TLS excitation as a function of cavity drive time and amplitude, respectively. }\label{Figs4}
\end{figure*}

\section{TLS control at zero-flux point}
\label{sec: TLS_drive}
At the zero-flux point, precise control over the TLS is achievable using a strong TLS drive pulse. The frequency of the TLS, determined as $\omega_{\mathrm{TLS}}/2\pi=2.8814$ GHz through TLS spectroscopy (depicted in Fig.~\ref{Figs4}(a)), agrees with the results obtained from the avoided crossing. Fig.~\ref{Figs4}(b) illustrates the exploration of time and amplitude parameters of a drive pulse to manipulate the TLS state.  Utilizing the qubit as an ancilla, we successfully read out the TLS state by employing a selective $X^{e}_{\pi}$ ($X^{g}_{\pi}$) pulse. This pulse is applied at the frequency of ~$\overline{\omega_{01}}-\delta_{b}/2$ ($\overline{\omega_{01}}+\delta_{b}/2$), with a 4-$\mu$s Gaussian envelope, to effectively map the population of the TLS excited (ground) state to the population of the qubit excited state. 
The challenge impeding direct TLS driving is the limited drive strength, resulting from using the cavity input port for both qubit and TLS drive.
On the other hand, we unexpectedly observe an increase in the probability of the lower branch occurrence with extended readout lengths and higher amplitudes, as deduced from qubit spectroscopy measurements. Here, we partially 
excite TLS 
by applying a long drive at the cavity frequency. In this context, we vary the time and amplitude of the cavity drive as function of qubit and TLS excitation (refer to Figs.~\ref{Figs4}(c) and (d)). A clear distinction emerges when comparing Fig.~\ref{Figs4}(c) to Fig.~\ref{Figs4}(d), indicating that the TLS can indeed be driven by this cavity drive. Following the consideration of the impact of the high-level state, we employ a $2.4$ ms-long cavity drive to excite the TLS for the measurements of TLS lifetime shown in the main text. That explains the appearance of lower branch in main text Fig.~1(a) while the TLS thermal population is around 1\%.

\section{Joint TLS - parity measurement}
\subsection{Distinguish joint TLS - parity state}
\label{subsec:mapping}

\begin{figure*}[tbp]
	\centering
	\includegraphics[width=14cm]{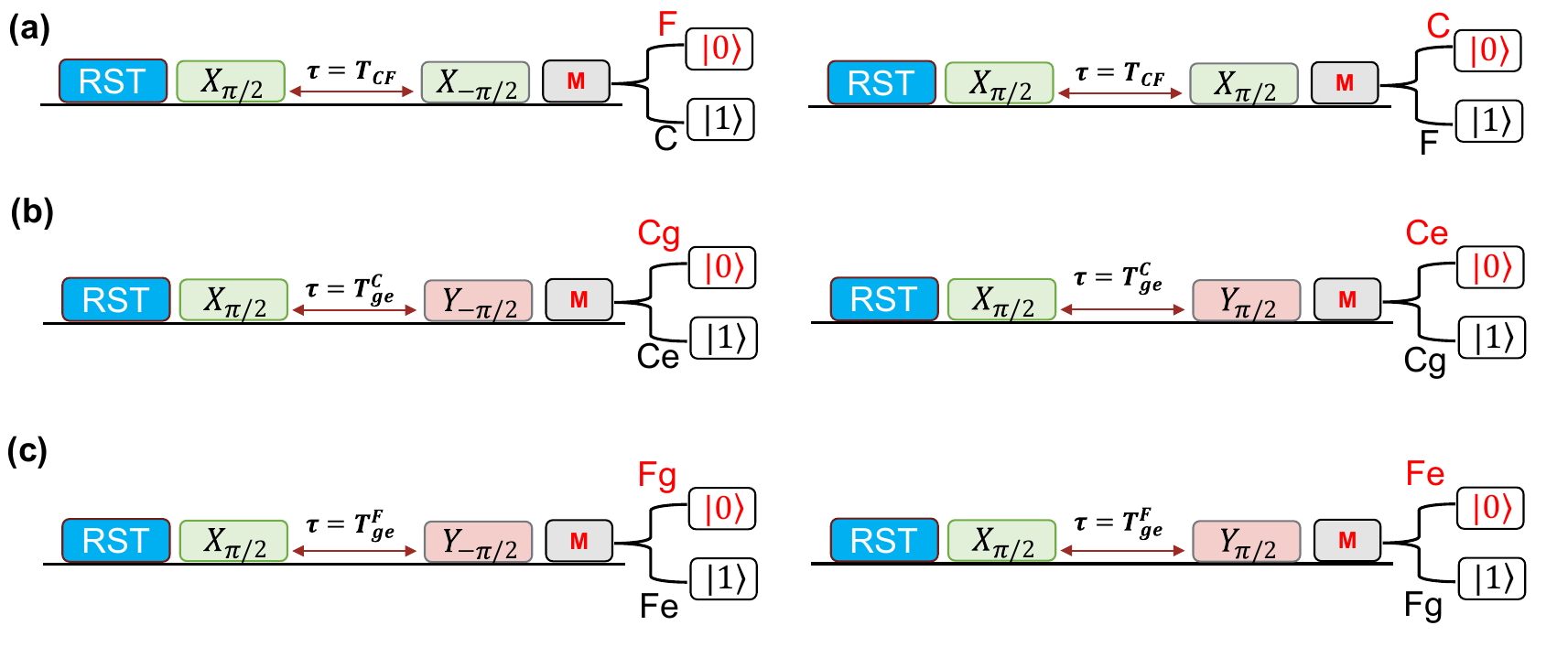}
	\caption{Illustrates the mapping protocol for distinguishing four branch states. A Ramsey-like sequence is employed to map the qubit to C-F (close-far) and TLS g-e (ground-excited) detectors. These pulses are comparable to a qubit flip conditioned on: \textbf{(a)} $F$ or $C$ states, \textbf{(b)} $Cg$ or $Ce$ states, and \textbf{(c)} $Fg$ or $Fe$ states.}\label{Fig:mapping}
\end{figure*}

By selecting the time when $n_g$ is close to 0 in the zero-flux condition, we have $\delta_b/3 \approx \delta_C \cos(2\pi n_g)$. We set the drive at their average frequency, where the far states $F (gO, eE)$ detune from the drive by $\Delta_F = \frac{1}{2}\delta_b + \frac{1}{2}\delta_C \approx 2 \delta_C$, and the close states $C (gE, eO)$ detune from the drive by $\Delta_C = \frac{1}{2}\delta_b - \frac{1}{2}\delta_C \approx \delta_C$. With the condition $\Delta_F \approx 2\Delta_C$, we distinguish the close states $C$ versus the far states $F$ using the $X_{\pi/2}$ - idle - $X_{\pi/2}$ (or $-X_{\pi/2}$) pulse sequence, denoted as $D_C$ (or $D_F$), with a drive frequency at $\overline{\omega_{01}}$ and idle time $T_{CF} = \frac{1}{2\times\Delta_F}$ (Fig.~\ref{Fig:mapping} (a)). During the idle time, the far states $F(gE, eO)$ rotate by $2\pi$ while close states rotate by $\pi$. We map the close states to $|0\rangle$ (or $|1\rangle$) and the far states to $|1\rangle$ (or $|0\rangle$) through the second $X_{\pi/2}$ pulse (or $-X_{\pi/2}$). With the $C/F$ state determined, the distinction between TLS in $g$ $(gO, gE)$ versus $e$ $(eO, eE)$ is similar to a regular parity measurement through a Ramsey-like sequence. The different sequences are labeled as $D_g$ (or $D_e$), with idle time $T_{ge}^C = \frac{1}{4\times\Delta_C}$ and $T_{ge}^F = \frac{1}{4\times\Delta_F}$ for close and far states, respectively, as shown in Fig.~\ref{Fig:mapping} (b, c). During the idle time, the states for which the TLS is in $g$ rotate $+\pi/4$, and $-\pi/4$ for the TLS being in $e$. We map TLS in $g$ to $|0\rangle$ (or $|1\rangle$) and TLS in $e$ to $|1\rangle$ (or $|0\rangle$) through the second $Y_{\pi/2}$ pulse (or $-Y_{\pi/2}$). Combining the results from the above measurements allows us to distinguish the joint TLS-parity state.

Given the stability of the qubit $|0\rangle$ state compared to the $|1\rangle$ state, our inclination is to map the TLS-parity state to the qubit $|0\rangle$ state to mitigate mapping errors attributable to qubit decay. With the FPGA protocol, we have the capability to dynamically adjust the measurement protocol in real-time, ensuring that our measurements consistently map to the qubit $|0\rangle$ state. Specifically, in the measurement aimed at distinguishing the close state $C(gE, eO)$ versus the far state $F(gO, eE)$, if we initiate with the $D_C$ pulse and find the qubit in the $|1\rangle$ state, signifying a tendency towards far states in TLS-parity, we reset the qubit to the $|0\rangle$ state and switch to the measuring protocol $D_F$ pulse. Consequently, the current TLS-parity state, representing far states, is effectively mapped to the qubit $|0\rangle$ state.

\subsection{Joint TLS - parity measurement and analyze}

\begin{figure*}[tbp]
	\centering
	\includegraphics[width=17cm]{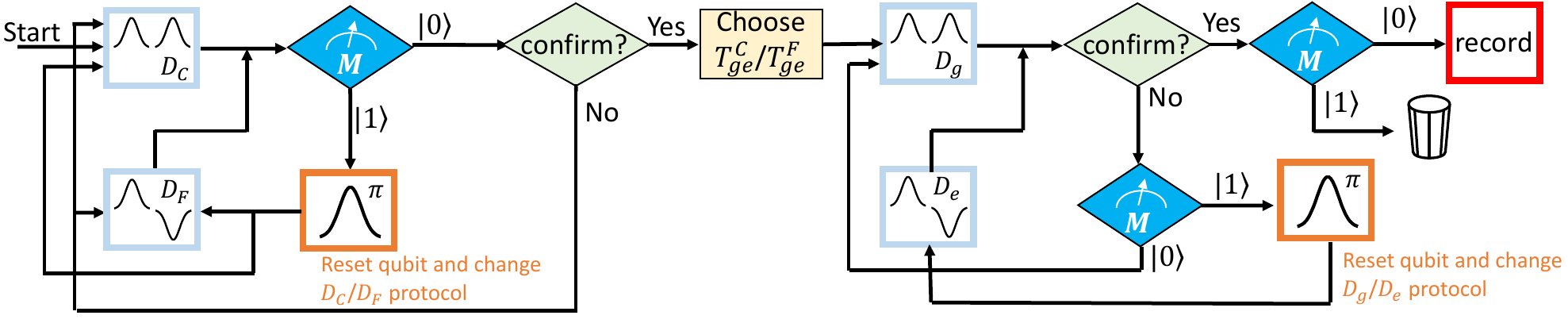}
	\caption{Flowchart illustrating the joint TLS-parity measurement process. The procedure begins by distinguishing the $C/F$ state, initiating with the $D_C$ pulse. If the readout yields $\ket{0}$, the same $D_C$ protocol is maintained for the confirmation step. Upon obtaining another $\ket{0}$ during confirmation, the $C$ state is confirmed, progressing to the subsequent step. Detection of a $\ket{1}$ during the $C/F$ distinguishing prompts a reset of the qubit to $\ket{0}$ and an active switch between $D_C$ and $D_F$ protocols, initiating a restart of the measure-and-confirm process. Following the confirmation of the $C/F$ state, the delay time $T_{ge}^C / T_{ge}^F$ is determined in the $D_g$ (or $D_e$) pulse. The TLS in $g/e$ state is then to be distinguished, starting with the $D_g$ pulse. Depending on the result being $\ket{0}$ or $\ket{1}$, the state is confirmed with the same $D_g$ protocol or a switch is made to $D_e$. The final step involves verifying if the confirmation process yields a $\ket{0}$ readout consistent with the same TLS state as the initial measurement. Results are recorded when the agreement is affirmative, and results are discarded otherwise. }\label{Fig:delay_flowchart}
\end{figure*}

With the protocols for mapping and changing logic, shown in Sec.~\ref{subsec:mapping}, we establish the joint TLS - parity monitoring as shown in the Fig.~\ref{Fig:delay_flowchart}.  We first distinguish between the $C/F$ state by actively changing between the $D_C$ and $D_F$ measurement protocol depending on the readout measurement result. If a qubit $\ket{1}$ is detected, the protocol is swapped and confirmed by a measurement of a qubit $|0\rangle$. With the confirmed $C/F$ state, the delay time, $T_{ge}^C / T_{ge}^F$, is chosen for the $D_g$ and $D_e$ pulse sequences. The TLS being in $g/e$ state with 
a similar measure-and-confirm process. To avoid false positives, the TLS state is confirmed, checking if a repeated process results in a qubit $\ket{0}$ readout, agreeing with the first TLS state measurement. Upon confirmation, the $C/F$ and the TLS $g/e$ results are then recorded. If the confirmation failed, the results are discarded. 

\begin{figure*}[tbp]
	\centering
	\includegraphics[width=17.5cm]{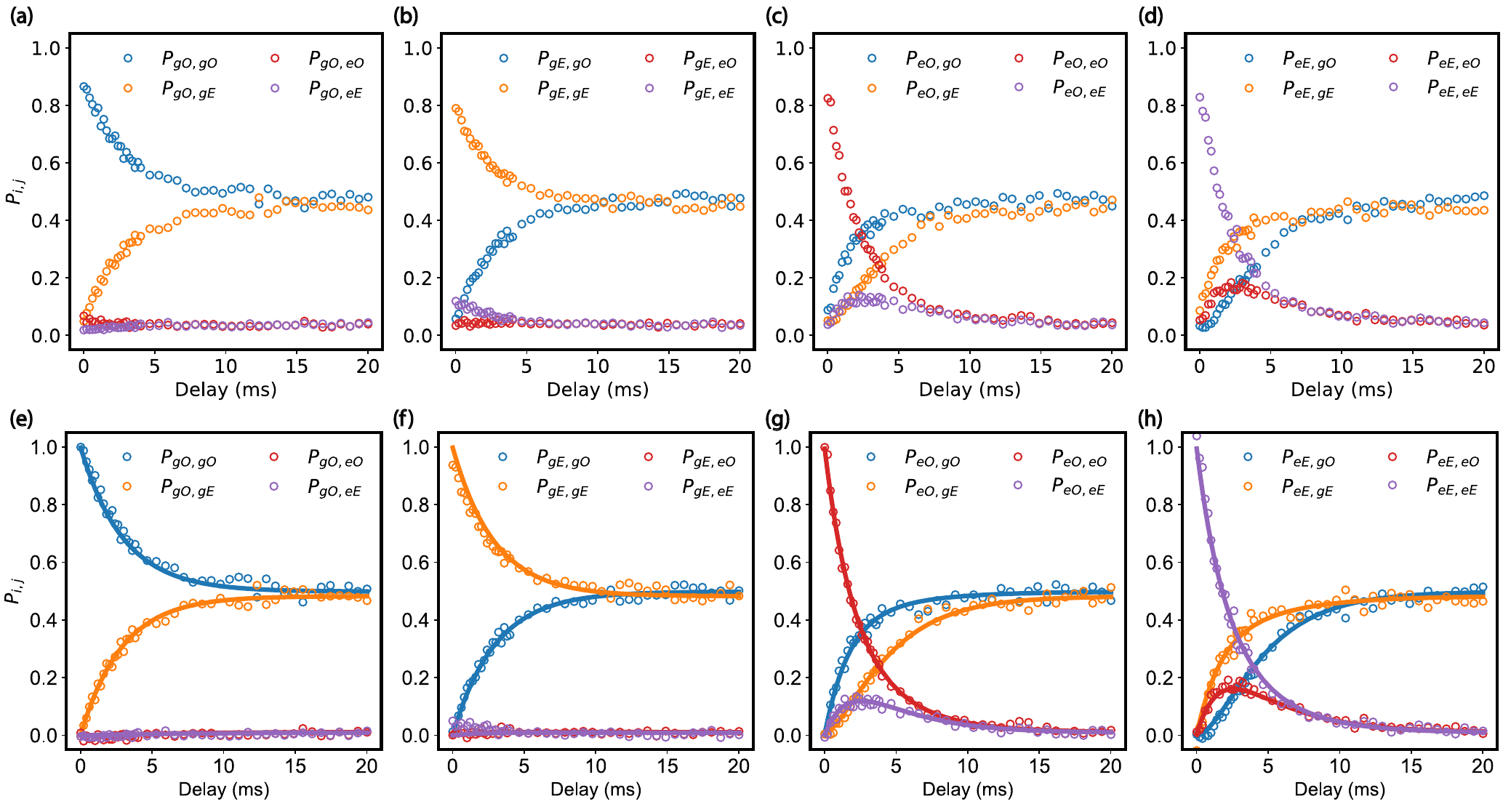}
	\caption{The conditional probability $P_{ij}$ ($P_{ij}$ is also noted as $P_{SS'}$). \textbf{(a-d)} plot the raw $P_{ij}$, with $i$  ($S$)  the initial measurement and $j$  ($S'$) the second measurement result, directly derived from the measurement results. With data in each initial state plotted in the same figure. \textbf{(e-h)} display the corrected $P_{ij}$ given by the confusion matrix $M$. The solid line is the fitting result with the transition-rate matrix $\boldsymbol{\Gamma}$.}\label{Fig:Pij}
\end{figure*}

With all of the recorded $S$ and $S'$ states, we note a matrix $A$ with $A_{ij}(t)$ representing the possibility of finding $S$ in state $i$ and $S'$ in state $j$ at any given delay time $t$, with $i,j$ in $\{gO,gE,eO,eE\}$, and $\sum_{i,j} A_{ij}(t) = 1$. The conditional probabilities $P_{SS'}$ (which we noted here as $P_{ij}$ for simplicity and consistency with other notations) of measuring $S'$ under initial state of $S$ after variable delay (Fig.~\ref{Fig:Pij}(a)-(d)), can be easily computed with:
\begin{equation}
\label{eq:normalize_Pij}
\begin{aligned}
P_{ij} = \frac{A_{ij}}{\sum_j A_{ij}},
\end{aligned}
\end{equation}
taking the assumption that for the $t = 0$ point no state transition will happen, which is reasonable since it's two consecutive measurements with the measurement duration being around 7 $\mu s$, much smaller than the parity jumping and TLS decay/excitation time scales. In this case, the real intrinsic value for $A$, which we note as $A_r$, contains only diagonal terms and follows:
\begin{equation}
\begin{aligned}
A_r(0) = diag(\boldsymbol{\rho}),
\end{aligned}
\end{equation}
with $\boldsymbol{\rho}$ is a 4-vector representing the population in $\{gO, gE, eO, eE\}$.

Considering the readout errors that could lead to the misassignment of states, denoted by the confusion matrix $M$ where $M_{ij}$ represents the probability of state $i$ being assigned to state $j$, the raw $A$ and the read intrinsic $A_{r}$ can be readily linked through the confusion matrix $M$ as follows:
\begin{equation}
\label{eq:cal_matrix}
\begin{aligned}
A(0) = M^T A_r(0)M.
\end{aligned}
\end{equation}

So that the off-diagonal terms show up and the $P(0)$ deviates from identity matrix, as shown in the Fig.~\ref{Fig:Pij}(a)-(d). 
Directly solving this equation for the $M$ matrix is relatively hard, thus the assumption $M_{ij} = M_{ji}$ is applied for simplification, allowing $M^T = M$ and ignoring the double assignment error terms due to the relatively small error rate. We could therefore obtain the off-diagonal terms of $M$ ($i\neq j$):
\begin{equation}
\begin{aligned}
M_{ij} = \frac{1}{2}\times\frac{A_{ij}(0) + A_{ji}(0)}{A_{ii}(0) + A_{ij}(0)+ A_{ji}(0)+ A_{jj}(0)},
\end{aligned}
\end{equation}
and use $\sum_i M_{ij} = 1$ to constrain the value of the diagonal terms $M_{ii}$.

With the obtained confusion matrix $M$, we could extend the method to the real intrinsic state $A_r$ at any delay time $t$:
\begin{equation}
\begin{aligned}
A_{r} = (M^T)^{-1} AM^{-1},
\end{aligned}
\end{equation}
thus the corrected $P_{ij}$ can be easily obtained from $A_r$ following the Eq.~\eqref{eq:normalize_Pij} and plot in Fig.~\ref{Fig:Pij}(e-h).

The next step is to fit $P$ and extract the $4\times 4$ transition rate matrix $\Gamma$. Note that there are 12 free parameters in this matrix, $\Gamma_{ij}$ for $i\neq j$, with $\Gamma_{ii} = \sum_j \Gamma_{ij,(i\neq j)}$ constrained by the rest. With the master equation containing the $4\times 4$ transition-rate matrix $\boldsymbol{\Gamma}$: 
\begin{equation}
\dot{\boldsymbol{\rho}}=\boldsymbol{\Gamma} \boldsymbol{\rho},
\end{equation}
where $\boldsymbol{\rho}$ is the vector representing the population in $\{gO, gE, eO, eE\}$. Thus we can simply derive the population evolution as:
\begin{equation}
\label{eq:transition_fitting}
\boldsymbol{\rho} = e^{\boldsymbol{\Gamma}t}\boldsymbol{\rho}_0.
\end{equation}

Fitting each curve $P_{i,j}$ with the $\boldsymbol{\rho}_j$ derived from initial state $\boldsymbol{\rho}_0$ being a unit population in state $i$, we obtain the full $4\times 4$ transition-rate matrix $\boldsymbol{\Gamma}$ with the fitting results plotted as solid lines in Fig.~\ref{Fig:Pij}(e-h).

Repeating the above process $7$ times with the data obtained at different days, we obtain $7$ sets of transition rates to form the diagram in Fig. 5(d) of main text.

\subsection{TLS decay under increasing non-equilibrium QP density}
\begin{figure*}[tbp]
	\centering
	\includegraphics[width=17.5cm]{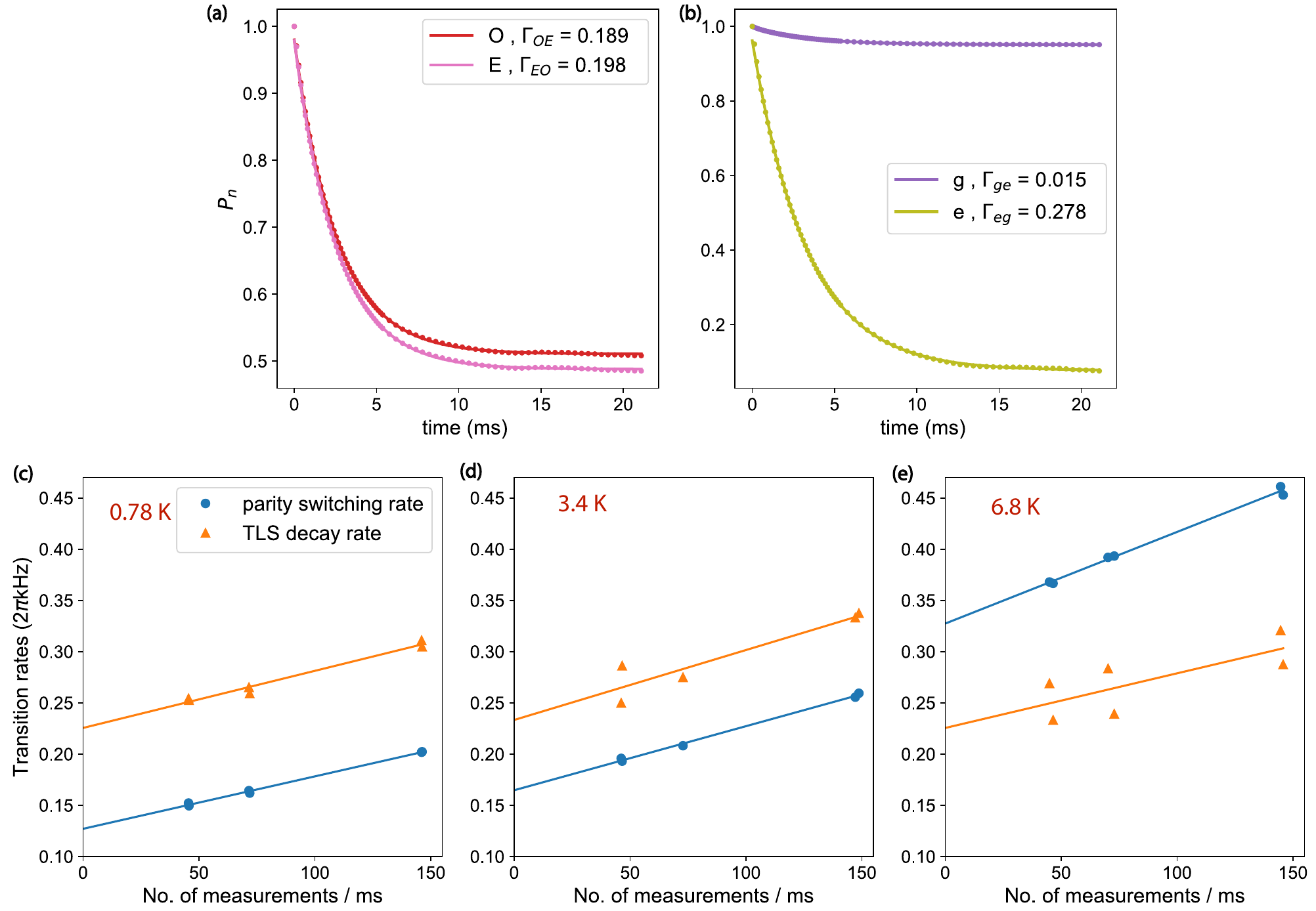}
	\caption{The TLS-parity state switching rate under radiation.  The probability of the \textbf{(a)} parity state stay in odd (O) and even (E) during the repeated measurement and \textbf{(b)} TLS state stay in ground (g) or excited (e). The solid lines are exponential fit to it and the transition rate is extracted from the exponential decay rate and the offset. \textbf{(c-e)} the parity switching rate (blue dot) and TLS decay rate (orange triangle) vs measurement rate under \textbf{(c)} 0.78 K, \textbf{(d)} 3.4 K, \textbf{(e)} 6.8 K radiation temperature, the solid line is a linear extrapolation of transition rates to the 0 measurement situation.  }\label{Fig:radiator}
\end{figure*}
The radiator, a resistive heater to apply IR radiation, is installed on the still plate and thermalized to it through Tolron 4203, a poor thermalization material, to make sure that increasing the radiator temperature is not strongly affecting the still plate and MXC plate temperature. The radiator temperature was measured by using a Cernox 1050 sensor. We intentionally increase the QP density by about 3 fold to check the TLS decay rate. The various QP density is achieved by maintaining various radiator temperature, and indicated through qubit charge parity jump rate. This measurement is to obtain the TLS - parity state in 1 million continuous measurements, which is done by applying the measurement protocol aforementioned repeatedly. 

For parity jump rate calculation, we record the charge state at even (odd) parity as 1 and the other state as 0 thus get a list of parity states, $S_p$. The probability of parity state being in even (odd) after $n$ measurements is calculated through:
\begin{equation}
P_n = \frac{\sum_{i = 0}^{N - n} S_{p,i}S_{p,i+n}}{\sum_{i = 0}^{N - n} S_{p,i}S_{p,i}}.
\end{equation}
The $P_n$ over time, for both even and odd parity, can be fit to an exponential function to obtain a characteristic time $\tau_{E(O)}$ and the stable probability as tail offset $s_{E(O)}$ (Fig.~\ref{Fig:radiator}(a)), and the state jump rate $\Gamma_{EO(OE)} = \frac{s_{E(O)}}{\tau_{E(O)}}$. The parity jump rate is then calculated as the average of even to add transition rate and odd to even transition rate, $\Gamma_p = \frac{1}{2}(\Gamma_{EO} + \Gamma_{OE})$. The TLS decay rate $\Gamma_T$ is calculated similarly by recording the state at TLS excited (ground) state $S_T$, and fitting the probability function to obtain a characteristic time $\tau_{e(g)}$ and the stable probability as tail offset $s_{e(g)}$, $\Gamma_T =\Gamma_{eg} = \frac{s_e}{\tau_e}$, as shown in Fig.~\ref{Fig:radiator}(b).

As we show in Sec.~\ref{sec: TLS_drive} that the readout drive affects the TLS drive and what we observe later that the charge parity state is affected by frequent readout pulse as well, we choose to use the long delay measurement protocol in our main text TLS - parity measurement to avoid this effect.
While in this measurement, we mitigate this influence by reducing the measurement rate. This is achieved by introducing extra delay time after each measurement, reducing the effects from readout pulses. We repeat the protocol to acquire the $\Gamma_p$ and $\Gamma_T$ for various measurement rates and plot them in Fig.~\ref{Fig:radiator}(c-e). Linearly extrapolating them to the point where the measurement rate is 0, we determine the intrinsic parity switching rate $\Gamma_p$ and TLS decay rate $\Gamma_T$, as illustrated in Fig.~\ref{Fig:radiator}(c-e) and shown in main text Fig.~5(e). Observing the trends of $\Gamma_p$ and $\Gamma_T$ with increasing radiation temperature, we note a rapid increase in the parity switching rate $\Gamma_p$ which indicates the QP density increasing, while the TLS decay rate $\Gamma_T$ remains stable. 
This test shows that the TLS lifetime even at millisecond level is not limited by QPs in the electrodes. 

\section{Discussions on random telegraphing noise in superconducting qubits}

\begin{figure}[tbp]
	\centering
	\includegraphics[width=13cm]{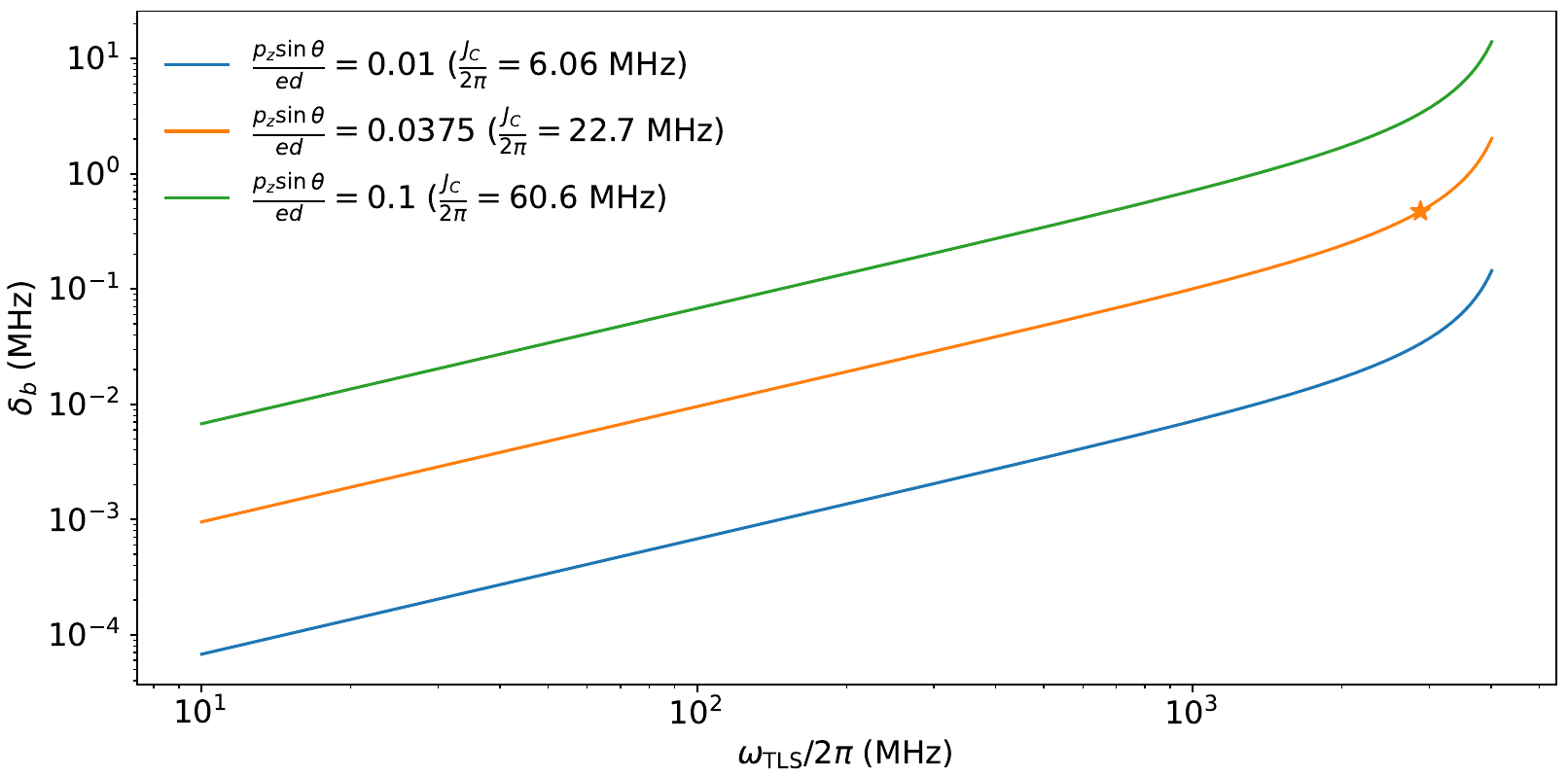}
	\caption{Dispersive frequency shift $\delta_b$ in the qubit-TLS charge coupling model as a function of the TLS frequency.  Different TLS dipole moments are considered as examples, represented in a dimensionless term $\frac{p_z\sin{\theta}}{ed}$, where $d$ represents the barrier thickness.  The transmon qubit parameters are fixed to the zero-flux values of the device in this study (see Table~\ref{table1}).  The star symbol corresponds to the TLS in our experiment. } \label{Figs3}
\end{figure}

Observations of random telegraphing switching (RTS) between two qubit frequencies is very common in superconducting qubits.  One mechanism is charge-parity switching in qubits with non-vanishing offset-charge sensitivity, which can be explicitly modeled and taken into account as in this study.  Other mechanisms may include flux-type of noise potentially from discrete spins or vortices.  However, even for qubits known to be insensitive to both dc charge and flux noise, RTS remains widely observed and has been an active topic of study~\cite{sPaik2011,sSteffen2019Correlating,slevine2023demonstrating,sgertler2021protecting}. 

In particular, fixed-frequency transmons in 3D cavities have long coherence times and kHz-level of long-term frequency stability, but still occasionally display prominent RTS of few kHz to 100's of KHz.  This happens for one in approximately every 5-10 qubits $\times$ thermal cycles based on the authors' experience, which is also consistent with the impression across multiple circuit QED labs based on a recent conference discussions within the C$^2$QA center.  This type of RTS was noted in the original 3D transmon work~\cite{sPaik2011}, ``on subsequent thermal cycling of two devices, a slow telegraph switching behavior (with $\delta f\sim$ 5–50 kHz) was also observed".  These RTS tend to have switching rates faster than the measurement integration time of a few seconds, showing up as beating patterns in Ramsey oscillations, and tend to be persistent with relatively consistent $\delta f$ and branching ratios for weeks or even months.  They are usually removed (stochastically) in small-scale science experiments by thermal cycling of the qubit devices and therefore get neglected (or relegated to a footnote) in the literature, but will pose a challenge in scaling.  We noted one such RTS in recent years in Ref.~\cite{sgertler2021protecting}, where the fixed-frequency transmon ``displays a random switching behaviour between two values of $\omega_q/2\pi$ that are 40 kHz apart, with a dwell-time split of approximately $85\%:15\%$".  This RTS behavior was consistent for the entire duration of the main experiment ($\sim 3$ months).  

In this section, we argue that these type of pronounced RTS in otherwise highly coherent and stable qubits are caused by strongly-coupled but far-detuned TLS in the GHz to sub-GHz frequency range.  
Now that our charged TLS-qubit coupling model (Eq.~\eqref{SS1}) has been quantitatively verified, we use it to calculate the transmon-TLS dispersive frequency shifts $\delta_b$ as a function of TLS frequency, as shown in Fig.~\ref{Figs3}.  Here we consider three realistic cases of dimensionless TLS tunneling dipole moments, ${p_z\sin{\theta}}/{ed}$ (where $d$ represents the barrier thickness), which is directly proportional to the charge coupling energy $J_C$ as in Eq.~\eqref{SS1}.  
The TLS in our experiment is marked by the star symbol.  The result shows that in the low-frequency regime ($\omega_{TLS}\ll\omega_{01}-\omega_{TLS}$) where it is crucial to take into account the counter-rotating terms, $\delta_b$ is proportional to $\omega_{TLS}$.  Potential TLS in the junction barrier in the $0.1$-$1$ GHz range (assuming a similar dipole to the one in this study) are expected to induced dispersive shift of 10's of kHz.  In this frequency range, the TLS can be expected to have even longer $T_1$ times and significant thermal excitations to produce RTS of qubit frequencies as observed in other experiments.  For example, assuming a temperature of $\sim$20 mK, a TLS at about 700 MHz with a similar coupling dipole as in this study would explain the magnitude and the dwell-time ratio of the RTS observed in Ref.~\cite{sgertler2021protecting} very well.  In this mechanism, since the TLS is multi-GHz detuned from the qubit, spectral diffusion of the TLS on the order of MHz or 10's of MHz often observed in dedicated TLS investigations would not lead to notable changes to the RTS behavior or notable low-frequency noise to the qubit.  Therefore, the qubit can have excellent coherence times (e.g., as measured with a spin echo) and long-term frequency stability despite the RTS.

We note that the qubit-TLS charge coupling model (Eq.~\eqref{SS1}) verified in this work is the more basic and more direct model than the TLS-TF model.  The TLS-TF model, also known as the interacting TLS model presuming the TLS and TF are fundamentally similar, has been developed in recent years~\cite{sburnett2014evidence, sKlimov2018Fluctuations} to explain low-frequency fluctuations in qubits and resonators, where the RTS is explained by the switching of individual TF~\cite{sSteffen2019Correlating}.  Direct dispersive coupling from thermally-accessible TLS was not considered a source of RTS partially due to the generally short $T_1$ time of coherently-interacting TLS found in previous experiments (no more than a few $\mu$s~\cite{sneeley2008process,sShalibo2010Lifetime,sLisenfeld2010Measuring}).  Our study presents a TLS several orders of magnitude longer-lived than previously known, and hence making direct dispersive TLS coupling a viable model for the RTS phenomena.  

Indeed, it is possible that both direct contributions from low-frequency coherent TLS and resonant-TLS-mediated contributions from TF are at play for different qubits.  Reported studies of RTS of charge-insensitive qubits in the literature, primarily in planar devices, tend to have long switching time scales of many seconds to over minutes ~\cite{sburnett2019decoherence,sSteffen2019Correlating,slevine2023demonstrating} and are often accompanied by multiple varying RTS components and a strong 1/$f$ noise background, as expected from the TLS-TF model.  Given that we have in the past primarily observed stable RTS with sub-second switching time in 3D-style transmons, it is worth exploring further whether there is statistically a notable difference between 3D and planar qubits in RTS behaviors.  One potential difference is that the electric field distribution of our 3D-style transmons results in fewer TLS with qubit coupling rate on the order of $\sim$100 kHz (corresponding to those near the edge of the planar capacitor, as analyzed in Ref.~\cite{sSteffen2019Correlating}) that can mediate prominent RTS from TF.  It is also possible that the unexpectedly long $T_1$ time of our TLS is related to accidental suppression of phononic dissipation due to the 3D cavity packaging scheme.


\begin{thebibliography}{50}%
\makeatletter
\providecommand \@ifxundefined [1]{%
 \@ifx{#1\undefined}
}%
\providecommand \@ifnum [1]{%
 \ifnum #1\expandafter \@firstoftwo
 \else \expandafter \@secondoftwo
 \fi
}%
\providecommand \@ifx [1]{%
 \ifx #1\expandafter \@firstoftwo
 \else \expandafter \@secondoftwo
 \fi
}%
\providecommand \natexlab [1]{#1}%
\providecommand \enquote  [1]{``#1''}%
\providecommand \bibnamefont  [1]{#1}%
\providecommand \bibfnamefont [1]{#1}%
\providecommand \citenamefont [1]{#1}%
\providecommand \href@noop [0]{\@secondoftwo}%
\providecommand \href [0]{\begingroup \@sanitize@url \@href}%
\providecommand \@href[1]{\@@startlink{#1}\@@href}%
\providecommand \@@href[1]{\endgroup#1\@@endlink}%
\providecommand \@sanitize@url [0]{\catcode `\\12\catcode `\$12\catcode
  `\&12\catcode `\#12\catcode `\^12\catcode `\_12\catcode `\%12\relax}%
\providecommand \@@startlink[1]{}%
\providecommand \@@endlink[0]{}%
\providecommand \url  [0]{\begingroup\@sanitize@url \@url }%
\providecommand \@url [1]{\endgroup\@href {#1}{\urlprefix }}%
\providecommand \urlprefix  [0]{URL }%
\providecommand \Eprint [0]{\href }%
\providecommand \doibase [0]{http://dx.doi.org/}%
\providecommand \selectlanguage [0]{\@gobble}%
\providecommand \bibinfo  [0]{\@secondoftwo}%
\providecommand \bibfield  [0]{\@secondoftwo}%
\providecommand \translation [1]{[#1]}%
\providecommand \BibitemOpen [0]{}%
\providecommand \bibitemStop [0]{}%
\providecommand \bibitemNoStop [0]{.\EOS\space}%
\providecommand \EOS [0]{\spacefactor3000\relax}%
\providecommand \BibitemShut  [1]{\csname bibitem#1\endcsname}%
\let\auto@bib@innerbib\@empty
\bibitem [{\citenamefont {De~Leon}\ \emph {et~al.}(2021)\citenamefont
  {De~Leon}, \citenamefont {Itoh}, \citenamefont {Kim}, \citenamefont {Mehta},
  \citenamefont {Northup}, \citenamefont {Paik}, \citenamefont {Palmer},
  \citenamefont {Samarth}, \citenamefont {Sangtawesin},\ and\ \citenamefont
  {Steuerman}}]{de2021materials}%
  \BibitemOpen
  \bibfield  {author} {\bibinfo {author} {\bibfnamefont {Nathalie~P}\
  \bibnamefont {De~Leon}}, \bibinfo {author} {\bibfnamefont {Kohei~M}\
  \bibnamefont {Itoh}}, \bibinfo {author} {\bibfnamefont {Dohun}\ \bibnamefont
  {Kim}}, \bibinfo {author} {\bibfnamefont {Karan~K}\ \bibnamefont {Mehta}},
  \bibinfo {author} {\bibfnamefont {Tracy~E}\ \bibnamefont {Northup}}, \bibinfo
  {author} {\bibfnamefont {Hanhee}\ \bibnamefont {Paik}}, \bibinfo {author}
  {\bibfnamefont {BS}~\bibnamefont {Palmer}}, \bibinfo {author} {\bibfnamefont
  {Nitin}\ \bibnamefont {Samarth}}, \bibinfo {author} {\bibfnamefont {Sorawis}\
  \bibnamefont {Sangtawesin}}, \ and\ \bibinfo {author} {\bibfnamefont
  {David~W}\ \bibnamefont {Steuerman}},\ }\bibfield  {title} {\enquote
  {\bibinfo {title} {Materials challenges and opportunities for quantum
  computing hardware},}\ }\href
  {{https://www.science.org/doi/abs/10.1126/science.abb2823}} {\bibfield
  {journal} {\bibinfo  {journal} {Science}\ }\textbf {\bibinfo {volume}
  {372}},\ \bibinfo {pages} {eabb2823} (\bibinfo {year} {2021})}\BibitemShut
  {NoStop}%
\bibitem [{\citenamefont {Paladino}\ \emph {et~al.}(2014)\citenamefont
  {Paladino}, \citenamefont {Galperin}, \citenamefont {Falci},\ and\
  \citenamefont {Altshuler}}]{Paladino2014noise}%
  \BibitemOpen
  \bibfield  {author} {\bibinfo {author} {\bibfnamefont {E.}~\bibnamefont
  {Paladino}}, \bibinfo {author} {\bibfnamefont {Y.~M.}\ \bibnamefont
  {Galperin}}, \bibinfo {author} {\bibfnamefont {G.}~\bibnamefont {Falci}}, \
  and\ \bibinfo {author} {\bibfnamefont {B.~L.}\ \bibnamefont {Altshuler}},\
  }\bibfield  {title} {\enquote {\bibinfo {title} {1/f noise: Implications for
  solid-state quantum information},}\ }\href {\doibase
  10.1103/RevModPhys.86.361} {\bibfield  {journal} {\bibinfo  {journal} {Rev.
  Mod. Phys.}\ }\textbf {\bibinfo {volume} {86}},\ \bibinfo {pages} {361--418}
  (\bibinfo {year} {2014})}\BibitemShut {NoStop}%
\bibitem [{\citenamefont {M{\"u}ller}\ \emph {et~al.}(2019)\citenamefont
  {M{\"u}ller}, \citenamefont {Cole},\ and\ \citenamefont
  {Lisenfeld}}]{muller2019towards}%
  \BibitemOpen
  \bibfield  {author} {\bibinfo {author} {\bibfnamefont {Clemens}\ \bibnamefont
  {M{\"u}ller}}, \bibinfo {author} {\bibfnamefont {Jared~H}\ \bibnamefont
  {Cole}}, \ and\ \bibinfo {author} {\bibfnamefont {J{\"u}rgen}\ \bibnamefont
  {Lisenfeld}},\ }\bibfield  {title} {\enquote {\bibinfo {title} {Towards
  understanding two-level-systems in amorphous solids: insights from quantum
  circuits},}\ }\href
  {https://iopscience.iop.org/article/10.1088/1361-6633/ab3a7e} {\bibfield
  {journal} {\bibinfo  {journal} {Reports on Progress in Physics}\ }\textbf
  {\bibinfo {volume} {82}},\ \bibinfo {pages} {124501} (\bibinfo {year}
  {2019})}\BibitemShut {NoStop}%
\bibitem [{\citenamefont {Martinis}\ \emph {et~al.}(2005)\citenamefont
  {Martinis}, \citenamefont {Cooper}, \citenamefont {McDermott}, \citenamefont
  {Steffen}, \citenamefont {Ansmann}, \citenamefont {Osborn}, \citenamefont
  {Cicak}, \citenamefont {Oh}, \citenamefont {Pappas}, \citenamefont
  {Simmonds},\ and\ \citenamefont {Yu}}]{martinis2005decoherence}%
  \BibitemOpen
  \bibfield  {author} {\bibinfo {author} {\bibfnamefont {John~M.}\ \bibnamefont
  {Martinis}}, \bibinfo {author} {\bibfnamefont {K.~B.}\ \bibnamefont
  {Cooper}}, \bibinfo {author} {\bibfnamefont {R.}~\bibnamefont {McDermott}},
  \bibinfo {author} {\bibfnamefont {Matthias}\ \bibnamefont {Steffen}},
  \bibinfo {author} {\bibfnamefont {Markus}\ \bibnamefont {Ansmann}}, \bibinfo
  {author} {\bibfnamefont {K.~D.}\ \bibnamefont {Osborn}}, \bibinfo {author}
  {\bibfnamefont {K.}~\bibnamefont {Cicak}}, \bibinfo {author} {\bibfnamefont
  {Seongshik}\ \bibnamefont {Oh}}, \bibinfo {author} {\bibfnamefont {D.~P.}\
  \bibnamefont {Pappas}}, \bibinfo {author} {\bibfnamefont {R.~W.}\
  \bibnamefont {Simmonds}}, \ and\ \bibinfo {author} {\bibfnamefont {Clare~C.}\
  \bibnamefont {Yu}},\ }\bibfield  {title} {\enquote {\bibinfo {title}
  {Decoherence in josephson qubits from dielectric loss},}\ }\href {\doibase
  10.1103/PhysRevLett.95.210503} {\bibfield  {journal} {\bibinfo  {journal}
  {Phys. Rev. Lett.}\ }\textbf {\bibinfo {volume} {95}},\ \bibinfo {pages}
  {210503} (\bibinfo {year} {2005})}\BibitemShut {NoStop}%
\bibitem [{\citenamefont {Barends}\ \emph {et~al.}(2013)\citenamefont
  {Barends}, \citenamefont {Kelly}, \citenamefont {Megrant}, \citenamefont
  {Sank}, \citenamefont {Jeffrey}, \citenamefont {Chen}, \citenamefont {Yin},
  \citenamefont {Chiaro}, \citenamefont {Mutus}, \citenamefont {Neill},
  \citenamefont {O'Malley}, \citenamefont {Roushan}, \citenamefont {Wenner},
  \citenamefont {White}, \citenamefont {Cleland},\ and\ \citenamefont
  {Martinis}}]{Barends2013Coherent}%
  \BibitemOpen
  \bibfield  {author} {\bibinfo {author} {\bibfnamefont {R.}~\bibnamefont
  {Barends}}, \bibinfo {author} {\bibfnamefont {J.}~\bibnamefont {Kelly}},
  \bibinfo {author} {\bibfnamefont {A.}~\bibnamefont {Megrant}}, \bibinfo
  {author} {\bibfnamefont {D.}~\bibnamefont {Sank}}, \bibinfo {author}
  {\bibfnamefont {E.}~\bibnamefont {Jeffrey}}, \bibinfo {author} {\bibfnamefont
  {Y.}~\bibnamefont {Chen}}, \bibinfo {author} {\bibfnamefont {Y.}~\bibnamefont
  {Yin}}, \bibinfo {author} {\bibfnamefont {B.}~\bibnamefont {Chiaro}},
  \bibinfo {author} {\bibfnamefont {J.}~\bibnamefont {Mutus}}, \bibinfo
  {author} {\bibfnamefont {C.}~\bibnamefont {Neill}}, \bibinfo {author}
  {\bibfnamefont {P.}~\bibnamefont {O'Malley}}, \bibinfo {author}
  {\bibfnamefont {P.}~\bibnamefont {Roushan}}, \bibinfo {author} {\bibfnamefont
  {J.}~\bibnamefont {Wenner}}, \bibinfo {author} {\bibfnamefont {T.~C.}\
  \bibnamefont {White}}, \bibinfo {author} {\bibfnamefont {A.~N.}\ \bibnamefont
  {Cleland}}, \ and\ \bibinfo {author} {\bibfnamefont {John~M.}\ \bibnamefont
  {Martinis}},\ }\bibfield  {title} {\enquote {\bibinfo {title} {Coherent
  josephson qubit suitable for scalable quantum integrated circuits},}\ }\href
  {\doibase 10.1103/PhysRevLett.111.080502} {\bibfield  {journal} {\bibinfo
  {journal} {Phys. Rev. Lett.}\ }\textbf {\bibinfo {volume} {111}},\ \bibinfo
  {pages} {080502} (\bibinfo {year} {2013})}\BibitemShut {NoStop}%
\bibitem [{\citenamefont {Carroll}\ \emph {et~al.}(2022)\citenamefont
  {Carroll}, \citenamefont {Rosenblatt}, \citenamefont {Jurcevic},
  \citenamefont {Lauer},\ and\ \citenamefont {Kandala}}]{carroll2022dynamics}%
  \BibitemOpen
  \bibfield  {author} {\bibinfo {author} {\bibfnamefont {Malcolm}\ \bibnamefont
  {Carroll}}, \bibinfo {author} {\bibfnamefont {Sami}\ \bibnamefont
  {Rosenblatt}}, \bibinfo {author} {\bibfnamefont {Petar}\ \bibnamefont
  {Jurcevic}}, \bibinfo {author} {\bibfnamefont {Isaac}\ \bibnamefont {Lauer}},
  \ and\ \bibinfo {author} {\bibfnamefont {Abhinav}\ \bibnamefont {Kandala}},\
  }\bibfield  {title} {\enquote {\bibinfo {title} {Dynamics of superconducting
  qubit relaxation times},}\ }\href
  {https://doi.org/10.1038/s41534-022-00643-y} {\bibfield  {journal} {\bibinfo
  {journal} {npj Quantum Information}\ }\textbf {\bibinfo {volume} {8}},\
  \bibinfo {pages} {132} (\bibinfo {year} {2022})}\BibitemShut {NoStop}%
\bibitem [{\citenamefont {Wang}\ \emph {et~al.}(2015)\citenamefont {Wang},
  \citenamefont {Axline}, \citenamefont {Gao}, \citenamefont {Brecht},
  \citenamefont {Chu}, \citenamefont {Frunzio}, \citenamefont {Devoret},\ and\
  \citenamefont {Schoelkopf}}]{wang2015surface}%
  \BibitemOpen
  \bibfield  {author} {\bibinfo {author} {\bibfnamefont {Chen}\ \bibnamefont
  {Wang}}, \bibinfo {author} {\bibfnamefont {Christopher}\ \bibnamefont
  {Axline}}, \bibinfo {author} {\bibfnamefont {Yvonne~Y}\ \bibnamefont {Gao}},
  \bibinfo {author} {\bibfnamefont {Teresa}\ \bibnamefont {Brecht}}, \bibinfo
  {author} {\bibfnamefont {Yiwen}\ \bibnamefont {Chu}}, \bibinfo {author}
  {\bibfnamefont {Luigi}\ \bibnamefont {Frunzio}}, \bibinfo {author}
  {\bibfnamefont {MH}~\bibnamefont {Devoret}}, \ and\ \bibinfo {author}
  {\bibfnamefont {Robert~J}\ \bibnamefont {Schoelkopf}},\ }\bibfield  {title}
  {\enquote {\bibinfo {title} {Surface participation and dielectric loss in
  superconducting qubits},}\ }\href {https://doi.org/10.1063/1.4934486}
  {\bibfield  {journal} {\bibinfo  {journal} {Applied Physics Letters}\
  }\textbf {\bibinfo {volume} {107}} (\bibinfo {year} {2015})}\BibitemShut
  {NoStop}%
\bibitem [{\citenamefont {Weissman}(1988)}]{Weissman1988}%
  \BibitemOpen
  \bibfield  {author} {\bibinfo {author} {\bibfnamefont {M.~B.}\ \bibnamefont
  {Weissman}},\ }\bibfield  {title} {\enquote {\bibinfo {title} {$\frac{1}{f}$
  noise and other slow, nonexponential kinetics in condensed matter},}\ }\href
  {\doibase 10.1103/RevModPhys.60.537} {\bibfield  {journal} {\bibinfo
  {journal} {Rev. Mod. Phys.}\ }\textbf {\bibinfo {volume} {60}},\ \bibinfo
  {pages} {537--571} (\bibinfo {year} {1988})}\BibitemShut {NoStop}%
\bibitem [{\citenamefont {Ku}\ and\ \citenamefont {Yu}(2005)}]{Ku2005}%
  \BibitemOpen
  \bibfield  {author} {\bibinfo {author} {\bibfnamefont {Li-Chung}\
  \bibnamefont {Ku}}\ and\ \bibinfo {author} {\bibfnamefont {Clare~C.}\
  \bibnamefont {Yu}},\ }\bibfield  {title} {\enquote {\bibinfo {title}
  {Decoherence of a josephson qubit due to coupling to two-level systems},}\
  }\href {\doibase 10.1103/PhysRevB.72.024526} {\bibfield  {journal} {\bibinfo
  {journal} {Phys. Rev. B}\ }\textbf {\bibinfo {volume} {72}},\ \bibinfo
  {pages} {024526} (\bibinfo {year} {2005})}\BibitemShut {NoStop}%
\bibitem [{\citenamefont {Pourkabirian}\ \emph {et~al.}(2014)\citenamefont
  {Pourkabirian}, \citenamefont {Gustafsson}, \citenamefont {Johansson},
  \citenamefont {Clarke},\ and\ \citenamefont {Delsing}}]{Pourkabirian2014}%
  \BibitemOpen
  \bibfield  {author} {\bibinfo {author} {\bibfnamefont {A.}~\bibnamefont
  {Pourkabirian}}, \bibinfo {author} {\bibfnamefont {M.~V.}\ \bibnamefont
  {Gustafsson}}, \bibinfo {author} {\bibfnamefont {G.}~\bibnamefont
  {Johansson}}, \bibinfo {author} {\bibfnamefont {J.}~\bibnamefont {Clarke}}, \
  and\ \bibinfo {author} {\bibfnamefont {P.}~\bibnamefont {Delsing}},\
  }\bibfield  {title} {\enquote {\bibinfo {title} {Nonequilibrium probing of
  two-level charge fluctuators using the step response of a single-electron
  transistor},}\ }\href {\doibase 10.1103/PhysRevLett.113.256801} {\bibfield
  {journal} {\bibinfo  {journal} {Phys. Rev. Lett.}\ }\textbf {\bibinfo
  {volume} {113}},\ \bibinfo {pages} {256801} (\bibinfo {year}
  {2014})}\BibitemShut {NoStop}%
\bibitem [{\citenamefont {Simkins}\ \emph {et~al.}(2009)\citenamefont
  {Simkins}, \citenamefont {Rees}, \citenamefont {Glasson}, \citenamefont
  {Antonov}, \citenamefont {Collin}, \citenamefont {Frayne}, \citenamefont
  {Meeson},\ and\ \citenamefont {Lea}}]{simkins2009thermal}%
  \BibitemOpen
  \bibfield  {author} {\bibinfo {author} {\bibfnamefont {Luke~R}\ \bibnamefont
  {Simkins}}, \bibinfo {author} {\bibfnamefont {David~G}\ \bibnamefont {Rees}},
  \bibinfo {author} {\bibfnamefont {Philip~H}\ \bibnamefont {Glasson}},
  \bibinfo {author} {\bibfnamefont {Vladimir}\ \bibnamefont {Antonov}},
  \bibinfo {author} {\bibfnamefont {Eddy}\ \bibnamefont {Collin}}, \bibinfo
  {author} {\bibfnamefont {Peter~G}\ \bibnamefont {Frayne}}, \bibinfo {author}
  {\bibfnamefont {Philip~J}\ \bibnamefont {Meeson}}, \ and\ \bibinfo {author}
  {\bibfnamefont {Michael~J}\ \bibnamefont {Lea}},\ }\bibfield  {title}
  {\enquote {\bibinfo {title} {Thermal excitation of large charge offsets in a
  single-cooper-pair transistor},}\ }\href {https://doi.org/10.1063/1.3266012}
  {\bibfield  {journal} {\bibinfo  {journal} {Journal of Applied Physics}\
  }\textbf {\bibinfo {volume} {106}} (\bibinfo {year} {2009})}\BibitemShut
  {NoStop}%
\bibitem [{\citenamefont {Gustafsson}\ \emph {et~al.}(2013)\citenamefont
  {Gustafsson}, \citenamefont {Pourkabirian}, \citenamefont {Johansson},
  \citenamefont {Clarke},\ and\ \citenamefont {Delsing}}]{Gustafsson2013}%
  \BibitemOpen
  \bibfield  {author} {\bibinfo {author} {\bibfnamefont {Martin~V.}\
  \bibnamefont {Gustafsson}}, \bibinfo {author} {\bibfnamefont {Arsalan}\
  \bibnamefont {Pourkabirian}}, \bibinfo {author} {\bibfnamefont {G\"oran}\
  \bibnamefont {Johansson}}, \bibinfo {author} {\bibfnamefont {John}\
  \bibnamefont {Clarke}}, \ and\ \bibinfo {author} {\bibfnamefont {Per}\
  \bibnamefont {Delsing}},\ }\bibfield  {title} {\enquote {\bibinfo {title}
  {Thermal properties of charge noise sources},}\ }\href {\doibase
  10.1103/PhysRevB.88.245410} {\bibfield  {journal} {\bibinfo  {journal} {Phys.
  Rev. B}\ }\textbf {\bibinfo {volume} {88}},\ \bibinfo {pages} {245410}
  (\bibinfo {year} {2013})}\BibitemShut {NoStop}%
\bibitem [{\citenamefont {De~Graaf}\ \emph {et~al.}(2018)\citenamefont
  {De~Graaf}, \citenamefont {Faoro}, \citenamefont {Burnett}, \citenamefont
  {Adamyan}, \citenamefont {Tzalenchuk}, \citenamefont {Kubatkin},
  \citenamefont {Lindstr{\"o}m},\ and\ \citenamefont
  {Danilov}}]{de2018suppression}%
  \BibitemOpen
  \bibfield  {author} {\bibinfo {author} {\bibfnamefont {SE}~\bibnamefont
  {De~Graaf}}, \bibinfo {author} {\bibfnamefont {L}~\bibnamefont {Faoro}},
  \bibinfo {author} {\bibfnamefont {J}~\bibnamefont {Burnett}}, \bibinfo
  {author} {\bibfnamefont {AA}~\bibnamefont {Adamyan}}, \bibinfo {author}
  {\bibfnamefont {A~Ya}\ \bibnamefont {Tzalenchuk}}, \bibinfo {author}
  {\bibfnamefont {SE}~\bibnamefont {Kubatkin}}, \bibinfo {author}
  {\bibfnamefont {T}~\bibnamefont {Lindstr{\"o}m}}, \ and\ \bibinfo {author}
  {\bibfnamefont {AV}~\bibnamefont {Danilov}},\ }\bibfield  {title} {\enquote
  {\bibinfo {title} {Suppression of low-frequency charge noise in
  superconducting resonators by surface spin desorption},}\ }\href
  {https://doi.org/10.1038/s41467-018-03577-2} {\bibfield  {journal} {\bibinfo
  {journal} {Nature communications}\ }\textbf {\bibinfo {volume} {9}},\
  \bibinfo {pages} {1143} (\bibinfo {year} {2018})}\BibitemShut {NoStop}%
\bibitem [{\citenamefont {Phillips}(1987)}]{phillips1987two}%
  \BibitemOpen
  \bibfield  {author} {\bibinfo {author} {\bibfnamefont {William~A}\
  \bibnamefont {Phillips}},\ }\bibfield  {title} {\enquote {\bibinfo {title}
  {Two-level states in glasses},}\ }\href
  {https://iopscience.iop.org/article/10.1088/0034-4885/50/12/003} {\bibfield
  {journal} {\bibinfo  {journal} {Reports on Progress in Physics}\ }\textbf
  {\bibinfo {volume} {50}},\ \bibinfo {pages} {1657} (\bibinfo {year}
  {1987})}\BibitemShut {NoStop}%
\bibitem [{\citenamefont {Phillips}(1972)}]{phillips1972tunneling}%
  \BibitemOpen
  \bibfield  {author} {\bibinfo {author} {\bibfnamefont {William~A}\
  \bibnamefont {Phillips}},\ }\bibfield  {title} {\enquote {\bibinfo {title}
  {Tunneling states in amorphous solids},}\ }\href
  {https://link.springer.com/article/10.1007/bf00660072} {\bibfield  {journal}
  {\bibinfo  {journal} {Journal of low temperature physics}\ }\textbf {\bibinfo
  {volume} {7}},\ \bibinfo {pages} {351--360} (\bibinfo {year}
  {1972})}\BibitemShut {NoStop}%
\bibitem [{\citenamefont {Koch}\ \emph {et~al.}(2007)\citenamefont {Koch},
  \citenamefont {Yu}, \citenamefont {Gambetta}, \citenamefont {Houck},
  \citenamefont {Schuster}, \citenamefont {Majer}, \citenamefont {Blais},
  \citenamefont {Devoret}, \citenamefont {Girvin},\ and\ \citenamefont
  {Schoelkopf}}]{Koch2007Charge}%
  \BibitemOpen
  \bibfield  {author} {\bibinfo {author} {\bibfnamefont {Jens}\ \bibnamefont
  {Koch}}, \bibinfo {author} {\bibfnamefont {Terri~M.}\ \bibnamefont {Yu}},
  \bibinfo {author} {\bibfnamefont {Jay}\ \bibnamefont {Gambetta}}, \bibinfo
  {author} {\bibfnamefont {A.~A.}\ \bibnamefont {Houck}}, \bibinfo {author}
  {\bibfnamefont {D.~I.}\ \bibnamefont {Schuster}}, \bibinfo {author}
  {\bibfnamefont {J.}~\bibnamefont {Majer}}, \bibinfo {author} {\bibfnamefont
  {Alexandre}\ \bibnamefont {Blais}}, \bibinfo {author} {\bibfnamefont {M.~H.}\
  \bibnamefont {Devoret}}, \bibinfo {author} {\bibfnamefont {S.~M.}\
  \bibnamefont {Girvin}}, \ and\ \bibinfo {author} {\bibfnamefont {R.~J.}\
  \bibnamefont {Schoelkopf}},\ }\bibfield  {title} {\enquote {\bibinfo {title}
  {Charge-insensitive qubit design derived from the cooper pair box},}\ }\href
  {\doibase 10.1103/PhysRevA.76.042319} {\bibfield  {journal} {\bibinfo
  {journal} {Phys. Rev. A}\ }\textbf {\bibinfo {volume} {76}},\ \bibinfo
  {pages} {042319} (\bibinfo {year} {2007})}\BibitemShut {NoStop}%
\bibitem [{\citenamefont {McRae}\ \emph {et~al.}(2020)\citenamefont {McRae},
  \citenamefont {Wang}, \citenamefont {Gao}, \citenamefont {Vissers},
  \citenamefont {Brecht}, \citenamefont {Dunsworth}, \citenamefont {Pappas},\
  and\ \citenamefont {Mutus}}]{mcrae2020materials}%
  \BibitemOpen
  \bibfield  {author} {\bibinfo {author} {\bibfnamefont {Corey Rae~Harrington}\
  \bibnamefont {McRae}}, \bibinfo {author} {\bibfnamefont {Haozhi}\
  \bibnamefont {Wang}}, \bibinfo {author} {\bibfnamefont {Jiansong}\
  \bibnamefont {Gao}}, \bibinfo {author} {\bibfnamefont {Michael~R}\
  \bibnamefont {Vissers}}, \bibinfo {author} {\bibfnamefont {Teresa}\
  \bibnamefont {Brecht}}, \bibinfo {author} {\bibfnamefont {Andrew}\
  \bibnamefont {Dunsworth}}, \bibinfo {author} {\bibfnamefont {David~P}\
  \bibnamefont {Pappas}}, \ and\ \bibinfo {author} {\bibfnamefont {Josh}\
  \bibnamefont {Mutus}},\ }\bibfield  {title} {\enquote {\bibinfo {title}
  {Materials loss measurements using superconducting microwave resonators},}\
  }\href {https://doi.org/10.1063/5.0017378} {\bibfield  {journal} {\bibinfo
  {journal} {Review of Scientific Instruments}\ }\textbf {\bibinfo {volume}
  {91}} (\bibinfo {year} {2020})}\BibitemShut {NoStop}%
\bibitem [{\citenamefont {Crowley}\ \emph {et~al.}(2023)\citenamefont
  {Crowley}, \citenamefont {McLellan}, \citenamefont {Dutta}, \citenamefont
  {Shumiya}, \citenamefont {Place}, \citenamefont {Le}, \citenamefont {Gang},
  \citenamefont {Madhavan}, \citenamefont {Bland}, \citenamefont {Chang},
  \citenamefont {Khedkar}, \citenamefont {Feng}, \citenamefont {Umbarkar},
  \citenamefont {Gui}, \citenamefont {Rodgers}, \citenamefont {Jia},
  \citenamefont {Feldman}, \citenamefont {Lyon}, \citenamefont {Liu},
  \citenamefont {Cava}, \citenamefont {Houck},\ and\ \citenamefont
  {de~Leon}}]{Crowley2023}%
  \BibitemOpen
  \bibfield  {author} {\bibinfo {author} {\bibfnamefont {Kevin~D.}\
  \bibnamefont {Crowley}}, \bibinfo {author} {\bibfnamefont {Russell~A.}\
  \bibnamefont {McLellan}}, \bibinfo {author} {\bibfnamefont {Aveek}\
  \bibnamefont {Dutta}}, \bibinfo {author} {\bibfnamefont {Nana}\ \bibnamefont
  {Shumiya}}, \bibinfo {author} {\bibfnamefont {Alexander P.~M.}\ \bibnamefont
  {Place}}, \bibinfo {author} {\bibfnamefont {Xuan~Hoang}\ \bibnamefont {Le}},
  \bibinfo {author} {\bibfnamefont {Youqi}\ \bibnamefont {Gang}}, \bibinfo
  {author} {\bibfnamefont {Trisha}\ \bibnamefont {Madhavan}}, \bibinfo {author}
  {\bibfnamefont {Matthew~P.}\ \bibnamefont {Bland}}, \bibinfo {author}
  {\bibfnamefont {Ray}\ \bibnamefont {Chang}}, \bibinfo {author} {\bibfnamefont
  {Nishaad}\ \bibnamefont {Khedkar}}, \bibinfo {author} {\bibfnamefont
  {Yiming~Cady}\ \bibnamefont {Feng}}, \bibinfo {author} {\bibfnamefont
  {Esha~A.}\ \bibnamefont {Umbarkar}}, \bibinfo {author} {\bibfnamefont {Xin}\
  \bibnamefont {Gui}}, \bibinfo {author} {\bibfnamefont {Lila V.~H.}\
  \bibnamefont {Rodgers}}, \bibinfo {author} {\bibfnamefont {Yichen}\
  \bibnamefont {Jia}}, \bibinfo {author} {\bibfnamefont {Mayer~M.}\
  \bibnamefont {Feldman}}, \bibinfo {author} {\bibfnamefont {Stephen~A.}\
  \bibnamefont {Lyon}}, \bibinfo {author} {\bibfnamefont {Mingzhao}\
  \bibnamefont {Liu}}, \bibinfo {author} {\bibfnamefont {Robert~J.}\
  \bibnamefont {Cava}}, \bibinfo {author} {\bibfnamefont {Andrew~A.}\
  \bibnamefont {Houck}}, \ and\ \bibinfo {author} {\bibfnamefont {Nathalie~P.}\
  \bibnamefont {de~Leon}},\ }\bibfield  {title} {\enquote {\bibinfo {title}
  {Disentangling losses in tantalum superconducting circuits},}\ }\href
  {\doibase 10.1103/PhysRevX.13.041005} {\bibfield  {journal} {\bibinfo
  {journal} {Phys. Rev. X}\ }\textbf {\bibinfo {volume} {13}},\ \bibinfo
  {pages} {041005} (\bibinfo {year} {2023})}\BibitemShut {NoStop}%
\bibitem [{\citenamefont {Neeley}\ \emph {et~al.}(2008)\citenamefont {Neeley},
  \citenamefont {Ansmann}, \citenamefont {Bialczak}, \citenamefont {Hofheinz},
  \citenamefont {Katz}, \citenamefont {Lucero}, \citenamefont {O’connell},
  \citenamefont {Wang}, \citenamefont {Cleland},\ and\ \citenamefont
  {Martinis}}]{neeley2008process}%
  \BibitemOpen
  \bibfield  {author} {\bibinfo {author} {\bibfnamefont {Matthew}\ \bibnamefont
  {Neeley}}, \bibinfo {author} {\bibfnamefont {Markus}\ \bibnamefont
  {Ansmann}}, \bibinfo {author} {\bibfnamefont {Radoslaw~C}\ \bibnamefont
  {Bialczak}}, \bibinfo {author} {\bibfnamefont {Max}\ \bibnamefont
  {Hofheinz}}, \bibinfo {author} {\bibfnamefont {Nadav}\ \bibnamefont {Katz}},
  \bibinfo {author} {\bibfnamefont {Erik}\ \bibnamefont {Lucero}}, \bibinfo
  {author} {\bibfnamefont {A}~\bibnamefont {O’connell}}, \bibinfo {author}
  {\bibfnamefont {Haohua}\ \bibnamefont {Wang}}, \bibinfo {author}
  {\bibfnamefont {Andrew~N}\ \bibnamefont {Cleland}}, \ and\ \bibinfo {author}
  {\bibfnamefont {John~M}\ \bibnamefont {Martinis}},\ }\bibfield  {title}
  {\enquote {\bibinfo {title} {Process tomography of quantum memory in a
  josephson-phase qubit coupled to a two-level state},}\ }\href
  {https://doi.org/10.1038/nphys972} {\bibfield  {journal} {\bibinfo  {journal}
  {Nature Physics}\ }\textbf {\bibinfo {volume} {4}},\ \bibinfo {pages}
  {523--526} (\bibinfo {year} {2008})}\BibitemShut {NoStop}%
\bibitem [{\citenamefont {Shalibo}\ \emph {et~al.}(2010)\citenamefont
  {Shalibo}, \citenamefont {Rofe}, \citenamefont {Shwa}, \citenamefont
  {Zeides}, \citenamefont {Neeley}, \citenamefont {Martinis},\ and\
  \citenamefont {Katz}}]{Shalibo2010Lifetime}%
  \BibitemOpen
  \bibfield  {author} {\bibinfo {author} {\bibfnamefont {Yoni}\ \bibnamefont
  {Shalibo}}, \bibinfo {author} {\bibfnamefont {Ya'ara}\ \bibnamefont {Rofe}},
  \bibinfo {author} {\bibfnamefont {David}\ \bibnamefont {Shwa}}, \bibinfo
  {author} {\bibfnamefont {Felix}\ \bibnamefont {Zeides}}, \bibinfo {author}
  {\bibfnamefont {Matthew}\ \bibnamefont {Neeley}}, \bibinfo {author}
  {\bibfnamefont {John~M.}\ \bibnamefont {Martinis}}, \ and\ \bibinfo {author}
  {\bibfnamefont {Nadav}\ \bibnamefont {Katz}},\ }\bibfield  {title} {\enquote
  {\bibinfo {title} {Lifetime and coherence of two-level defects in a josephson
  junction},}\ }\href {\doibase 10.1103/PhysRevLett.105.177001} {\bibfield
  {journal} {\bibinfo  {journal} {Phys. Rev. Lett.}\ }\textbf {\bibinfo
  {volume} {105}},\ \bibinfo {pages} {177001} (\bibinfo {year}
  {2010})}\BibitemShut {NoStop}%
\bibitem [{\citenamefont {Lisenfeld}\ \emph {et~al.}(2010)\citenamefont
  {Lisenfeld}, \citenamefont {M\"uller}, \citenamefont {Cole}, \citenamefont
  {Bushev}, \citenamefont {Lukashenko}, \citenamefont {Shnirman},\ and\
  \citenamefont {Ustinov}}]{Lisenfeld2010Measuring}%
  \BibitemOpen
  \bibfield  {author} {\bibinfo {author} {\bibfnamefont {J.}~\bibnamefont
  {Lisenfeld}}, \bibinfo {author} {\bibfnamefont {C.}~\bibnamefont {M\"uller}},
  \bibinfo {author} {\bibfnamefont {J.~H.}\ \bibnamefont {Cole}}, \bibinfo
  {author} {\bibfnamefont {P.}~\bibnamefont {Bushev}}, \bibinfo {author}
  {\bibfnamefont {A.}~\bibnamefont {Lukashenko}}, \bibinfo {author}
  {\bibfnamefont {A.}~\bibnamefont {Shnirman}}, \ and\ \bibinfo {author}
  {\bibfnamefont {A.~V.}\ \bibnamefont {Ustinov}},\ }\bibfield  {title}
  {\enquote {\bibinfo {title} {Measuring the temperature dependence of
  individual two-level systems by direct coherent control},}\ }\href {\doibase
  10.1103/PhysRevLett.105.230504} {\bibfield  {journal} {\bibinfo  {journal}
  {Phys. Rev. Lett.}\ }\textbf {\bibinfo {volume} {105}},\ \bibinfo {pages}
  {230504} (\bibinfo {year} {2010})}\BibitemShut {NoStop}%
\bibitem [{\citenamefont {Thorbeck}\ \emph {et~al.}(2023)\citenamefont
  {Thorbeck}, \citenamefont {Eddins}, \citenamefont {Lauer}, \citenamefont
  {McClure},\ and\ \citenamefont {Carroll}}]{Thorbeck2023Two}%
  \BibitemOpen
  \bibfield  {author} {\bibinfo {author} {\bibfnamefont {Ted}\ \bibnamefont
  {Thorbeck}}, \bibinfo {author} {\bibfnamefont {Andrew}\ \bibnamefont
  {Eddins}}, \bibinfo {author} {\bibfnamefont {Isaac}\ \bibnamefont {Lauer}},
  \bibinfo {author} {\bibfnamefont {Douglas~T.}\ \bibnamefont {McClure}}, \
  and\ \bibinfo {author} {\bibfnamefont {Malcolm}\ \bibnamefont {Carroll}},\
  }\bibfield  {title} {\enquote {\bibinfo {title} {Two-level-system dynamics in
  a superconducting qubit due to background ionizing radiation},}\ }\href
  {\doibase 10.1103/PRXQuantum.4.020356} {\bibfield  {journal} {\bibinfo
  {journal} {PRX Quantum}\ }\textbf {\bibinfo {volume} {4}},\ \bibinfo {pages}
  {020356} (\bibinfo {year} {2023})}\BibitemShut {NoStop}%
\bibitem [{\citenamefont {Abdurakhimov}\ \emph {et~al.}(2022)\citenamefont
  {Abdurakhimov}, \citenamefont {Mahboob}, \citenamefont {Toida}, \citenamefont
  {Kakuyanagi}, \citenamefont {Matsuzaki},\ and\ \citenamefont
  {Saito}}]{Abdurakhimov222Identification}%
  \BibitemOpen
  \bibfield  {author} {\bibinfo {author} {\bibfnamefont {Leonid~V.}\
  \bibnamefont {Abdurakhimov}}, \bibinfo {author} {\bibfnamefont {Imran}\
  \bibnamefont {Mahboob}}, \bibinfo {author} {\bibfnamefont {Hiraku}\
  \bibnamefont {Toida}}, \bibinfo {author} {\bibfnamefont {Kosuke}\
  \bibnamefont {Kakuyanagi}}, \bibinfo {author} {\bibfnamefont {Yuichiro}\
  \bibnamefont {Matsuzaki}}, \ and\ \bibinfo {author} {\bibfnamefont {Shiro}\
  \bibnamefont {Saito}},\ }\bibfield  {title} {\enquote {\bibinfo {title}
  {Identification of different types of high-frequency defects in
  superconducting qubits},}\ }\href {\doibase 10.1103/PRXQuantum.3.040332}
  {\bibfield  {journal} {\bibinfo  {journal} {PRX Quantum}\ }\textbf {\bibinfo
  {volume} {3}},\ \bibinfo {pages} {040332} (\bibinfo {year}
  {2022})}\BibitemShut {NoStop}%
\bibitem [{\citenamefont {Spiecker}\ \emph {et~al.}(2023)\citenamefont
  {Spiecker}, \citenamefont {Paluch}, \citenamefont {Gosling}, \citenamefont
  {Drucker}, \citenamefont {Matityahu}, \citenamefont {Gusenkova},
  \citenamefont {G{\"u}nzler}, \citenamefont {Rieger}, \citenamefont
  {Takmakov}, \citenamefont {Valenti} \emph {et~al.}}]{spiecker2023two}%
  \BibitemOpen
  \bibfield  {author} {\bibinfo {author} {\bibfnamefont {Martin}\ \bibnamefont
  {Spiecker}}, \bibinfo {author} {\bibfnamefont {Patrick}\ \bibnamefont
  {Paluch}}, \bibinfo {author} {\bibfnamefont {Nicolas}\ \bibnamefont
  {Gosling}}, \bibinfo {author} {\bibfnamefont {Niv}\ \bibnamefont {Drucker}},
  \bibinfo {author} {\bibfnamefont {Shlomi}\ \bibnamefont {Matityahu}},
  \bibinfo {author} {\bibfnamefont {Daria}\ \bibnamefont {Gusenkova}}, \bibinfo
  {author} {\bibfnamefont {Simon}\ \bibnamefont {G{\"u}nzler}}, \bibinfo
  {author} {\bibfnamefont {Dennis}\ \bibnamefont {Rieger}}, \bibinfo {author}
  {\bibfnamefont {Ivan}\ \bibnamefont {Takmakov}}, \bibinfo {author}
  {\bibfnamefont {Francesco}\ \bibnamefont {Valenti}},  \emph {et~al.},\
  }\bibfield  {title} {\enquote {\bibinfo {title} {Two-level system
  hyperpolarization using a quantum szilard engine},}\ }\href
  {https://doi.org/10.1038/s41567-023-02082-8} {\bibfield  {journal} {\bibinfo
  {journal} {Nature Physics}\ ,\ \bibinfo {pages} {1--6}} (\bibinfo {year}
  {2023})}\BibitemShut {NoStop}%
\bibitem [{\citenamefont {Carruzzo}\ \emph {et~al.}(2021)\citenamefont
  {Carruzzo}, \citenamefont {Bilmes}, \citenamefont {Lisenfeld}, \citenamefont
  {Yu}, \citenamefont {Wang}, \citenamefont {Wan}, \citenamefont {Schmidt},\
  and\ \citenamefont {Yu}}]{Carruzzo2021}%
  \BibitemOpen
  \bibfield  {author} {\bibinfo {author} {\bibfnamefont {Herve~M.}\
  \bibnamefont {Carruzzo}}, \bibinfo {author} {\bibfnamefont {Alexander}\
  \bibnamefont {Bilmes}}, \bibinfo {author} {\bibfnamefont {J\"urgen}\
  \bibnamefont {Lisenfeld}}, \bibinfo {author} {\bibfnamefont {Zheng}\
  \bibnamefont {Yu}}, \bibinfo {author} {\bibfnamefont {Bu}~\bibnamefont
  {Wang}}, \bibinfo {author} {\bibfnamefont {Zhongyi}\ \bibnamefont {Wan}},
  \bibinfo {author} {\bibfnamefont {J.~R.}\ \bibnamefont {Schmidt}}, \ and\
  \bibinfo {author} {\bibfnamefont {Clare~C.}\ \bibnamefont {Yu}},\ }\bibfield
  {title} {\enquote {\bibinfo {title} {Distribution of two-level system
  couplings to strain and electric fields in glasses at low temperatures},}\
  }\href {\doibase 10.1103/PhysRevB.104.134203} {\bibfield  {journal} {\bibinfo
   {journal} {Phys. Rev. B}\ }\textbf {\bibinfo {volume} {104}},\ \bibinfo
  {pages} {134203} (\bibinfo {year} {2021})}\BibitemShut {NoStop}%
\bibitem [{\citenamefont {Constantin}\ \emph {et~al.}(2009)\citenamefont
  {Constantin}, \citenamefont {Yu},\ and\ \citenamefont
  {Martinis}}]{Constantin2009}%
  \BibitemOpen
  \bibfield  {author} {\bibinfo {author} {\bibfnamefont {Magdalena}\
  \bibnamefont {Constantin}}, \bibinfo {author} {\bibfnamefont {Clare~C.}\
  \bibnamefont {Yu}}, \ and\ \bibinfo {author} {\bibfnamefont {John~M.}\
  \bibnamefont {Martinis}},\ }\bibfield  {title} {\enquote {\bibinfo {title}
  {Saturation of two-level systems and charge noise in josephson junction
  qubits},}\ }\href {\doibase 10.1103/PhysRevB.79.094520} {\bibfield  {journal}
  {\bibinfo  {journal} {Phys. Rev. B}\ }\textbf {\bibinfo {volume} {79}},\
  \bibinfo {pages} {094520} (\bibinfo {year} {2009})}\BibitemShut {NoStop}%
\bibitem [{\citenamefont {Sarabi}\ \emph {et~al.}(2016)\citenamefont {Sarabi},
  \citenamefont {Ramanayaka}, \citenamefont {Burin}, \citenamefont
  {Wellstood},\ and\ \citenamefont {Osborn}}]{Sarabi2016}%
  \BibitemOpen
  \bibfield  {author} {\bibinfo {author} {\bibfnamefont {B.}~\bibnamefont
  {Sarabi}}, \bibinfo {author} {\bibfnamefont {A.~N.}\ \bibnamefont
  {Ramanayaka}}, \bibinfo {author} {\bibfnamefont {A.~L.}\ \bibnamefont
  {Burin}}, \bibinfo {author} {\bibfnamefont {F.~C.}\ \bibnamefont
  {Wellstood}}, \ and\ \bibinfo {author} {\bibfnamefont {K.~D.}\ \bibnamefont
  {Osborn}},\ }\bibfield  {title} {\enquote {\bibinfo {title} {Projected dipole
  moments of individual two-level defects extracted using circuit quantum
  electrodynamics},}\ }\href {\doibase 10.1103/PhysRevLett.116.167002}
  {\bibfield  {journal} {\bibinfo  {journal} {Phys. Rev. Lett.}\ }\textbf
  {\bibinfo {volume} {116}},\ \bibinfo {pages} {167002} (\bibinfo {year}
  {2016})}\BibitemShut {NoStop}%
\bibitem [{\citenamefont {Oh}\ \emph {et~al.}(2006)\citenamefont {Oh},
  \citenamefont {Cicak}, \citenamefont {Kline}, \citenamefont {Sillanp\"a\"a},
  \citenamefont {Osborn}, \citenamefont {Whittaker}, \citenamefont {Simmonds},\
  and\ \citenamefont {Pappas}}]{Seongshik2006}%
  \BibitemOpen
  \bibfield  {author} {\bibinfo {author} {\bibfnamefont {Seongshik}\
  \bibnamefont {Oh}}, \bibinfo {author} {\bibfnamefont {Katarina}\ \bibnamefont
  {Cicak}}, \bibinfo {author} {\bibfnamefont {Jeffrey~S.}\ \bibnamefont
  {Kline}}, \bibinfo {author} {\bibfnamefont {Mika~A.}\ \bibnamefont
  {Sillanp\"a\"a}}, \bibinfo {author} {\bibfnamefont {Kevin~D.}\ \bibnamefont
  {Osborn}}, \bibinfo {author} {\bibfnamefont {Jed~D.}\ \bibnamefont
  {Whittaker}}, \bibinfo {author} {\bibfnamefont {Raymond~W.}\ \bibnamefont
  {Simmonds}}, \ and\ \bibinfo {author} {\bibfnamefont {David~P.}\ \bibnamefont
  {Pappas}},\ }\bibfield  {title} {\enquote {\bibinfo {title} {Elimination of
  two level fluctuators in superconducting quantum bits by an epitaxial tunnel
  barrier},}\ }\href {\doibase 10.1103/PhysRevB.74.100502} {\bibfield
  {journal} {\bibinfo  {journal} {Phys. Rev. B}\ }\textbf {\bibinfo {volume}
  {74}},\ \bibinfo {pages} {100502} (\bibinfo {year} {2006})}\BibitemShut
  {NoStop}%
\bibitem [{\citenamefont {de~Graaf}\ \emph {et~al.}(2021)\citenamefont
  {de~Graaf}, \citenamefont {Mahashabde}, \citenamefont {Kubatkin},
  \citenamefont {Tzalenchuk},\ and\ \citenamefont
  {Danilov}}]{Graaf2021Quantifying}%
  \BibitemOpen
  \bibfield  {author} {\bibinfo {author} {\bibfnamefont {S.~E.}\ \bibnamefont
  {de~Graaf}}, \bibinfo {author} {\bibfnamefont {S.}~\bibnamefont
  {Mahashabde}}, \bibinfo {author} {\bibfnamefont {S.~E.}\ \bibnamefont
  {Kubatkin}}, \bibinfo {author} {\bibfnamefont {A.~Ya.}\ \bibnamefont
  {Tzalenchuk}}, \ and\ \bibinfo {author} {\bibfnamefont {A.~V.}\ \bibnamefont
  {Danilov}},\ }\bibfield  {title} {\enquote {\bibinfo {title} {Quantifying
  dynamics and interactions of individual spurious low-energy fluctuators in
  superconducting circuits},}\ }\href {\doibase 10.1103/PhysRevB.103.174103}
  {\bibfield  {journal} {\bibinfo  {journal} {Phys. Rev. B}\ }\textbf {\bibinfo
  {volume} {103}},\ \bibinfo {pages} {174103} (\bibinfo {year}
  {2021})}\BibitemShut {NoStop}%
\bibitem [{\citenamefont {Burnett}\ \emph {et~al.}(2014)\citenamefont
  {Burnett}, \citenamefont {Faoro}, \citenamefont {Wisby}, \citenamefont
  {Gurtovoi}, \citenamefont {Chernykh}, \citenamefont {Mikhailov},
  \citenamefont {Tulin}, \citenamefont {Shaikhaidarov}, \citenamefont
  {Antonov}, \citenamefont {Meeson} \emph {et~al.}}]{burnett2014evidence}%
  \BibitemOpen
  \bibfield  {author} {\bibinfo {author} {\bibfnamefont {Jonathan}\
  \bibnamefont {Burnett}}, \bibinfo {author} {\bibfnamefont {Lara}\
  \bibnamefont {Faoro}}, \bibinfo {author} {\bibfnamefont {Ilana}\ \bibnamefont
  {Wisby}}, \bibinfo {author} {\bibfnamefont {Vladimir~L}\ \bibnamefont
  {Gurtovoi}}, \bibinfo {author} {\bibfnamefont {Alexey~V}\ \bibnamefont
  {Chernykh}}, \bibinfo {author} {\bibfnamefont {Gennady~M}\ \bibnamefont
  {Mikhailov}}, \bibinfo {author} {\bibfnamefont {Vyacheslav~A}\ \bibnamefont
  {Tulin}}, \bibinfo {author} {\bibfnamefont {Rais}\ \bibnamefont
  {Shaikhaidarov}}, \bibinfo {author} {\bibfnamefont {Vladimir}\ \bibnamefont
  {Antonov}}, \bibinfo {author} {\bibfnamefont {Phil~J}\ \bibnamefont
  {Meeson}},  \emph {et~al.},\ }\bibfield  {title} {\enquote {\bibinfo {title}
  {Evidence for interacting two-level systems from the 1/f noise of a
  superconducting resonator},}\ }\href {https://doi.org/10.1038/ncomms5119}
  {\bibfield  {journal} {\bibinfo  {journal} {Nature Communications}\ }\textbf
  {\bibinfo {volume} {5}},\ \bibinfo {pages} {4119} (\bibinfo {year}
  {2014})}\BibitemShut {NoStop}%
\bibitem [{\citenamefont {Grabovskij}\ \emph {et~al.}(2012)\citenamefont
  {Grabovskij}, \citenamefont {Peichl}, \citenamefont {Lisenfeld},
  \citenamefont {Weiss},\ and\ \citenamefont {Ustinov}}]{grabovskij2012strain}%
  \BibitemOpen
  \bibfield  {author} {\bibinfo {author} {\bibfnamefont {Grigorij~J}\
  \bibnamefont {Grabovskij}}, \bibinfo {author} {\bibfnamefont {Torben}\
  \bibnamefont {Peichl}}, \bibinfo {author} {\bibfnamefont {J{\"u}rgen}\
  \bibnamefont {Lisenfeld}}, \bibinfo {author} {\bibfnamefont {Georg}\
  \bibnamefont {Weiss}}, \ and\ \bibinfo {author} {\bibfnamefont {Alexey~V}\
  \bibnamefont {Ustinov}},\ }\bibfield  {title} {\enquote {\bibinfo {title}
  {Strain tuning of individual atomic tunneling systems detected by a
  superconducting qubit},}\ }\href {10.1126/science.1226487} {\bibfield
  {journal} {\bibinfo  {journal} {Science}\ }\textbf {\bibinfo {volume}
  {338}},\ \bibinfo {pages} {232--234} (\bibinfo {year} {2012})}\BibitemShut
  {NoStop}%
\bibitem [{\citenamefont {Hung}\ \emph {et~al.}(2022)\citenamefont {Hung},
  \citenamefont {Yu}, \citenamefont {Foroozani}, \citenamefont {Fritz},
  \citenamefont {Gerthsen},\ and\ \citenamefont {Osborn}}]{Hung2022}%
  \BibitemOpen
  \bibfield  {author} {\bibinfo {author} {\bibfnamefont {Chih-Chiao}\
  \bibnamefont {Hung}}, \bibinfo {author} {\bibfnamefont {Liuqi}\ \bibnamefont
  {Yu}}, \bibinfo {author} {\bibfnamefont {Neda}\ \bibnamefont {Foroozani}},
  \bibinfo {author} {\bibfnamefont {Stefan}\ \bibnamefont {Fritz}}, \bibinfo
  {author} {\bibfnamefont {Dagmar}\ \bibnamefont {Gerthsen}}, \ and\ \bibinfo
  {author} {\bibfnamefont {Kevin~D.}\ \bibnamefont {Osborn}},\ }\bibfield
  {title} {\enquote {\bibinfo {title} {Probing hundreds of individual quantum
  defects in polycrystalline and amorphous alumina},}\ }\href {\doibase
  10.1103/PhysRevApplied.17.034025} {\bibfield  {journal} {\bibinfo  {journal}
  {Phys. Rev. Appl.}\ }\textbf {\bibinfo {volume} {17}},\ \bibinfo {pages}
  {034025} (\bibinfo {year} {2022})}\BibitemShut {NoStop}%
\bibitem [{\citenamefont {Paik}\ \emph {et~al.}(2011)\citenamefont {Paik},
  \citenamefont {Schuster}, \citenamefont {Bishop}, \citenamefont {Kirchmair},
  \citenamefont {Catelani}, \citenamefont {Sears}, \citenamefont {Johnson},
  \citenamefont {Reagor}, \citenamefont {Frunzio}, \citenamefont {Glazman},
  \citenamefont {Girvin}, \citenamefont {Devoret},\ and\ \citenamefont
  {Schoelkopf}}]{Paik2011}%
  \BibitemOpen
  \bibfield  {author} {\bibinfo {author} {\bibfnamefont {Hanhee}\ \bibnamefont
  {Paik}}, \bibinfo {author} {\bibfnamefont {D.~I.}\ \bibnamefont {Schuster}},
  \bibinfo {author} {\bibfnamefont {Lev~S.}\ \bibnamefont {Bishop}}, \bibinfo
  {author} {\bibfnamefont {G.}~\bibnamefont {Kirchmair}}, \bibinfo {author}
  {\bibfnamefont {G.}~\bibnamefont {Catelani}}, \bibinfo {author}
  {\bibfnamefont {A.~P.}\ \bibnamefont {Sears}}, \bibinfo {author}
  {\bibfnamefont {B.~R.}\ \bibnamefont {Johnson}}, \bibinfo {author}
  {\bibfnamefont {M.~J.}\ \bibnamefont {Reagor}}, \bibinfo {author}
  {\bibfnamefont {L.}~\bibnamefont {Frunzio}}, \bibinfo {author} {\bibfnamefont
  {L.~I.}\ \bibnamefont {Glazman}}, \bibinfo {author} {\bibfnamefont {S.~M.}\
  \bibnamefont {Girvin}}, \bibinfo {author} {\bibfnamefont {M.~H.}\
  \bibnamefont {Devoret}}, \ and\ \bibinfo {author} {\bibfnamefont {R.~J.}\
  \bibnamefont {Schoelkopf}},\ }\bibfield  {title} {\enquote {\bibinfo {title}
  {Observation of high coherence in josephson junction qubits measured in a
  three-dimensional circuit qed architecture},}\ }\href {\doibase
  10.1103/PhysRevLett.107.240501} {\bibfield  {journal} {\bibinfo  {journal}
  {Phys. Rev. Lett.}\ }\textbf {\bibinfo {volume} {107}},\ \bibinfo {pages}
  {240501} (\bibinfo {year} {2011})}\BibitemShut {NoStop}%
\bibitem [{\citenamefont {Burnett}\ \emph {et~al.}(2019)\citenamefont
  {Burnett}, \citenamefont {Bengtsson}, \citenamefont {Scigliuzzo},
  \citenamefont {Niepce}, \citenamefont {Kudra}, \citenamefont {Delsing},\ and\
  \citenamefont {Bylander}}]{burnett2019decoherence}%
  \BibitemOpen
  \bibfield  {author} {\bibinfo {author} {\bibfnamefont {Jonathan~J}\
  \bibnamefont {Burnett}}, \bibinfo {author} {\bibfnamefont {Andreas}\
  \bibnamefont {Bengtsson}}, \bibinfo {author} {\bibfnamefont {Marco}\
  \bibnamefont {Scigliuzzo}}, \bibinfo {author} {\bibfnamefont {David}\
  \bibnamefont {Niepce}}, \bibinfo {author} {\bibfnamefont {Marina}\
  \bibnamefont {Kudra}}, \bibinfo {author} {\bibfnamefont {Per}\ \bibnamefont
  {Delsing}}, \ and\ \bibinfo {author} {\bibfnamefont {Jonas}\ \bibnamefont
  {Bylander}},\ }\bibfield  {title} {\enquote {\bibinfo {title} {Decoherence
  benchmarking of superconducting qubits},}\ }\href
  {https://doi.org/10.1038/s41534-019-0168-5} {\bibfield  {journal} {\bibinfo
  {journal} {npj Quantum Information}\ }\textbf {\bibinfo {volume} {5}},\
  \bibinfo {pages} {54} (\bibinfo {year} {2019})}\BibitemShut {NoStop}%
\bibitem [{\citenamefont {Schl\"or}\ \emph {et~al.}(2019)\citenamefont
  {Schl\"or}, \citenamefont {Lisenfeld}, \citenamefont {M\"uller},
  \citenamefont {Bilmes}, \citenamefont {Schneider}, \citenamefont {Pappas},
  \citenamefont {Ustinov},\ and\ \citenamefont
  {Weides}}]{Steffen2019Correlating}%
  \BibitemOpen
  \bibfield  {author} {\bibinfo {author} {\bibfnamefont {Steffen}\ \bibnamefont
  {Schl\"or}}, \bibinfo {author} {\bibfnamefont {J\"urgen}\ \bibnamefont
  {Lisenfeld}}, \bibinfo {author} {\bibfnamefont {Clemens}\ \bibnamefont
  {M\"uller}}, \bibinfo {author} {\bibfnamefont {Alexander}\ \bibnamefont
  {Bilmes}}, \bibinfo {author} {\bibfnamefont {Andre}\ \bibnamefont
  {Schneider}}, \bibinfo {author} {\bibfnamefont {David~P.}\ \bibnamefont
  {Pappas}}, \bibinfo {author} {\bibfnamefont {Alexey~V.}\ \bibnamefont
  {Ustinov}}, \ and\ \bibinfo {author} {\bibfnamefont {Martin}\ \bibnamefont
  {Weides}},\ }\bibfield  {title} {\enquote {\bibinfo {title} {Correlating
  decoherence in transmon qubits: Low frequency noise by single fluctuators},}\
  }\href {\doibase 10.1103/PhysRevLett.123.190502} {\bibfield  {journal}
  {\bibinfo  {journal} {Phys. Rev. Lett.}\ }\textbf {\bibinfo {volume} {123}},\
  \bibinfo {pages} {190502} (\bibinfo {year} {2019})}\BibitemShut {NoStop}%
\bibitem [{\citenamefont {Levine}\ \emph {et~al.}(2023)\citenamefont {Levine},
  \citenamefont {Haim}, \citenamefont {Hung}, \citenamefont {Alidoust},
  \citenamefont {Kalaee}, \citenamefont {DeLorenzo}, \citenamefont {Wollack},
  \citenamefont {Arriola}, \citenamefont {Khalajhedayati}, \citenamefont
  {Vaknin} \emph {et~al.}}]{levine2023demonstrating}%
  \BibitemOpen
  \bibfield  {author} {\bibinfo {author} {\bibfnamefont {Harry}\ \bibnamefont
  {Levine}}, \bibinfo {author} {\bibfnamefont {Arbel}\ \bibnamefont {Haim}},
  \bibinfo {author} {\bibfnamefont {Jimmy~SC}\ \bibnamefont {Hung}}, \bibinfo
  {author} {\bibfnamefont {Nasser}\ \bibnamefont {Alidoust}}, \bibinfo {author}
  {\bibfnamefont {Mahmoud}\ \bibnamefont {Kalaee}}, \bibinfo {author}
  {\bibfnamefont {Laura}\ \bibnamefont {DeLorenzo}}, \bibinfo {author}
  {\bibfnamefont {E~Alex}\ \bibnamefont {Wollack}}, \bibinfo {author}
  {\bibfnamefont {Patricio~Arrangoiz}\ \bibnamefont {Arriola}}, \bibinfo
  {author} {\bibfnamefont {Amirhossein}\ \bibnamefont {Khalajhedayati}},
  \bibinfo {author} {\bibfnamefont {Yotam}\ \bibnamefont {Vaknin}},  \emph
  {et~al.},\ }\bibfield  {title} {\enquote {\bibinfo {title} {Demonstrating a
  long-coherence dual-rail erasure qubit using tunable transmons},}\ }\href
  {https://doi.org/10.48550/arXiv.2307.08737} {\bibfield  {journal} {\bibinfo
  {journal} {arXiv preprint arXiv:2307.08737}\ } (\bibinfo {year}
  {2023})}\BibitemShut {NoStop}%
\bibitem [{\citenamefont {Gertler}\ \emph {et~al.}(2021)\citenamefont
  {Gertler}, \citenamefont {Baker}, \citenamefont {Li}, \citenamefont {Shirol},
  \citenamefont {Koch},\ and\ \citenamefont {Wang}}]{gertler2021protecting}%
  \BibitemOpen
  \bibfield  {author} {\bibinfo {author} {\bibfnamefont {Jeffrey~M}\
  \bibnamefont {Gertler}}, \bibinfo {author} {\bibfnamefont {Brian}\
  \bibnamefont {Baker}}, \bibinfo {author} {\bibfnamefont {Juliang}\
  \bibnamefont {Li}}, \bibinfo {author} {\bibfnamefont {Shruti}\ \bibnamefont
  {Shirol}}, \bibinfo {author} {\bibfnamefont {Jens}\ \bibnamefont {Koch}}, \
  and\ \bibinfo {author} {\bibfnamefont {Chen}\ \bibnamefont {Wang}},\
  }\bibfield  {title} {\enquote {\bibinfo {title} {Protecting a bosonic qubit
  with autonomous quantum error correction},}\ }\href
  {https://doi.org/10.1038/s41586-021-03257-0} {\bibfield  {journal} {\bibinfo
  {journal} {Nature}\ }\textbf {\bibinfo {volume} {590}},\ \bibinfo {pages}
  {243--248} (\bibinfo {year} {2021})}\BibitemShut {NoStop}%
\bibitem [{\citenamefont {Constantin}\ and\ \citenamefont
  {Yu}(2007)}]{Constantin2007}%
  \BibitemOpen
  \bibfield  {author} {\bibinfo {author} {\bibfnamefont {Magdalena}\
  \bibnamefont {Constantin}}\ and\ \bibinfo {author} {\bibfnamefont {Clare~C.}\
  \bibnamefont {Yu}},\ }\bibfield  {title} {\enquote {\bibinfo {title}
  {Microscopic model of critical current noise in josephson junctions},}\
  }\href {\doibase 10.1103/PhysRevLett.99.207001} {\bibfield  {journal}
  {\bibinfo  {journal} {Phys. Rev. Lett.}\ }\textbf {\bibinfo {volume} {99}},\
  \bibinfo {pages} {207001} (\bibinfo {year} {2007})}\BibitemShut {NoStop}%
\bibitem [{\citenamefont {Zaretskey}\ \emph {et~al.}(2013)\citenamefont
  {Zaretskey}, \citenamefont {Suri}, \citenamefont {Novikov}, \citenamefont
  {Wellstood},\ and\ \citenamefont {Palmer}}]{Zaretskey2013}%
  \BibitemOpen
  \bibfield  {author} {\bibinfo {author} {\bibfnamefont {V.}~\bibnamefont
  {Zaretskey}}, \bibinfo {author} {\bibfnamefont {B.}~\bibnamefont {Suri}},
  \bibinfo {author} {\bibfnamefont {S.}~\bibnamefont {Novikov}}, \bibinfo
  {author} {\bibfnamefont {F.~C.}\ \bibnamefont {Wellstood}}, \ and\ \bibinfo
  {author} {\bibfnamefont {B.~S.}\ \bibnamefont {Palmer}},\ }\bibfield  {title}
  {\enquote {\bibinfo {title} {Spectroscopy of a cooper-pair box coupled to a
  two-level system via charge and critical current},}\ }\href {\doibase
  10.1103/PhysRevB.87.174522} {\bibfield  {journal} {\bibinfo  {journal} {Phys.
  Rev. B}\ }\textbf {\bibinfo {volume} {87}},\ \bibinfo {pages} {174522}
  (\bibinfo {year} {2013})}\BibitemShut {NoStop}%
\bibitem [{\citenamefont {Nugroho}\ \emph {et~al.}(2013)\citenamefont
  {Nugroho}, \citenamefont {Orlyanchik},\ and\ \citenamefont
  {Van~Harlingen}}]{nugroho2013low}%
  \BibitemOpen
  \bibfield  {author} {\bibinfo {author} {\bibfnamefont {CD}~\bibnamefont
  {Nugroho}}, \bibinfo {author} {\bibfnamefont {V}~\bibnamefont {Orlyanchik}},
  \ and\ \bibinfo {author} {\bibfnamefont {DJ}~\bibnamefont {Van~Harlingen}},\
  }\bibfield  {title} {\enquote {\bibinfo {title} {Low frequency resistance and
  critical current fluctuations in al-based josephson junctions},}\ }\href
  {https://doi.org/10.1063/1.4801521} {\bibfield  {journal} {\bibinfo
  {journal} {Applied Physics Letters}\ }\textbf {\bibinfo {volume} {102}}
  (\bibinfo {year} {2013})}\BibitemShut {NoStop}%
\bibitem [{\citenamefont {Cole}\ \emph {et~al.}(2010)\citenamefont {Cole},
  \citenamefont {M{\"u}ller}, \citenamefont {Bushev}, \citenamefont
  {Grabovskij}, \citenamefont {Lisenfeld}, \citenamefont {Lukashenko},
  \citenamefont {Ustinov},\ and\ \citenamefont
  {Shnirman}}]{cole2010quantitative}%
  \BibitemOpen
  \bibfield  {author} {\bibinfo {author} {\bibfnamefont {Jared~H}\ \bibnamefont
  {Cole}}, \bibinfo {author} {\bibfnamefont {Clemens}\ \bibnamefont
  {M{\"u}ller}}, \bibinfo {author} {\bibfnamefont {Pavel}\ \bibnamefont
  {Bushev}}, \bibinfo {author} {\bibfnamefont {Grigorij~J}\ \bibnamefont
  {Grabovskij}}, \bibinfo {author} {\bibfnamefont {J{\"u}rgen}\ \bibnamefont
  {Lisenfeld}}, \bibinfo {author} {\bibfnamefont {Alexander}\ \bibnamefont
  {Lukashenko}}, \bibinfo {author} {\bibfnamefont {Alexey~V}\ \bibnamefont
  {Ustinov}}, \ and\ \bibinfo {author} {\bibfnamefont {Alexander}\ \bibnamefont
  {Shnirman}},\ }\bibfield  {title} {\enquote {\bibinfo {title} {Quantitative
  evaluation of defect-models in superconducting phase qubits},}\ }\href
  {https://doi.org/10.1063/1.3529457} {\bibfield  {journal} {\bibinfo
  {journal} {Applied Physics Letters}\ }\textbf {\bibinfo {volume} {97}}
  (\bibinfo {year} {2010})}\BibitemShut {NoStop}%
\bibitem [{\citenamefont {Rist{\`e}}\ \emph {et~al.}(2013)\citenamefont
  {Rist{\`e}}, \citenamefont {Bultink}, \citenamefont {Tiggelman},
  \citenamefont {Schouten}, \citenamefont {Lehnert},\ and\ \citenamefont
  {DiCarlo}}]{riste2013millisecond}%
  \BibitemOpen
  \bibfield  {author} {\bibinfo {author} {\bibfnamefont {Diego}\ \bibnamefont
  {Rist{\`e}}}, \bibinfo {author} {\bibfnamefont {CC}~\bibnamefont {Bultink}},
  \bibinfo {author} {\bibfnamefont {Marijn~J}\ \bibnamefont {Tiggelman}},
  \bibinfo {author} {\bibfnamefont {Raymond~N}\ \bibnamefont {Schouten}},
  \bibinfo {author} {\bibfnamefont {Konrad~W}\ \bibnamefont {Lehnert}}, \ and\
  \bibinfo {author} {\bibfnamefont {Leonardo}\ \bibnamefont {DiCarlo}},\
  }\bibfield  {title} {\enquote {\bibinfo {title} {Millisecond charge-parity
  fluctuations and induced decoherence in a superconducting transmon qubit},}\
  }\href {https://doi.org/10.1038/ncomms2936} {\bibfield  {journal} {\bibinfo
  {journal} {Nature communications}\ }\textbf {\bibinfo {volume} {4}},\
  \bibinfo {pages} {1913} (\bibinfo {year} {2013})}\BibitemShut {NoStop}%
\bibitem [{\citenamefont {Connolly}\ \emph {et~al.}(2023)\citenamefont
  {Connolly}, \citenamefont {Kurilovich}, \citenamefont {Diamond},
  \citenamefont {Nho}, \citenamefont {B{\o}ttcher}, \citenamefont {Glazman},
  \citenamefont {Fatemi},\ and\ \citenamefont
  {Devoret}}]{connolly2023coexistence}%
  \BibitemOpen
  \bibfield  {author} {\bibinfo {author} {\bibfnamefont {Thomas}\ \bibnamefont
  {Connolly}}, \bibinfo {author} {\bibfnamefont {Pavel~D}\ \bibnamefont
  {Kurilovich}}, \bibinfo {author} {\bibfnamefont {Spencer}\ \bibnamefont
  {Diamond}}, \bibinfo {author} {\bibfnamefont {Heekun}\ \bibnamefont {Nho}},
  \bibinfo {author} {\bibfnamefont {Charlotte~GL}\ \bibnamefont {B{\o}ttcher}},
  \bibinfo {author} {\bibfnamefont {Leonid~I}\ \bibnamefont {Glazman}},
  \bibinfo {author} {\bibfnamefont {Valla}\ \bibnamefont {Fatemi}}, \ and\
  \bibinfo {author} {\bibfnamefont {Michel~H}\ \bibnamefont {Devoret}},\
  }\bibfield  {title} {\enquote {\bibinfo {title} {Coexistence of
  nonequilibrium density and equilibrium energy distribution of quasiparticles
  in a superconducting qubit},}\ }\href
  {https://doi.org/10.48550/arXiv.2302.12330} {\bibfield  {journal} {\bibinfo
  {journal} {arXiv preprint arXiv:2302.12330}\ } (\bibinfo {year}
  {2023})}\BibitemShut {NoStop}%
\bibitem [{\citenamefont {Serniak}\ \emph {et~al.}(2018)\citenamefont
  {Serniak}, \citenamefont {Hays}, \citenamefont {de~Lange}, \citenamefont
  {Diamond}, \citenamefont {Shankar}, \citenamefont {Burkhart}, \citenamefont
  {Frunzio}, \citenamefont {Houzet},\ and\ \citenamefont
  {Devoret}}]{Serniak2012Hot}%
  \BibitemOpen
  \bibfield  {author} {\bibinfo {author} {\bibfnamefont {K.}~\bibnamefont
  {Serniak}}, \bibinfo {author} {\bibfnamefont {M.}~\bibnamefont {Hays}},
  \bibinfo {author} {\bibfnamefont {G.}~\bibnamefont {de~Lange}}, \bibinfo
  {author} {\bibfnamefont {S.}~\bibnamefont {Diamond}}, \bibinfo {author}
  {\bibfnamefont {S.}~\bibnamefont {Shankar}}, \bibinfo {author} {\bibfnamefont
  {L.~D.}\ \bibnamefont {Burkhart}}, \bibinfo {author} {\bibfnamefont
  {L.}~\bibnamefont {Frunzio}}, \bibinfo {author} {\bibfnamefont
  {M.}~\bibnamefont {Houzet}}, \ and\ \bibinfo {author} {\bibfnamefont {M.~H.}\
  \bibnamefont {Devoret}},\ }\bibfield  {title} {\enquote {\bibinfo {title}
  {Hot nonequilibrium quasiparticles in transmon qubits},}\ }\href {\doibase
  10.1103/PhysRevLett.121.157701} {\bibfield  {journal} {\bibinfo  {journal}
  {Phys. Rev. Lett.}\ }\textbf {\bibinfo {volume} {121}},\ \bibinfo {pages}
  {157701} (\bibinfo {year} {2018})}\BibitemShut {NoStop}%
\bibitem [{\citenamefont {Wilen}\ \emph {et~al.}(2021)\citenamefont {Wilen},
  \citenamefont {Abdullah}, \citenamefont {Kurinsky}, \citenamefont {Stanford},
  \citenamefont {Cardani}, \citenamefont {d’Imperio}, \citenamefont {Tomei},
  \citenamefont {Faoro}, \citenamefont {Ioffe}, \citenamefont {Liu} \emph
  {et~al.}}]{wilen2021correlated}%
  \BibitemOpen
  \bibfield  {author} {\bibinfo {author} {\bibfnamefont {Christopher~D}\
  \bibnamefont {Wilen}}, \bibinfo {author} {\bibfnamefont {S}~\bibnamefont
  {Abdullah}}, \bibinfo {author} {\bibfnamefont {NA}~\bibnamefont {Kurinsky}},
  \bibinfo {author} {\bibfnamefont {C}~\bibnamefont {Stanford}}, \bibinfo
  {author} {\bibfnamefont {L}~\bibnamefont {Cardani}}, \bibinfo {author}
  {\bibfnamefont {G}~\bibnamefont {d’Imperio}}, \bibinfo {author}
  {\bibfnamefont {C}~\bibnamefont {Tomei}}, \bibinfo {author} {\bibfnamefont
  {L}~\bibnamefont {Faoro}}, \bibinfo {author} {\bibfnamefont {LB}~\bibnamefont
  {Ioffe}}, \bibinfo {author} {\bibfnamefont {CH}~\bibnamefont {Liu}},  \emph
  {et~al.},\ }\bibfield  {title} {\enquote {\bibinfo {title} {Correlated charge
  noise and relaxation errors in superconducting qubits},}\ }\href
  {https://doi.org/10.1038/s41586-021-03557-5} {\bibfield  {journal} {\bibinfo
  {journal} {Nature}\ }\textbf {\bibinfo {volume} {594}},\ \bibinfo {pages}
  {369--373} (\bibinfo {year} {2021})}\BibitemShut {NoStop}%
\bibitem [{\citenamefont {Chen}\ \emph {et~al.}(2023)\citenamefont {Chen},
  \citenamefont {Owens}, \citenamefont {Putterman}, \citenamefont
  {Sch{\"a}fer},\ and\ \citenamefont {Painter}}]{chen2023phonon}%
  \BibitemOpen
  \bibfield  {author} {\bibinfo {author} {\bibfnamefont {Mo}~\bibnamefont
  {Chen}}, \bibinfo {author} {\bibfnamefont {John~Clai}\ \bibnamefont {Owens}},
  \bibinfo {author} {\bibfnamefont {Harald}\ \bibnamefont {Putterman}},
  \bibinfo {author} {\bibfnamefont {Max}\ \bibnamefont {Sch{\"a}fer}}, \ and\
  \bibinfo {author} {\bibfnamefont {Oskar}\ \bibnamefont {Painter}},\
  }\bibfield  {title} {\enquote {\bibinfo {title} {Phonon engineering of
  atomic-scale defects in superconducting quantum circuits},}\ }\href
  {https://doi.org/10.48550/arXiv.2310.03929} {\bibfield  {journal} {\bibinfo
  {journal} {arXiv preprint arXiv:2310.03929}\ } (\bibinfo {year}
  {2023})}\BibitemShut {NoStop}%
\bibitem [{sup()}]{supp}%
  \BibitemOpen
  \href@noop {} {\ }\bibinfo {note} {See supplemental materials at xxx for more
  details and data of the experimental device and setup, discrete dispersive
  frequency shift fitting, TLS control at the zero-flux point, joint TLS-parity
  measurement and discussions on random telegraphing noise in superconducting
  qubits.}\BibitemShut {Stop}%
\bibitem [{\citenamefont {Lisenfeld}\ \emph {et~al.}(2019)\citenamefont
  {Lisenfeld}, \citenamefont {Bilmes}, \citenamefont {Megrant}, \citenamefont
  {Barends}, \citenamefont {Kelly}, \citenamefont {Klimov}, \citenamefont
  {Weiss}, \citenamefont {Martinis},\ and\ \citenamefont
  {Ustinov}}]{lisenfeld2019electric}%
  \BibitemOpen
  \bibfield  {author} {\bibinfo {author} {\bibfnamefont {J{\"u}rgen}\
  \bibnamefont {Lisenfeld}}, \bibinfo {author} {\bibfnamefont {Alexander}\
  \bibnamefont {Bilmes}}, \bibinfo {author} {\bibfnamefont {Anthony}\
  \bibnamefont {Megrant}}, \bibinfo {author} {\bibfnamefont {Rami}\
  \bibnamefont {Barends}}, \bibinfo {author} {\bibfnamefont {Julian}\
  \bibnamefont {Kelly}}, \bibinfo {author} {\bibfnamefont {Paul}\ \bibnamefont
  {Klimov}}, \bibinfo {author} {\bibfnamefont {Georg}\ \bibnamefont {Weiss}},
  \bibinfo {author} {\bibfnamefont {John~M}\ \bibnamefont {Martinis}}, \ and\
  \bibinfo {author} {\bibfnamefont {Alexey~V}\ \bibnamefont {Ustinov}},\
  }\bibfield  {title} {\enquote {\bibinfo {title} {Electric field spectroscopy
  of material defects in transmon qubits},}\ }\href
  {https://doi.org/10.1038/s41534-019-0224-1} {\bibfield  {journal} {\bibinfo
  {journal} {npj Quantum Information}\ }\textbf {\bibinfo {volume} {5}},\
  \bibinfo {pages} {105} (\bibinfo {year} {2019})}\BibitemShut {NoStop}%
\bibitem [{\citenamefont {Odeh}\ \emph {et~al.}(2023)\citenamefont {Odeh},
  \citenamefont {Godeneli}, \citenamefont {Li}, \citenamefont {Tangirala},
  \citenamefont {Zhou}, \citenamefont {Zhang}, \citenamefont {Zhang},\ and\
  \citenamefont {Sipahigil}}]{odeh2023non}%
  \BibitemOpen
  \bibfield  {author} {\bibinfo {author} {\bibfnamefont {Mutasem}\ \bibnamefont
  {Odeh}}, \bibinfo {author} {\bibfnamefont {Kadircan}\ \bibnamefont
  {Godeneli}}, \bibinfo {author} {\bibfnamefont {Eric}\ \bibnamefont {Li}},
  \bibinfo {author} {\bibfnamefont {Rohin}\ \bibnamefont {Tangirala}}, \bibinfo
  {author} {\bibfnamefont {Haoxin}\ \bibnamefont {Zhou}}, \bibinfo {author}
  {\bibfnamefont {Xueyue}\ \bibnamefont {Zhang}}, \bibinfo {author}
  {\bibfnamefont {Zi-Huai}\ \bibnamefont {Zhang}}, \ and\ \bibinfo {author}
  {\bibfnamefont {Alp}\ \bibnamefont {Sipahigil}},\ }\bibfield  {title}
  {\enquote {\bibinfo {title} {Non-markovian dynamics of a superconducting
  qubit in a phononic bandgap},}\ }\href
  {https://doi.org/10.48550/arXiv.2312.01031} {\bibfield  {journal} {\bibinfo
  {journal} {arXiv preprint arXiv:2312.01031}\ } (\bibinfo {year}
  {2023})}\BibitemShut {NoStop}%
\bibitem [{\citenamefont {Bilmes}\ \emph {et~al.}(2017)\citenamefont {Bilmes},
  \citenamefont {Zanker}, \citenamefont {Heimes}, \citenamefont {Marthaler},
  \citenamefont {Sch\"on}, \citenamefont {Weiss}, \citenamefont {Ustinov},\
  and\ \citenamefont {Lisenfeld}}]{Bilmes2017electronic}%
  \BibitemOpen
  \bibfield  {author} {\bibinfo {author} {\bibfnamefont {Alexander}\
  \bibnamefont {Bilmes}}, \bibinfo {author} {\bibfnamefont {Sebastian}\
  \bibnamefont {Zanker}}, \bibinfo {author} {\bibfnamefont {Andreas}\
  \bibnamefont {Heimes}}, \bibinfo {author} {\bibfnamefont {Michael}\
  \bibnamefont {Marthaler}}, \bibinfo {author} {\bibfnamefont {Gerd}\
  \bibnamefont {Sch\"on}}, \bibinfo {author} {\bibfnamefont {Georg}\
  \bibnamefont {Weiss}}, \bibinfo {author} {\bibfnamefont {Alexey~V.}\
  \bibnamefont {Ustinov}}, \ and\ \bibinfo {author} {\bibfnamefont {J\"urgen}\
  \bibnamefont {Lisenfeld}},\ }\bibfield  {title} {\enquote {\bibinfo {title}
  {Electronic decoherence of two-level systems in a josephson junction},}\
  }\href {\doibase 10.1103/PhysRevB.96.064504} {\bibfield  {journal} {\bibinfo
  {journal} {Phys. Rev. B}\ }\textbf {\bibinfo {volume} {96}},\ \bibinfo
  {pages} {064504} (\bibinfo {year} {2017})}\BibitemShut {NoStop}%
\end{thebibliography}

\begin{thebibliography}{13}%
\makeatletter
\providecommand \@ifxundefined [1]{%
 \@ifx{#1\undefined}
}%
\providecommand \@ifnum [1]{%
 \ifnum #1\expandafter \@firstoftwo
 \else \expandafter \@secondoftwo
 \fi
}%
\providecommand \@ifx [1]{%
 \ifx #1\expandafter \@firstoftwo
 \else \expandafter \@secondoftwo
 \fi
}%
\providecommand \natexlab [1]{#1}%
\providecommand \enquote  [1]{``#1''}%
\providecommand \bibnamefont  [1]{#1}%
\providecommand \bibfnamefont [1]{#1}%
\providecommand \citenamefont [1]{#1}%
\providecommand \href@noop [0]{\@secondoftwo}%
\providecommand \href [0]{\begingroup \@sanitize@url \@href}%
\providecommand \@href[1]{\@@startlink{#1}\@@href}%
\providecommand \@@href[1]{\endgroup#1\@@endlink}%
\providecommand \@sanitize@url [0]{\catcode `\\12\catcode `\$12\catcode
  `\&12\catcode `\#12\catcode `\^12\catcode `\_12\catcode `\%12\relax}%
\providecommand \@@startlink[1]{}%
\providecommand \@@endlink[0]{}%
\providecommand \url  [0]{\begingroup\@sanitize@url \@url }%
\providecommand \@url [1]{\endgroup\@href {#1}{\urlprefix }}%
\providecommand \urlprefix  [0]{URL }%
\providecommand \Eprint [0]{\href }%
\providecommand \doibase [0]{http://dx.doi.org/}%
\providecommand \selectlanguage [0]{\@gobble}%
\providecommand \bibinfo  [0]{\@secondoftwo}%
\providecommand \bibfield  [0]{\@secondoftwo}%
\providecommand \translation [1]{[#1]}%
\providecommand \BibitemOpen [0]{}%
\providecommand \bibitemStop [0]{}%
\providecommand \bibitemNoStop [0]{.\EOS\space}%
\providecommand \EOS [0]{\spacefactor3000\relax}%
\providecommand \BibitemShut  [1]{\csname bibitem#1\endcsname}%
\let\auto@bib@innerbib\@empty
\bibitem [{\citenamefont {Paik}\ \emph {et~al.}(2011)\citenamefont {Paik},
  \citenamefont {Schuster}, \citenamefont {Bishop}, \citenamefont {Kirchmair},
  \citenamefont {Catelani}, \citenamefont {Sears}, \citenamefont {Johnson},
  \citenamefont {Reagor}, \citenamefont {Frunzio}, \citenamefont {Glazman},
  \citenamefont {Girvin}, \citenamefont {Devoret},\ and\ \citenamefont
  {Schoelkopf}}]{sPaik2011}%
  \BibitemOpen
  \bibfield  {author} {\bibinfo {author} {\bibfnamefont {H.}~\bibnamefont
  {Paik}}, \bibinfo {author} {\bibfnamefont {D.~I.}\ \bibnamefont {Schuster}},
  \bibinfo {author} {\bibfnamefont {L.~S.}\ \bibnamefont {Bishop}}, \bibinfo
  {author} {\bibfnamefont {G.}~\bibnamefont {Kirchmair}}, \bibinfo {author}
  {\bibfnamefont {G.}~\bibnamefont {Catelani}}, \bibinfo {author}
  {\bibfnamefont {A.~P.}\ \bibnamefont {Sears}}, \bibinfo {author}
  {\bibfnamefont {B.~R.}\ \bibnamefont {Johnson}}, \bibinfo {author}
  {\bibfnamefont {M.~J.}\ \bibnamefont {Reagor}}, \bibinfo {author}
  {\bibfnamefont {L.}~\bibnamefont {Frunzio}}, \bibinfo {author} {\bibfnamefont
  {L.~I.}\ \bibnamefont {Glazman}}, \bibinfo {author} {\bibfnamefont {S.~M.}\
  \bibnamefont {Girvin}}, \bibinfo {author} {\bibfnamefont {M.~H.}\
  \bibnamefont {Devoret}}, \ and\ \bibinfo {author} {\bibfnamefont {R.~J.}\
  \bibnamefont {Schoelkopf}},\ }\href {\doibase 10.1103/PhysRevLett.107.240501}
  {\bibfield  {journal} {\bibinfo  {journal} {Phys. Rev. Lett.}\ }\textbf
  {\bibinfo {volume} {107}},\ \bibinfo {pages} {240501} (\bibinfo {year}
  {2011})}\BibitemShut {NoStop}%
\bibitem [{\citenamefont {Abdurakhimov}\ \emph {et~al.}(2022)\citenamefont
  {Abdurakhimov}, \citenamefont {Mahboob}, \citenamefont {Toida}, \citenamefont
  {Kakuyanagi}, \citenamefont {Matsuzaki},\ and\ \citenamefont
  {Saito}}]{sAbdurakhimov222Identification}%
  \BibitemOpen
  \bibfield  {author} {\bibinfo {author} {\bibfnamefont {L.~V.}\ \bibnamefont
  {Abdurakhimov}}, \bibinfo {author} {\bibfnamefont {I.}~\bibnamefont
  {Mahboob}}, \bibinfo {author} {\bibfnamefont {H.}~\bibnamefont {Toida}},
  \bibinfo {author} {\bibfnamefont {K.}~\bibnamefont {Kakuyanagi}}, \bibinfo
  {author} {\bibfnamefont {Y.}~\bibnamefont {Matsuzaki}}, \ and\ \bibinfo
  {author} {\bibfnamefont {S.}~\bibnamefont {Saito}},\ }\href {\doibase
  10.1103/PRXQuantum.3.040332} {\bibfield  {journal} {\bibinfo  {journal} {PRX
  Quantum}\ }\textbf {\bibinfo {volume} {3}},\ \bibinfo {pages} {040332}
  (\bibinfo {year} {2022})}\BibitemShut {NoStop}%
\bibitem [{\citenamefont {Groszkowski}\ and\ \citenamefont
  {Koch}(2021)}]{sgroszkowski2021scqubits}%
  \BibitemOpen
  \bibfield  {author} {\bibinfo {author} {\bibfnamefont {P.}~\bibnamefont
  {Groszkowski}}\ and\ \bibinfo {author} {\bibfnamefont {J.}~\bibnamefont
  {Koch}},\ }\href {https://doi.org/10.22331/q-2021-11-17-583} {\bibfield
  {journal} {\bibinfo  {journal} {Quantum}\ }\textbf {\bibinfo {volume} {5}},\
  \bibinfo {pages} {583} (\bibinfo {year} {2021})}\BibitemShut {NoStop}%
\bibitem [{\citenamefont {M{\"u}ller}\ \emph {et~al.}(2019)\citenamefont
  {M{\"u}ller}, \citenamefont {Cole},\ and\ \citenamefont
  {Lisenfeld}}]{smuller2019towards}%
  \BibitemOpen
  \bibfield  {author} {\bibinfo {author} {\bibfnamefont {C.}~\bibnamefont
  {M{\"u}ller}}, \bibinfo {author} {\bibfnamefont {J.~H.}\ \bibnamefont
  {Cole}}, \ and\ \bibinfo {author} {\bibfnamefont {J.}~\bibnamefont
  {Lisenfeld}},\ }\href
  {https://iopscience.iop.org/article/10.1088/1361-6633/ab3a7e} {\bibfield
  {journal} {\bibinfo  {journal} {Reports on Progress in Physics}\ }\textbf
  {\bibinfo {volume} {82}},\ \bibinfo {pages} {124501} (\bibinfo {year}
  {2019})}\BibitemShut {NoStop}%
\bibitem [{\citenamefont {Schl\"or}\ \emph {et~al.}(2019)\citenamefont
  {Schl\"or}, \citenamefont {Lisenfeld}, \citenamefont {M\"uller},
  \citenamefont {Bilmes}, \citenamefont {Schneider}, \citenamefont {Pappas},
  \citenamefont {Ustinov},\ and\ \citenamefont
  {Weides}}]{sSteffen2019Correlating}%
  \BibitemOpen
  \bibfield  {author} {\bibinfo {author} {\bibfnamefont {S.}~\bibnamefont
  {Schl\"or}}, \bibinfo {author} {\bibfnamefont {J.}~\bibnamefont {Lisenfeld}},
  \bibinfo {author} {\bibfnamefont {C.}~\bibnamefont {M\"uller}}, \bibinfo
  {author} {\bibfnamefont {A.}~\bibnamefont {Bilmes}}, \bibinfo {author}
  {\bibfnamefont {A.}~\bibnamefont {Schneider}}, \bibinfo {author}
  {\bibfnamefont {D.~P.}\ \bibnamefont {Pappas}}, \bibinfo {author}
  {\bibfnamefont {A.~V.}\ \bibnamefont {Ustinov}}, \ and\ \bibinfo {author}
  {\bibfnamefont {M.}~\bibnamefont {Weides}},\ }\href {\doibase
  10.1103/PhysRevLett.123.190502} {\bibfield  {journal} {\bibinfo  {journal}
  {Phys. Rev. Lett.}\ }\textbf {\bibinfo {volume} {123}},\ \bibinfo {pages}
  {190502} (\bibinfo {year} {2019})}\BibitemShut {NoStop}%
\bibitem [{\citenamefont {Levine}\ \emph {et~al.}(2023)\citenamefont {Levine},
  \citenamefont {Haim}, \citenamefont {Hung}, \citenamefont {Alidoust},
  \citenamefont {Kalaee}, \citenamefont {DeLorenzo}, \citenamefont {Wollack},
  \citenamefont {Arriola}, \citenamefont {Khalajhedayati}, \citenamefont
  {Vaknin} \emph {et~al.}}]{slevine2023demonstrating}%
  \BibitemOpen
  \bibfield  {author} {\bibinfo {author} {\bibfnamefont {H.}~\bibnamefont
  {Levine}}, \bibinfo {author} {\bibfnamefont {A.}~\bibnamefont {Haim}},
  \bibinfo {author} {\bibfnamefont {J.~S.}\ \bibnamefont {Hung}}, \bibinfo
  {author} {\bibfnamefont {N.}~\bibnamefont {Alidoust}}, \bibinfo {author}
  {\bibfnamefont {M.}~\bibnamefont {Kalaee}}, \bibinfo {author} {\bibfnamefont
  {L.}~\bibnamefont {DeLorenzo}}, \bibinfo {author} {\bibfnamefont {E.~A.}\
  \bibnamefont {Wollack}}, \bibinfo {author} {\bibfnamefont {P.~A.}\
  \bibnamefont {Arriola}}, \bibinfo {author} {\bibfnamefont {A.}~\bibnamefont
  {Khalajhedayati}}, \bibinfo {author} {\bibfnamefont {Y.}~\bibnamefont
  {Vaknin}},  \emph {et~al.},\ }\href
  {https://doi.org/10.48550/arXiv.2307.08737} {\bibfield  {journal} {\bibinfo
  {journal} {arXiv preprint arXiv:2307.08737}\ } (\bibinfo {year}
  {2023})}\BibitemShut {NoStop}%
\bibitem [{\citenamefont {Gertler}\ \emph {et~al.}(2021)\citenamefont
  {Gertler}, \citenamefont {Baker}, \citenamefont {Li}, \citenamefont {Shirol},
  \citenamefont {Koch},\ and\ \citenamefont {Wang}}]{sgertler2021protecting}%
  \BibitemOpen
  \bibfield  {author} {\bibinfo {author} {\bibfnamefont {J.~M.}\ \bibnamefont
  {Gertler}}, \bibinfo {author} {\bibfnamefont {B.}~\bibnamefont {Baker}},
  \bibinfo {author} {\bibfnamefont {J.}~\bibnamefont {Li}}, \bibinfo {author}
  {\bibfnamefont {S.}~\bibnamefont {Shirol}}, \bibinfo {author} {\bibfnamefont
  {J.}~\bibnamefont {Koch}}, \ and\ \bibinfo {author} {\bibfnamefont
  {C.}~\bibnamefont {Wang}},\ }\href
  {https://doi.org/10.1038/s41586-021-03257-0} {\bibfield  {journal} {\bibinfo
  {journal} {Nature}\ }\textbf {\bibinfo {volume} {590}},\ \bibinfo {pages}
  {243} (\bibinfo {year} {2021})}\BibitemShut {NoStop}%
\bibitem [{\citenamefont {Burnett}\ \emph {et~al.}(2014)\citenamefont
  {Burnett}, \citenamefont {Faoro}, \citenamefont {Wisby}, \citenamefont
  {Gurtovoi}, \citenamefont {Chernykh}, \citenamefont {Mikhailov},
  \citenamefont {Tulin}, \citenamefont {Shaikhaidarov}, \citenamefont
  {Antonov}, \citenamefont {Meeson} \emph {et~al.}}]{sburnett2014evidence}%
  \BibitemOpen
  \bibfield  {author} {\bibinfo {author} {\bibfnamefont {J.}~\bibnamefont
  {Burnett}}, \bibinfo {author} {\bibfnamefont {L.}~\bibnamefont {Faoro}},
  \bibinfo {author} {\bibfnamefont {I.}~\bibnamefont {Wisby}}, \bibinfo
  {author} {\bibfnamefont {V.~L.}\ \bibnamefont {Gurtovoi}}, \bibinfo {author}
  {\bibfnamefont {A.~V.}\ \bibnamefont {Chernykh}}, \bibinfo {author}
  {\bibfnamefont {G.~M.}\ \bibnamefont {Mikhailov}}, \bibinfo {author}
  {\bibfnamefont {V.~A.}\ \bibnamefont {Tulin}}, \bibinfo {author}
  {\bibfnamefont {R.}~\bibnamefont {Shaikhaidarov}}, \bibinfo {author}
  {\bibfnamefont {V.}~\bibnamefont {Antonov}}, \bibinfo {author} {\bibfnamefont
  {P.~J.}\ \bibnamefont {Meeson}},  \emph {et~al.},\ }\href
  {https://doi.org/10.1038/ncomms5119} {\bibfield  {journal} {\bibinfo
  {journal} {Nature Communications}\ }\textbf {\bibinfo {volume} {5}},\
  \bibinfo {pages} {4119} (\bibinfo {year} {2014})}\BibitemShut {NoStop}%
\bibitem [{\citenamefont {Klimov}\ \emph {et~al.}(2018)\citenamefont {Klimov},
  \citenamefont {Kelly}, \citenamefont {Chen}, \citenamefont {Neeley},
  \citenamefont {Megrant}, \citenamefont {Burkett}, \citenamefont {Barends},
  \citenamefont {Arya}, \citenamefont {Chiaro}, \citenamefont {Chen} \emph
  {et~al.}}]{sKlimov2018Fluctuations}%
  \BibitemOpen
  \bibfield  {author} {\bibinfo {author} {\bibfnamefont {P.}~\bibnamefont
  {Klimov}}, \bibinfo {author} {\bibfnamefont {J.}~\bibnamefont {Kelly}},
  \bibinfo {author} {\bibfnamefont {Z.}~\bibnamefont {Chen}}, \bibinfo {author}
  {\bibfnamefont {M.}~\bibnamefont {Neeley}}, \bibinfo {author} {\bibfnamefont
  {A.}~\bibnamefont {Megrant}}, \bibinfo {author} {\bibfnamefont
  {B.}~\bibnamefont {Burkett}}, \bibinfo {author} {\bibfnamefont
  {R.}~\bibnamefont {Barends}}, \bibinfo {author} {\bibfnamefont
  {K.}~\bibnamefont {Arya}}, \bibinfo {author} {\bibfnamefont {B.}~\bibnamefont
  {Chiaro}}, \bibinfo {author} {\bibfnamefont {Y.}~\bibnamefont {Chen}},  \emph
  {et~al.},\ }\href {https://link.aps.org/doi/10.1103/PhysRevLett.121.090502}
  {\bibfield  {journal} {\bibinfo  {journal} {Physical review letters}\
  }\textbf {\bibinfo {volume} {121}},\ \bibinfo {pages} {090502} (\bibinfo
  {year} {2018})}\BibitemShut {NoStop}%
\bibitem [{\citenamefont {Neeley}\ \emph {et~al.}(2008)\citenamefont {Neeley},
  \citenamefont {Ansmann}, \citenamefont {Bialczak}, \citenamefont {Hofheinz},
  \citenamefont {Katz}, \citenamefont {Lucero}, \citenamefont {O’connell},
  \citenamefont {Wang}, \citenamefont {Cleland},\ and\ \citenamefont
  {Martinis}}]{sneeley2008process}%
  \BibitemOpen
  \bibfield  {author} {\bibinfo {author} {\bibfnamefont {M.}~\bibnamefont
  {Neeley}}, \bibinfo {author} {\bibfnamefont {M.}~\bibnamefont {Ansmann}},
  \bibinfo {author} {\bibfnamefont {R.~C.}\ \bibnamefont {Bialczak}}, \bibinfo
  {author} {\bibfnamefont {M.}~\bibnamefont {Hofheinz}}, \bibinfo {author}
  {\bibfnamefont {N.}~\bibnamefont {Katz}}, \bibinfo {author} {\bibfnamefont
  {E.}~\bibnamefont {Lucero}}, \bibinfo {author} {\bibfnamefont
  {A.}~\bibnamefont {O’connell}}, \bibinfo {author} {\bibfnamefont
  {H.}~\bibnamefont {Wang}}, \bibinfo {author} {\bibfnamefont {A.~N.}\
  \bibnamefont {Cleland}}, \ and\ \bibinfo {author} {\bibfnamefont {J.~M.}\
  \bibnamefont {Martinis}},\ }\href {https://doi.org/10.1038/nphys972}
  {\bibfield  {journal} {\bibinfo  {journal} {Nature Physics}\ }\textbf
  {\bibinfo {volume} {4}},\ \bibinfo {pages} {523} (\bibinfo {year}
  {2008})}\BibitemShut {NoStop}%
\bibitem [{\citenamefont {Shalibo}\ \emph {et~al.}(2010)\citenamefont
  {Shalibo}, \citenamefont {Rofe}, \citenamefont {Shwa}, \citenamefont
  {Zeides}, \citenamefont {Neeley}, \citenamefont {Martinis},\ and\
  \citenamefont {Katz}}]{sShalibo2010Lifetime}%
  \BibitemOpen
  \bibfield  {author} {\bibinfo {author} {\bibfnamefont {Y.}~\bibnamefont
  {Shalibo}}, \bibinfo {author} {\bibfnamefont {Y.}~\bibnamefont {Rofe}},
  \bibinfo {author} {\bibfnamefont {D.}~\bibnamefont {Shwa}}, \bibinfo {author}
  {\bibfnamefont {F.}~\bibnamefont {Zeides}}, \bibinfo {author} {\bibfnamefont
  {M.}~\bibnamefont {Neeley}}, \bibinfo {author} {\bibfnamefont {J.~M.}\
  \bibnamefont {Martinis}}, \ and\ \bibinfo {author} {\bibfnamefont
  {N.}~\bibnamefont {Katz}},\ }\href {\doibase 10.1103/PhysRevLett.105.177001}
  {\bibfield  {journal} {\bibinfo  {journal} {Phys. Rev. Lett.}\ }\textbf
  {\bibinfo {volume} {105}},\ \bibinfo {pages} {177001} (\bibinfo {year}
  {2010})}\BibitemShut {NoStop}%
\bibitem [{\citenamefont {Lisenfeld}\ \emph {et~al.}(2010)\citenamefont
  {Lisenfeld}, \citenamefont {M\"uller}, \citenamefont {Cole}, \citenamefont
  {Bushev}, \citenamefont {Lukashenko}, \citenamefont {Shnirman},\ and\
  \citenamefont {Ustinov}}]{sLisenfeld2010Measuring}%
  \BibitemOpen
  \bibfield  {author} {\bibinfo {author} {\bibfnamefont {J.}~\bibnamefont
  {Lisenfeld}}, \bibinfo {author} {\bibfnamefont {C.}~\bibnamefont {M\"uller}},
  \bibinfo {author} {\bibfnamefont {J.~H.}\ \bibnamefont {Cole}}, \bibinfo
  {author} {\bibfnamefont {P.}~\bibnamefont {Bushev}}, \bibinfo {author}
  {\bibfnamefont {A.}~\bibnamefont {Lukashenko}}, \bibinfo {author}
  {\bibfnamefont {A.}~\bibnamefont {Shnirman}}, \ and\ \bibinfo {author}
  {\bibfnamefont {A.~V.}\ \bibnamefont {Ustinov}},\ }\href {\doibase
  10.1103/PhysRevLett.105.230504} {\bibfield  {journal} {\bibinfo  {journal}
  {Phys. Rev. Lett.}\ }\textbf {\bibinfo {volume} {105}},\ \bibinfo {pages}
  {230504} (\bibinfo {year} {2010})}\BibitemShut {NoStop}%
\bibitem [{\citenamefont {Burnett}\ \emph {et~al.}(2019)\citenamefont
  {Burnett}, \citenamefont {Bengtsson}, \citenamefont {Scigliuzzo},
  \citenamefont {Niepce}, \citenamefont {Kudra}, \citenamefont {Delsing},\ and\
  \citenamefont {Bylander}}]{sburnett2019decoherence}%
  \BibitemOpen
  \bibfield  {author} {\bibinfo {author} {\bibfnamefont {J.~J.}\ \bibnamefont
  {Burnett}}, \bibinfo {author} {\bibfnamefont {A.}~\bibnamefont {Bengtsson}},
  \bibinfo {author} {\bibfnamefont {M.}~\bibnamefont {Scigliuzzo}}, \bibinfo
  {author} {\bibfnamefont {D.}~\bibnamefont {Niepce}}, \bibinfo {author}
  {\bibfnamefont {M.}~\bibnamefont {Kudra}}, \bibinfo {author} {\bibfnamefont
  {P.}~\bibnamefont {Delsing}}, \ and\ \bibinfo {author} {\bibfnamefont
  {J.}~\bibnamefont {Bylander}},\ }\href
  {https://doi.org/10.1038/s41534-019-0168-5} {\bibfield  {journal} {\bibinfo
  {journal} {npj Quantum Information}\ }\textbf {\bibinfo {volume} {5}},\
  \bibinfo {pages} {54} (\bibinfo {year} {2019})}\BibitemShut {NoStop}%
\end{thebibliography}
\end{document}